\documentclass[rmp,aps,twocolumn]{revtex4-2}
\usepackage{amsmath}
\usepackage{amssymb} 
\usepackage{mathtools}
\usepackage{graphicx}
\usepackage{amsfonts}
\usepackage{color}
\usepackage{microtype} 
\usepackage{subcaption} 
\usepackage{threeparttable}
\usepackage{verbatim} 
\usepackage{braket}
\usepackage{physics}
\usepackage{bm}

\usepackage{float}

\makeindex
\usepackage{amsthm}
\usepackage{thmtools}

\usepackage{tikz}
\usetikzlibrary{positioning, calc, shapes.geometric, backgrounds,fit}

\usepackage{titlesec}

\makeatletter
\renewcommand{\p@subsection}{}
\renewcommand{\p@subsubsection}{}
\makeatother
\titleformat*{\paragraph}{\bfseries}

\usepackage{hyperref} 
\usepackage[capitalize]{cleveref} 
\hypersetup{
	colorlinks = true,
	linkcolor =blue,
	citecolor=blue, 
	urlcolor=blue 
}

\newtheorem{remark}{Remark}[section]
\newtheorem{theorem}{Theorem}[section]

\theoremstyle{definition} 
\newtheorem{definition}{Definition}[section]

\newcommand{\Renyi}{R\'{e}nyi} 
\newcommand{\mbf}[1]{\mathbf{#1}}

\newcommand{\term}[1]{\textbf{#1}} 
\newcommand{\defvar}{\coloneqq} 

\newcommand{\pending}[1]{{#1}} 

\newcommand{\ideal}{ideal}

\newcommand{\infPOVM}[1][]{\vec{\Gamma}_{#1}} 
\newcommand{\finPOVM}[1][]{\vec{F}_{#1}} 

\newcommand{\Hmin}{H_\mathrm{min}}

\newcommand{\eps}{\varepsilon} 
\newcommand{\esecure}{\eps^\mathrm{secure}}
\newcommand{\ecorr}{\eps^\mathrm{correct}}
\newcommand{\esecret}{\eps^\mathrm{secret}}

\newcommand{\eAT}{\eps_\mathrm{AT}} 
\newcommand{\esmooth}{\overline{\eps}} 
\newcommand{\ecom}{\eps^\mathrm{com}}

\newcommand{\Oacc}{\Omega_\mathrm{acc}} 
\newcommand{\Ol}{\Omega_{\lkey}} 

\newcommand{\Odiff}{\Omega_\mathrm{diff}}
\newcommand{\OEV}{\Omega_\mathrm{EV}}

\newcommand{\CEC}{C_\text{EC}}
\newcommand{\leak}{\lambda_\text{EC}}
\newcommand{\lkey}{\ell} 
\newcommand{\lfixed}{\ell_{\mathrm{fixed}}} 
\newcommand{\PAstring}{\mbf{S}} 
\newcommand{\KA}{K_A} 
\newcommand{\KB}{K_B} 
\newcommand{\En}{\mbf{E}}
\newcommand{\EprePA}{E_{\mathrm{prePA}}} 
\newcommand{\Efinal}{E_{\mathrm{fin}}} 

\newcommand{\nK}{n_K}
\newcommand{\nT}{n_T}

\tikzset{
  pics/detector/.style args={#1,#2,#3}{
    code={
      \draw[line width=0.5mm] (0,-1) -- (0,1);
      \draw[line width=0.5mm] (0,1) -- ++(0.1,0);
      \draw[line width=0.5mm] (0,-1) -- ++(0.1,0);
      \draw[line width=0.5mm] (0.1,1) arc (90:-90:1); 
      \node[#3, align=center] (#1) at (0.5,0) {#2};
    }
  }
}

\tikzset{
  pics/source/.style args={#1,#2,#3,#4}{
    code={
    \coordinate (A) at (0,1);
    \coordinate (B) at (0,-1);
    \coordinate (C) at (3,-1);
    \coordinate (D) at (4,0);
    \coordinate (E) at (3,1);
    
    \draw[line width=0.4mm] (A) -- (B) -- (C) -- (D) -- (E) -- cycle; 
    \node[transform shape,scale=#4,#3, align=center] (#1) at (1.5,0) {#2};
    }
  }
}

\tikzstyle{process} = [rectangle, line width=0.3mm, minimum width=1cm, minimum height=1cm, text centered, text width=1cm, draw=black]

\renewcommand{\selectlanguage}[1]{} 
\UseRawInputEncoding 

\begin{document}
	\title{Security proofs for practical QKD: variations, techniques, gaps and limitations}

 \author{Devashish Tupkary}
\email{djtupkary@uwaterloo.ca}
	 \affiliation{Institute for Quantum Computing and Department of Physics and Astronomy, University of Waterloo, Waterloo, Ontario, Canada, N2L 3G1}

  \author{Ernest Y.-Z. Tan}
	 \email{yzetan@uwaterloo.ca}
	 \affiliation{Institute for Quantum Computing and Department of Physics and Astronomy, University of Waterloo, Waterloo, Ontario, Canada, N2L 3G1}

 \author{Shlok Nahar}
	 \email{sanahar@uwaterloo.ca}
	 \affiliation{Institute for Quantum Computing and Department of Physics and Astronomy, University of Waterloo, Waterloo, Ontario, Canada, N2L 3G1}

 \author{Lars Kamin}
\email{lars.kamin@outlook.com}
	 \affiliation{Institute for Quantum Computing and Department of Physics and Astronomy, University of Waterloo, Waterloo, Ontario, Canada, N2L 3G1}

	 \author{Norbert L\"utkenhaus}
	 \email{lutkenhaus.office@uwaterloo.ca}
	 \affiliation{Institute for Quantum Computing and Department of Physics and Astronomy, University of Waterloo, Waterloo, Ontario, Canada, N2L 3G1}

\begin{abstract}
We review the current status of security proofs for practical decoy-state Quantum Key Distribution using the BB84 protocol, focusing on optical implementations with weak coherent pulses and threshold photodetectors. The primary aim of this review is to highlight gaps in the existing literature. Such gaps may arise, for instance, from mismatches between detailed protocol specifications and proof technique elements, reliance on earlier results based on different assumptions, or protocol choices that overlook real world requirements. While substantial progress has been made, our overview draws attention to the details that still demand careful attention.
\end{abstract}

\begin{titlepage}
    \maketitle     
    \tableofcontents 
\end{titlepage}

\section{Introduction}
Quantum Key Distribution (QKD) is a maturing technology. Since its initial conception in 1984 with the Bennett-Brassard BB84 \cite{Bennett_2014} protocol, QKD has progressed from theoretical foundations to experimental implementations, and eventually to commercial deployment. On the protocol side, key milestones include the use of weak coherent laser pulses and the development of decoy-state methods. On the theoretical side, important advances include the formalization of a rigorous security definition for QKD protocols and the development of proof techniques -- initially for asymptotic scenarios, and later extended to the scenario of finite block length of signals. A further development was the recognition of the quantum optical nature of realistic implementations that relies on optical modes rather than abstract qubits as fundamental objects of description. 

\subsection{Motivation}
QKD is now a well-established field, with widely studied practical protocols such as the decoy-state BB84 protocol using weak coherent pulses \cite{Hwang_qkdhighloss_2003,Lo_decoystate_2005,Wang_beating_2005,Ma_practicaldecoy_2005}. There have also been a variety of experimental implementations of these protocols \cite{Zhao_experimentalQKD_withdecoystates,Rosenberg_LongDistance_2007,manderbarch_experimentalfreespaceqkd_2007,Peng_experimental_2007,boaron_secure_2018}. As the community begins to define the certification landscape of QKD technologies, it becomes clear that numerous design choices must be made in specifying a concrete QKD protocol. Importantly, complete security proofs must accurately account for these precise choices. Historically, proof techniques have evolved in myriad ways, resulting in a body of work that is often a patchwork of partial results. While each contribution addresses a critical component of a full security proof, the patches do not always integrate seamlessly -- either due to mismatched assumptions across different works, or because the results apply to differing protocol specifications. 

Thus, as the field moves towards certification, it is important to develop rigorous and comprehensive security proofs for practical QKD protocols. In particular, we need security proofs for practical QKD protocols that
\begin{itemize}
    \item specify complete protocols that are relevant to practical implementations,
    \item specify the exact assumptions about the devices used in the protocol,
    \item specify the exact security definition,
    \item and deliver a mathematical proof that the given protocol with the given assumptions satisfies the given security definition with given security parameters.
\end{itemize}
Moreover, the proof should be such that
\begin{itemize}
    \item the validity of the proof can be verified by mathematically trained evaluators,
    \item the achievable secure key generation rate should be practically relevant, not unnecessarily restricted by proof technique or protocol choices,
    \item the proof is as modular as possible, so that it can be confidently modified to protocol choices other than the originally specified ones,
    \item it uses proof techniques that allow the extension of the protocol security to implementation security, where implementation imperfections can be taken into account.
\end{itemize}
From this review it will become clear that currently no single publication satisfies all these criteria \cite{bundPositionPaper} \footnote{ After the initial arXiv posting of this work, Refs.~\cite{mizutani2025protocolleveldescriptionselfcontainedsecurity,inprep_BDR3}  have appeared on arXiv, promising a rigorous security proof for a fully specified decoy-state BB84 protocol. In doing so, these works attempt to meet a significantly higher standard than the works considered in this review. They are undergoing scrutiny from the community, and will require time before they are thoroughly assessed.}, though a tremendous progress of knowledge and insight has been achieved over the last few decades. Thus, the task of assembling the knowledge to provide such a protocol specification and associated security proof is still left open.

\subsection{Focus of this review} \label{subsec:focusofreview}
Numerous reviews of QKD exist, covering the nature of the technology, its implementation, experimental progress, and practical devices. For example, Ref.~\cite{portmann_security_2022} provides a pedagogical overview of the composable security framework for QKD; Refs.~\cite{Zapatero_review_di_2023,Primaatmaja2023securityofdevice} focuses on recent advances in device-independent QKD;
Ref.~\cite{Xu_review_2020} examines imperfections in QKD implementations; and Ref.~\cite{zapatero2023implementationsecurityquantumkey} offers a succinct discussion on implementation security. 
For foundational reviews from earlier stages of the field, see Refs.~\cite{Scarani_2009,Gisin_review_2002}.

The focus of this review is  different. It is heavily centered on the \textit{mathematical arguments and tools required to construct a rigorous security proof for QKD}. In particular, it aims to identify and analyze the existing gaps in the results pertaining to protocol security in the literature, and to highlight what would constitute a complete, modular, and verifiable proof of security for a realistic QKD protocol. Thus, we always assume that a mathematical model of the protocol implementation has already been specified, and that the goal is to prove security for this mathematical model. If one is instead interested in the security of a physical implementation, then the first step is to construct an appropriate mathematical model and verify that it reflects the real system to the desired level of accuracy. This modelling process is separate from the mathematical proof techniques discussed in this review. Only once such a mathematical model has been established does one turn to the tools and arguments presented here to construct a rigorous proof of security.

\subsection{Structure of this review}

This review is structured as follows: In \cref{sec:defs}, we set up common definitions and notations. In \cref{sec:structural_elements}, we describe the QKD protocol and highlight the variety of choices that can be made in its design. These design choices directly impact how the security proof is constructed and where potential gaps may arise. In \cref{sec:concepts}, we discuss the mathematical tools and arguments that underlie a QKD security proof, and we identify several common misconceptions and errors. \cref{sec:commonGaps} elaborates on these gaps and mistakes in existing proofs, while referring back to \cref{sec:concepts} for the conceptual details. In \cref{sec:families}, we present a critical review of several major works on security proofs for decoy-state BB84 (Refs.~\cite{lim_concise_2014,rusca_finite-key_2018,tupkary_phase_2024,wiesemann_consolidated_2024,hayashi_security_2014,kamin_finite-size_2024,kamin2025renyisecurityframeworkcoherent,Kamin2025}). We summarize the key contributions of each work while identifying their gaps, limitations, and open issues.

There are several proof techniques currently available to theorists for establishing QKD security. These can be broadly categorized into the following four classes:
\begin{enumerate}
\item Proofs based on entropic uncertainty relations (EUR) \cite{tomamichel_uncertainty_2011},
\item Proofs that employ the postselection technique \cite{nahar_postselection_2024,christandl_postselection_2009} followed by a security analysis against IID collective attacks,
\item Phase error correction-based proofs \cite{koashi_simple_2005},
\item Proofs relying on various entropy accumulation theorems (EAT) \cite{dupuis_entropy_2020,metger_generalised_2022,fawzi_additivity_2025,arqand_marginal_2025}.
\end{enumerate}

In \cref{sec:comparison}, we compare these proof approaches, candidly assessing their respective strengths and weaknesses (see \cref{tab:comparison}). This comparative analysis lays the groundwork for making informed choices regarding proof techniques in future security analyses.

\subsection{How to read this review}
This review is written in a modular fashion and is extensively cross-referenced, allowing for multiple reading paths depending on the reader's background and interests. Readers who are new to the field and wish to learn about security proofs are encouraged to begin with the structural elements of the QKD protocol in \cref{sec:structural_elements}, and the conceptual overview in \cref{sec:concepts}, before proceeding to the discussion of proof techniques in \cref{sec:families}. Readers already familiar with security proofs may find the discussion of common gaps in \cref{sec:commonGaps}, and the detailed examination of specific proofs in \cref{sec:families}, the most interesting. We expect the comparison of proof technique approaches in \cref{sec:comparison} to be of interest to both newcomers and experts alike.

\section{Definitions and Notation}
\label{sec:defs}

\subsection{Concepts from quantum information theory}
\label{subsec:basicquantum}

We do not aim to give a detailed introduction to the framework of quantum information theory within this work. However, we will need to refer to a number of concepts and properties from that framework. Here, we shall briefly outline the qualitative properties of some such concepts, deferring detailed definitions to suitable references --- see for instance Refs.~\cite{nielsen_quantum_2010,tomamichel_quantum_2016}. \\

\paragraph*{Quantum states.} A quantum state describing some register(s) $A$ is formalized as a positive semidefinite operator $\rho_A$, and we say that it is \term{normalized} if it has trace $1$, while it is \term{subnormalized} if its trace is at most $1$~\cite{tomamichel_quantum_2016}. However, we highlight that in this work we will mostly only refer to normalized states, only considering subnormalized states if explicitly specified. We may omit the subscript specifying the register(s) if it is clear from context. \\

\paragraph*{Quantum conditional states.} We will also often need to refer to the notion of a quantum state conditioned on some event. 
We defer the formal definitions to e.g.~Ref.~\cite[Section~2.2]{tomamichel_largely_2017}, here simply noting that this notion is formalized by requiring that the quantum state contains a classical register that determines whether the event occurred. Note that the conditional state can be defined in either a normalized or subnormalized manner --- the two possibilities are often denoted as $\rho_{A|\Omega}$ and $\rho_{A \land \Omega}$ respectively (given an event $\Omega$), though this work will mainly focus on normalized conditional states, as mentioned above.\\

\paragraph*{Trace distance.} The trace distance is a metric on the set of quantum states, which we shall denote as $d(\rho,\sigma)$ for any pair of quantum states $\rho,\sigma$. Here we follow the definition in e.g. Ref.~\cite[Definition 3.4]{tomamichel_quantum_2016}, which covers subnormalized states. Critically, trace distance has an operational implication in terms of quantifying how well the states can be distinguished by \emph{any} possible physical process, without computational limitations. 
\\

\paragraph*{Entropies.} In this work, we will refer to several notions of (conditional) entropy of a state $\rho_{AB}$, following the definitions in e.g. Ref.~\cite{tomamichel_quantum_2016}. We highlight the following in particular: 
\begin{itemize}
\item Von Neumann entropy $H(A|B)_\rho$
\item Min-entropy $\Hmin(A|B)_\rho$ 
\item Smooth min-entropy $\Hmin^{\esmooth}(A|B)_\rho$
\end{itemize}
Informally, these entropies can be viewed as describing how much ``uncertainty'' there is about the register $A$ from the perspective of a party holding register $B$. In particular, if $A$ is classical, the min-entropy $\Hmin(A|B)_\rho$ has an operational interpretation in terms of how well the value of $A$ can be guessed given access to the register $B$; the smooth min-entropy $\Hmin^{\esmooth}(A|B)_\rho$ is an extension of this concept that considers states in a ``neighbourhood'' of $\rho$.
We will also briefly mention {\Renyi} entropies (as described in e.g. Ref.~\cite{tomamichel_quantum_2016}) at some points, though we highlight that there are multiple distinct definitions of the concept, and when writing a security proof it is important to specify which definition is being used.

\subsection{Classes of QKD protocols}

In this work, we will mostly focus on what are known as \term{prepare-and-measure} (PM) QKD protocols, in which Alice prepares quantum states and sends them to Bob, who measures them. This is to be contrasted with \term{entanglement-based} (EB) QKD protocols, in which an entangled quantum state is shared between Alice and Bob, who measure their respective shares of the state.

Furthermore, in this work we will often separately discuss the following classes of QKD protocols:
\begin{itemize}
\item \term{Fixed-length protocols}: This refers to protocols that make a binary decision of whether to accept or abort, and output a final key of some fixed length $\lfixed$ (determined before the protocol begins) when they accept. 
\item \term{Variable-length protocols}: This refers to protocols where the length of the final key can be variable (possibly including zero), depending on the observations made during the protocol.
\end{itemize}

In principle, fixed-length protocols can in fact be viewed as a special case of variable-length protocols, where the only possible final key lengths are $0$ or the predetermined value $\lfixed$. The security definitions we present below for fixed-length and variable-length protocols are also all consistent with this interpretation. However, in this work it will often be convenient to discuss the two classes separately, as some proof techniques are better suited for fixed-length protocols specifically. 

\subsection{Security definitions}
\label{subsec:securitydefn}

\newcommand{\QKDblurb}{Consider a QKD protocol such that at the end of the protocol, Alice and Bob hold final keys $\KA$ and $\KB$ respectively, while Eve holds a final side-information register $\Efinal$.}

Current security proofs in QKD are mostly centred around a particular security definition involving trace distance. The justification behind this choice of security definition is that it yields \term{composable security}, which informally means that the keys generated from a QKD protocol can be safely used as part of a larger protocol.\footnote{A common form of such composition is if the keys generated by a QKD protocol are used for authentication in a subsequent QKD protocol. As discussed in Refs.~\cite{mullerquade2009composability,portmann_key_2014}, a composable security framework lets us conclude that for multiple QKD protocols composed in this manner, the overall security of the \emph{entire} sequence of protocols can be described in terms of the sum of the security parameters of the individual protocols. This does require the individual security parameters to be chosen sufficiently small, such that the sum of their values is still an acceptably small value.}
We defer a detailed discussion of this point to\footnote{These works present somewhat different perspectives on composable security, but they coincide in terms of the desired security definitions in terms of trace distance.} Refs.~\cite{part1,portmann_security_2022,benor2004universalcomposablesecurityquantum,broadbent_karvonen_2023}; here, we simply restate the security definitions from those works. 

\begin{definition}\label{def:security}
(Security)\footnote{This property is sometimes instead referred to as \term{soundness}, for instance in Ref.~\cite{portmann_security_2022}.} Let $\esecure\in[0,1]$. 
{\QKDblurb} 
Let 
$
\tau^\lkey_{\KA\KB} \defvar \sum_{k \in \{0,1\}^{\lkey}} \frac{1}{2^{\lkey}} \ketbra{kk}{kk}_{\KA \KB}
$
denote a maximally correlated key of length $\lkey$ across registers $\KA \KB$.

For a fixed-length protocol, we say it is $\esecure$-secure if the output state always satisfies
\begin{equation} \label{eq:securefixed}
\Pr[\Oacc]\, d\left(\rho_{\KA \KB \Efinal | \Oacc} \,,\,  \tau^{\lfixed}_{\KA\KB} \otimes\rho_{\Efinal | \Oacc}\right) \leq \esecure,
\end{equation}
where $\Oacc$ denotes the event that the protocol accepts. 

For a variable-length protocol, we say it is $\esecure$-secure if the output state always satisfies
\begin{equation}\label{eq:securevar}
\sum_{\lkey}
\Pr[\Ol]\, d\left(\rho_{\KA \KB \Efinal | \Ol} \,,\,  \tau^{\lkey}_{\KA\KB} \otimes\rho_{\Efinal | \Ol}\right) \leq \esecure,
\end{equation}
where $\Ol$ denotes the event that the protocol produces a key of length $\lkey$. 
\end{definition}

\begin{remark} \label{remark:securityconditionedonaccept}
We emphasize that the security definition does \emph{not} state that 
the trace distance term for the normalized conditional states in \cref{eq:securefixed} (for fixed-length protocols) is itself upper bounded by $\esecure$. This is because such an upper bound simply \emph{does not hold} in general; in particular, it is usually straightforward to construct attacks such that
\begin{equation}\label{eq:trdistcounterexample}
d\left(\rho_{\KA \KB \Efinal | \Oacc} \,,\,  \tau^{\lfixed}_{\KA\KB} \otimes\rho_{\Efinal | \Oacc}\right) \gg \esecure, 
\end{equation}
as we discuss later in \cref{subsec:estimating_entropy,subsubsec:securityconditionedonaccept}.
Similarly, for variable-length protocols, one cannot claim that the individual trace distance terms in \cref{eq:securevar} (excluding the $\Pr(\Omega_\ell)$ term) are all upper bounded by $\esecure$.

Critically, however, \cref{def:security} is already sufficient to ensure composable security, and hence this fact \emph{does not affect} the composable security of QKD protocols. We defer to Refs.~\cite{part1,portmann_security_2022} 
for more extensive discussion of this point.
\end{remark}

\subsubsection{Correctness and Secrecy}
As described in Refs.~\cite{part1,portmann_security_2022,benor2004universalcomposablesecurityquantum}, it is often convenient to instead study a pair of simpler conditions, referred to as \term{correctness} and \term{secrecy}:\footnote{The choice to use $\KA$ rather than $\KB$ in the secrecy definition is essentially a matter of convenience; analogous results can be derived under either choice.}
\begin{definition}\label{def:correctsecret}
(Correctness and secrecy) Let $\ecorr,\esecret\in[0,1]$. 
{\QKDblurb} 
Let 
$
\omega^\lkey_{\KA} \defvar \sum_{k \in \{0,1\}^{\lkey}} \frac{1}{2^{\lkey}} \ketbra{k}{k}_{\KA}
$
denote a uniformly random key of length $\lkey$ on register $\KA$.

For a fixed-length protocol, we say it is $\ecorr$-correct if the output state always satisfies
\begin{equation}\label{eq:correctfixed}
\Pr[\KA \neq \KB \text{ and protocol accepts}]\leq \ecorr,
\end{equation}
and we say it is $\esecret$-secret if the output state always satisfies
\begin{equation}\label{eq:secretfixed}
\Pr[\Oacc]\, d\left(\rho_{\KA \Efinal | \Oacc} \,,\, 
\omega^{\lfixed}_{\KA} \otimes\rho_{\Efinal | \Oacc}\right) \leq \esecret,
\end{equation}
where $\Oacc$ denotes the event that the protocol accepts. 

For a variable-length protocol, we say it is $\ecorr$-correct if the output state always satisfies
\begin{equation}\label{eq:correctvar}
\Pr[\KA \neq \KB \text{ and final key length is nonzero}]\leq \ecorr,
\end{equation}
and we say it is $\esecret$-secret if the output state always satisfies
\begin{equation}\label{eq:secretvar}
\sum_{\lkey}
\Pr[\Ol]\, d\left(\rho_{\KA \Efinal | \Ol} \,,\,  \omega^{\lkey}_{\KA} \otimes\rho_{\Efinal | \Ol}\right) \leq \esecret,
\end{equation}
where $\Ol$ denotes the event that the protocol produces a key of length $\lkey$. 
\end{definition}

Importantly, correctness and secrecy together imply security, as shown in Refs.~\cite{part1,portmann_security_2022}:
\begin{theorem} \label{theorem:secrecycorrectness}
\cite[Theorem~2]{portmann_security_2022} If a QKD protocol is $\ecorr$-correct and $\esecret$-secret, then it is $(\ecorr+\esecret)$-secure.
\end{theorem}
Hence to prove a protocol satisfies the security definition, it suffices to show it satisfies the correctness and secrecy definitions, and add the corresponding parameters.

\subsection{Completeness and expected key rate}
\label{subsec:completeness}

The above definitions suffice to formulate what it means for a QKD protocol to be ``secure'', in the sense that it can be ``safely'' used as part of a larger protocol. 
However, they do not fully address the question of whether the protocol is ``useful'', informally speaking. For instance, 
observe that if a fixed-length protocol has a high output key length $\lfixed$ when it accepts, but almost never actually accepts,
then this would not be a very practically useful protocol.\footnote{While this might perhaps appear artificial, we highlight that if the conditions for the accept/abort decision are not designed with sufficient ``tolerance'', then one would indeed end up with a protocol that almost always aborts. Some early works in QKD inadvertently constructed such protocols by having overly restrictive accept conditions; we briefly discuss this in \cref{subsec:gap_AT} under the term ``unique-acceptance''.}
More technically, 
inspecting \cref{def:security} shows that any fixed-length protocol in which the accept probability is \emph{never} greater than $\esecure$ would trivially fulfill that definition, but this would not seem to be very useful either.

To address this point, one usually imposes a condition that the protocol can perform ``nontrivially'' in at least some circumstances. Typically, this is done by specifying some assumed \term{honest  behaviour} of the protocol when Eve does not attempt an attack (note that this behaviour may still have some forms of noise or loss). For prepare-and-measure protocols, this is sometimes referred to as the (honest) \term{channel model}, as it mostly focuses on describing how the quantum channel between Alice and Bob (see \cref{sec:structural_elements}) behaves in the honest case.

Given such an honest behaviour or channel model, there are two common approaches to formalize the idea that the protocol has ``nontrivial'' performance:\\

\paragraph*{Completeness.} This approach is only relevant to fixed-length protocols, and is described in e.g. Ref.~\cite{portmann_security_2022}. For $\ecom\in[0,1]$, we say that a fixed-length protocol is \term{$\ecom$-complete} if the probability of it aborting in the honest case is at most $\ecom$. Requiring that $\ecom$ is small will ensure the protocol is ``nontrivial'' in that it has high probability of accepting (thus generating a key) in the honest case, at least.\\

\paragraph*{Expected key rates.} This approach can be used for both fixed-length and variable-length protocols. The idea is to compute the \emph{expected value} of the key rate under the honest behaviour, which serves as a way to quantify how much key we expect to get in that case. 
In particular, for fixed-length protocols the expected key rate is simply $\Pr[\Oacc] \lfixed$ (with $\Pr[\Oacc]$ computed on the honest behaviour), which is lower than the key length $\lfixed$ it produces when it accepts.\\

In light of the above, we highlight that to quantify the performance of a fixed-length protocol, it does \emph{not} suffice to focus solely on the key length $\lfixed$ it produces when it accepts, since that does not address the question of whether it is likely to ever accept. Instead, one has to account for the accept probability in the honest case in some form, for instance via the completeness parameter or the expected key rate. We discuss these aspects in greater detail in \cref{subsubsec:choosingacceptancethreshold}. 

\begin{remark}
We emphasize that the assumed honest behaviour is \emph{only} relevant for discussing completeness or expected key rates. It does not affect in any way the \emph{security} properties discussed in \cref{subsec:securitydefn}, which are completely independent of this honest behaviour. 
\end{remark}

\section{Structural elements of QKD protocols} \label{sec:structural_elements}
In this section, we outline the various steps involved in a QKD protocol, emphasizing the range of choices available at each stage of the protocol design. These choices often have direct implications for the types of mistakes, gaps, or proof techniques discussed in later sections. Each step, along with its possible variations is explored in detail in the individual subsections.

For a PM-QKD protocol, it is typically assumed that Alice and Bob have access to the following resources:
\begin{itemize}
\item Ideal RNG devices (``locally'' for each party, i.e.~they do not have access to shared randomness).
\item An ``untrusted'' quantum channel, where an eavesdropper Eve can freely manipulate states sent along the channel.
\item An \term{authenticated public classical channel}.
In much of the existing QKD literature, this is assumed to mean a channel where Eve can read all messages sent along the channel, but cannot modify them. We highlight however that in practice, one can typically only construct a weaker version of this resource, in which again Eve can read all messages, but she additionally has the option of causing the recipient to abort rather than receiving the unmodified message.\footnote{
As shown in~\cite{portmann_key_2014}, it is indeed possible to obtain an information-theoretically $\eps_{\text{auth}}$-secure construction (in the sense of a composable security framework) of such a resource, by using 
an unauthenticated public classical channel together with a short pre-shared key, via Wegman-Carter authentication~\cite{wegman_new_1981}. 
Hence in principle, this resource requirement could  be replaced with an unauthenticated public classical channel and a short pre-shared key. However, we will not focus on this within this work.} (With this version, the natural way for the QKD protocol to handle such abort outcomes is to also abort the overall QKD protocol.) This is a weaker property than the claim that Eve \emph{cannot} modify the messages, and hence some care is needed when analyzing this channel in the security proof. For now we note that the existing QKD literature, especially the usual security definition in \cref{def:security}, is focused on the case where the authentication cannot abort.  We discuss this further in \cref{subsubsec:gap_abort_auth}.
\item A tacit assumption in many QKD protocols is the existence of reliable time synchronization or time-tagging mechanisms for messages. This ensures that all parties agree on when specific phases of the protocol end and others begin, and that the sequence of messages is clear and lost messages are identified.
\end{itemize}

We now lay out a rough outline of the structure that most PM-QKD protocols follow.  
In the outline, all quantum states are implicitly meant to be sent using the ``open'' quantum channel, while all classical communication takes place over the authenticated public classical channel, with the protocol aborting whenever the authentication aborts.

\begin{enumerate}
\item \textbf{State preparation and measurement:} Alice generates states (referred to as \term{signal states}) over some number of rounds, and sends them to Bob, who measures them. Alice records her signal state choices in a classical register, and similarly Bob records his measurement outcomes (and \pending{basis} choices, if relevant) in another classical register. We refer to these registers as Alice and Bob's \term{raw data}. 
\item \textbf{Public announcements and \pending{processing}:} Alice and Bob perform various classical operations on the data gathered in the previous steps, usually assisted by announcing some classical values computed from that data. 
For instance, a common procedure in this step is \term{sifting}, in which Alice and Bob announce information such as their basis choices, then use this to erase or discard data from some rounds, such as rounds where they had different basis choices. 
They may also announce some data for use in later steps such as the acceptance test.
At the end of this step, one of the parties holds a string $\PAstring$ that they will use as the input to the final privacy amplification step (see below), which we will refer to as the \term{pre-amplification string}. Note that this pre-amplification string $\PAstring$ will not be modified in any way until the privacy amplification step.
\item \textbf{Acceptance test or variable-length decision:} The procedure implemented here depends on whether it is a fixed-length protocol, which outputs a key of fixed length whenever it accepts, or a variable-length protocol, which may output a key of variable length. For the former, this step consists of an \term{acceptance test} in which Alice and Bob decide whether to abort the protocol. (Note that the protocol may also abort in some later steps as well, such as error verification; see below.) For the latter, this step consists of a \term{variable-length decision} where Alice and Bob choose the length of final key that they will output (this length may be zero, effectively corresponding to an abort). They may also choose the parameters of their error correction protocol at this stage. 
\item \textbf{Error correction and error verification:} In \term{error correction}, the parties perform some further public communication, such that the party not already holding a pre-amplification string can produce their own pre-amplification string, which should ideally match the other's with high probability.\footnote{Some earlier works described this step as ``information reconciliation'' and allowed \emph{both} parties to modify their data to attempt to obtain some common shared string. For this work, we choose to instead group any such operations into the processing step, and hence there is no loss of generality in only having one party aim to match the other party's pre-amplification string in this step.} This may be followed by \term{error verification}, where Alice and Bob attempt to check whether their pre-amplification strings match, and abort if the check fails.
\item \textbf{Privacy amplification:} If the protocol has not aborted thus far, Alice and Bob finally perform a procedure that generates a final key from their pre-amplification strings. Most commonly, this is implemented by applying a hashing procedure to those strings, though other procedures (in general referred to as \term{randomness extraction}) are possible.
\end{enumerate}
In the following sections, we give qualitative descriptions of the purpose of each step, then elaborate further on details and possible variations. 
There is some flexibility in the ordering of various steps in the outline; we briefly mention alternative possibilities as they arise in the subsequent discussions.

\subsection{State preparation and measurement}\label{subsec:step_PM}

For brevity, we may sometimes also refer to this as the prepare-and-measure step. This step allows Alice and Bob to generate the raw data that will eventually be distilled into a secret key over the subsequent steps, as well as attempt to detect whether Eve is eavesdropping. It consists of a sequence of rounds in which Alice prepares signal states to send to Bob, who in turn measures them. Alice's record of her choices of signal states, and Bob's record of his measurement outcomes (and basis choices --- see \cref{subsubsec:activepassive} below), will serve as their raw data for the subsequent steps. We now elaborate further on various aspects of this step.

\subsubsection{Active or passive basis choices} 
\label{subsubsec:activepassive}

Broadly speaking, there are two possible options for how Bob performs his measurements. One option is referred to as \term{active} basis choice, in which Bob uses his RNG device to choose the basis used for measuring. The other option is referred to as \term{passive} basis choice, in which Bob's measurement device itself determines the basis to measure in, for instance via a beamsplitter. Different proof techniques may be needed to handle each case.

\subsubsection{Coordinating decisions between Alice and Bob} 

During this step, Alice and Bob usually have to pick their basis choices for the signal states and measurements randomly according to some distribution, to prevent Eve from knowing the basis choices used before the state is measured (most protocols become trivially insecure if Eve can do so). Similarly, for decoy-state protocols in particular, Alice would need to randomly choose some rounds in which to use the decoy-state intensities. The question then arises of which of these two possibilities is used:
\begin{itemize}
\item Alice and Bob {jointly} make these decisions, i.e.~we allow their choices of signal states and measurements to be correlated.  
\item Alice and Bob {independently} make these decisions. 
\end{itemize}

The first option often yields a protocol that is theoretically easier to analyze, but is difficult to achieve in practice, since Alice and Bob must coordinate this in such a way that Eve cannot learn values such as the basis choices before the state is measured.\footnote{Some methods to achieve this for specific protocols were suggested in Ref.~\cite{tan_improved_2022}, by exploiting some amount of pre-shared key, but the techniques may not generalize to all protocols. }
The second option is practically easier as it requires no coordination between Alice and Bob (and is also more compatible with passive setups), but it results in them using different basis choices in some rounds. The protocol must thus clearly specify how to perform some subsequent processing steps (in particular, choosing test rounds in \cref{subsubsec:step_testvsgen}, and the sifting procedure in \cref{subsubsec:step_sift}), and the security proof needs to analyze them appropriately.

\subsubsection{Distribution of rounds of each type}
\label{subsubsec:round_type_distribution}

Another important aspect is the distributions Alice and Bob use for choosing their signal choices and measurements across the rounds. Broadly, most protocols follow one of two options:\footnote{One could also consider combinations of the two options (for instance, Alice uses the first option for basis choices but the second option for intensity choices); furthermore, either of them can be combined with either of the options discussed in the preceding section.}
\begin{itemize}
\item Out of all the rounds, Alice and/or Bob pick random subsets of {fixed} size in which to use each signal choice and/or measurement. 
\item In each individual round, {independently} of all other rounds, Alice and/or Bob make a random decision of which  signal choice and/or measurement to use.
\end{itemize}
Loosely speaking, these two options are respectively more compatible with the sampling-without-replacement and spot-checking approaches discussed in \cref{remark:spotcheck_vs_sampling} later; we defer a more detailed discussion to that point.

A protocol specification must clearly state which procedure is being used, and the security proof must analyze it using suitable theorems. We highlight in particular that for the second option, the number of rounds in which each signal choice or measurement is used is usually a random variable rather than a constant, which must be accounted for in the security proof; see \cref{subsubsec:gap_testing}.

\subsubsection{Total number of rounds}

As noted above, in some protocols, the number of rounds where each signal choice or measurement is used can potentially be variable, especially for the case where these choices are made independently in each round. This raises the question of whether the total number of rounds can also be variable. For instance, one can consider protocols where Alice continues sending states until a certain number of rounds have been sent in a particular basis, or until a certain number of detections\footnote{Note that this option would require announcing detections while the prepare-and-measure step is still ongoing; see \cref{remark:onthefly} and \cref{subsubsec:iterative} for possible issues.} have been obtained. However, security proofs for such protocols must then address this aspect appropriately, especially in terms of the statistical analysis --- see \cref{subsubsec:iterative} for further discussion.

\subsection{Public announcements and \pending{processing}}
\label{subsec:step_process}

In this step, the parties perform some processing of their raw data, usually with the aid of some public announcements. Qualitatively, this processing serves the purpose of transforming the data into a form that can yield a higher secret key rate in the end. For this work, we require that at the end of this step, one of the parties has already finalized the string that they will use as the input to the privacy amplification step (\cref{subsec:step_PA}), which we will refer to as the pre-amplification string. 
We now describe examples of various forms of such public announcements and processing. 

\subsubsection{Public announcements}
\label{subsubsec:step_announce}

In many protocols, Alice and Bob would announce their choices of signal states and measurement bases respectively. Bob would also announce whether he obtained a detection in each round. These announcements allow them to perform various processing operations described below, such as choosing test rounds (\cref{subsubsec:step_testvsgen}) and sifting (\cref{subsubsec:step_sift}). Note that further public announcements may take place during other processing operations as well.

\begin{remark}\label{remark:onthefly}
In the protocol structure outlined here, we have described the public announcements as taking place after the entire prepare-and-measure step has been completed. In principle, one could instead consider the possibility of making some announcements while that step is still ongoing, which we shall refer to as \term{on-the-fly announcements}.\footnote{It would usually not be safe to announce (say) the basis choices for a particular round before the state in that round has been measured, since Eve could then use that information to adapt her attack; however, one could consider for instance announcing the basis choices immediately after the measurement in that round has been completed.} However, many proof techniques are not straightforwardly compatible with this scenario; we briefly highlight the techniques that can accommodate it in \cref{subsubsec:iterative}.
\end{remark}

\subsubsection{Choosing test rounds and announcing test data}
\label{subsubsec:step_testvsgen}

Alice and Bob need to decide on some subset of rounds to label as \term{test rounds}, for which the raw data will later be used the acceptance test or variable-length decision. The remaining rounds will be referred to as \term{generation rounds}. 

There is considerable freedom in the procedure for choosing the test rounds, also depending on the distributions used in the prepare-and-measure step. As a simple example, the test rounds could be chosen based only on the raw data held by one party, for instance all the rounds in which Alice chose a particular basis (or a decoy-state intensity, for decoy-state protocols). Other options are to choose the test rounds to be those where \emph{both} parties used a particular basis, or those where \emph{at least one} used a particular basis, or to introduce some further randomness in choosing the test rounds --- there are a wide variety of possibilities, and a protocol specification must be clear about which is being used.

Once the test rounds have been chosen, Alice and Bob would then usually\footnote{In principle, the parties may not need to ``directly'' announce all the test-round data at this step --- there can be protocols~\cite{arnon-friedman_simple_2019,tan_improved_2022} in which some of this data is instead ``compressed'' into the data that Alice sends later in the error correction step. However, this comes at the cost of additional complications in the implementation and analysis; we briefly discuss this later in \cref{subsubsec:step_AT}.} announce the raw data values for those rounds.  These announcements provide the data required for the acceptance test (\cref{subsec:step_decision} below).

\begin{remark}\label{remark:spotcheck_vs_sampling}
The two options for signal and measurement choices previously discussed in \cref{subsubsec:round_type_distribution} (namely, using a fixed number of each type of round, or making an independent choice in each round) can have a significant impact on the resulting distribution of test rounds. For instance, if Alice uses the first option (using a fixed number of rounds for each signal choice) and the test rounds are chosen based \emph{only} on her signal choices, then the number of test rounds can be a fixed constant.\footnote{Note however that if Alice and Bob made independent choices of signal states and measurements, and the choice of test rounds depends on \emph{both} their data, then the resulting number of test rounds might also not be a constant; see \cref{subsubsec:gap_testing} for potential mistakes regarding this point.} Such scenarios are sometimes referred to as \term{sampling without replacement} (this can also be realized by any other method that results in a fixed number of test rounds). On the other hand, if the second option (making an independent choice in each round) is used, then it is often the case that each round is independently a test or generation round. Such scenarios are sometimes referred to as \term{spot-checking} or \term{infrequent-sampling}.
\end{remark}

\subsubsection{Random permutation}
\label{subsubsec:step_permute}

Some proof techniques (see for instance \cref{subsec:postselection} later on the postselection technique) require that Alice and Bob explicitly perform a random permutation of their raw data, in order to enforce some symmetry properties required for the proof. More precisely, this consists of one party (say, Alice) choosing some permutation using local randomness, publicly announcing it, and then having both parties apply this permutation to their raw data. Note that there exist simple algorithms to implement such permutations with reasonable efficiency, in terms of both computational complexity and the number of random bits required~\cite{nahar_postselection_2024}.

\subsubsection{Classical postprocessing}

In many protocols, Alice and Bob remap their raw data into new values, producing a transformed string that is used in subsequent stages of the protocol. This postprocessing can be motivated by various factors, including convenience or technical requirements of the proof technique.

For example, a common approach is for Bob to randomly assign double-click events (both detectors in a given basis clicking) to one of the two detectors. In proofs that rely on the Entropic Uncertainty Relation (EUR) (see \cref{subsec:eur}) and phase error correction (see \cref{subsec:phaseerror}) methods, such postprocessing is necessary to satisfy the basis-efficiency mismatch assumption in the canonical detector model. Another instance where postprocessing becomes essential is the use of squashing maps (see \cref{subsubsec:squashing}), as certain squashing maps only exist when specific postprocessing steps are applied.

It is worth noting, however, that in many scenarios, this postprocessing is not required for error correction. In such cases, Alice and Bob can fully utilize their raw data during the error correction phase, maximizing the use of available information.

\subsubsection{Sifting}
\label{subsubsec:step_sift}

A common form of classical postprocessing is \term{sifting}, in which the parties ``discard'' data from some rounds. For instance, they would usually discard rounds in which Bob did not receive a detection, because we informally expect that such rounds do not contribute meaningfully to generating secret key between Alice and Bob. Another form of sifting is when Alice and Bob make independent basis choices and discard rounds in which they picked different basis choices. They may also wish to omit the raw data from the test rounds in the pre-amplification string (after choosing them according to some procedure as discussed in \cref{subsubsec:step_testvsgen}), as such data have already been publicly announced and Eve knows the values in those rounds.

We highlight a subtlety in the notion of ``discarding'' rounds in the sifting procedure: in many protocols, this is taken to mean that such rounds are entirely removed from the data, i.e.~the resulting data string is genuinely shorter than before. However, this introduces some technical issues in applying existing proof techniques, which are more suited for analyzing procedures in which the raw data for such rounds is instead replaced with some fixed value, rather than being literally discarded. We defer a discussion of these issues, as well as how to overcome them, to~\cref{subsubsec:varlength_rawkey}.

\subsubsection{Noisy preprocessing} 

Another possible operation is \term{noisy preprocessing}, in which a small amount of trusted noise is added to one party's raw data (say, Alice's) in each round. It has been shown that this can in fact improve the key rate in some situations~\cite{kraus_lower_2005,renner_information-theoretic_2005,renes_noisy_2007} --- informally, the idea is that the added noise ``degrades'' Eve's knowledge about Alice's secret data more than it ``degrades'' Bob's.

\subsubsection{Advantage distillation}

For this work, we use the term \term{advantage distillation} to broadly refer to a wide variety of processing procedures involving two-way communication that can be used to improve the key rates in some situations~\cite{maurer_secret_1993,wolf_information-theoretically_1999,chau_practical_2002,gllp_security_2004,renner_security_2005,bae_key_2007}. One example of advantage distillation is the {repetition-code protocol}~\cite{maurer_secret_1993,renner_security_2005,bae_key_2007}, in which Alice and Bob divide their raw data into blocks and announce some data about each block, then discard blocks in which the announcements indicate the values in that block are more ``poorly correlated''. This can potentially improve the key rates by preserving only the blocks with ``stronger correlations''. {However, in this work we will mostly not discuss advantage distillation procedures in much detail.}

\subsection{Acceptance test or variable-length decision}
\label{subsec:step_decision}

For a fixed-length protocol, in this step Alice and Bob make a ``binary'' decision of whether the protocol accepts or aborts, which we refer to as an \term{acceptance test}. 
Qualitatively, the idea is that they are attempting to detect whether Eve is eavesdropping on their communication ``more strongly'' than some threshold amount, and aborting if they detect such eavesdropping.  If they accept, they will produce a key of fixed length in the final privacy amplification step (as long as the subsequent error verification step does not abort).
For a variable-length protocol, in this step they instead make a decision of what length of key they will produce at the final privacy amplification step, which we refer to as a \term{variable-length decision}. Analogously, the idea here is that they are attempting to quantify the amount of eavesdropping by Eve, which will allow them to compensate for it via privacy amplification.

\begin{remark}
Many procedures in this step are often based on computing the \term{frequency distribution} observed on some classical registers gathered during the protocol; in other words, the fraction of rounds in which each outcome was observed. We emphasize that when specifying a protocol and presenting its security proof, it must be clearly specified whether the denominator in this fraction is the total number of protocol rounds, or only the number of sifted rounds or test rounds. Different proof techniques may be compatible to different extents with these different possibilities, especially since values such as the number of sifted rounds may be random variables rather than constants. 
\end{remark}

\subsubsection{Acceptance test (fixed-length protocols)}
\label{subsubsec:step_AT}

Typically, an acceptance test consists of one or both of the parties computing the frequency distribution observed on some classical registers, and then accepting (for this step) if and only if this frequency distribution lies in some predetermined set, which we shall call an \term{acceptance set}. We stress that for a fixed-length protocol, the acceptance set must be fixed \emph{before} the protocol begins, and the protocol can only output a key of a fixed length whenever it accepts. Any other form of dependence of the key length on the observations in the protocol would make it a variable-length protocol instead.

For many protocols, the accept/abort decision is based entirely on the various registers publicly announced in the previous steps, such as the test-round data, in which case Alice and Bob can clearly come to a common decision in this step. Some protocols may instead have one of the parties make this decision based on some values that were not publicly announced, in which case that party would then need to announce their final accept/abort decision.\footnote{While such protocols are not common, a possible example would be protocols in which the test-round data is not ``directly announced'' during the public announcement step, and instead some additional data is sent during the error correction step for Bob to produce a guess for Alice's test-round data, which he then uses to make the acceptance test decision. Note that this requires error correction to be performed \emph{before} the acceptance test, rather than the step order outlined in this work. To prove security for such protocols, the acceptance-test analysis must then account for the possibility that his guess might be wrong; see~\cite{tan_improved_2022} for further details.} 

As we shall discuss later in~\cref{subsec:estimating_entropy}, some care needs to be taken in interpreting the output state of such a protocol. For instance, it is \emph{not} valid to say that conditioned on the protocol accepting, the probability of Eve knowing the final key is small --- we describe simple counterexamples in~\cref{subsec:estimating_entropy}. However, it is indeed possible to prove that the final output state satisfies the composable security definitions described in Refs.~\cite{part1,portmann_security_2022}, which in turn means that it satisfies the various desirable properties implied by that definition.

\subsubsection{Variable-length decision (variable-length protocols)}

In such protocols, in this step Alice and Bob come to a decision of the length of key they will output at the end. Depending on the protocol, at this point they might also decide on the error correction procedure they will use in the next step (\cref{subsec:step_ECandEV}); in particular, this means that the number of bits communicated in that step may potentially also be variable, and depend on the observations made during the protocol.

Such protocols can be more challenging to analyze when constructing security proofs, as the security definition is more elaborate (see \cref{def:security}). 
For instance, it is often important to specify whether the variable-length decision is based only on the public announcements before this step, rather than involving ``private information'' held only by one of the parties. Another technicality is that if the number of bits communicated for error correction is variable, then this raises questions regarding how to account for this when choosing the final key length --- it may not be valid to assume one can simply choose the length of the final key depending on the number of bits actually communicated in error correction. Broadly speaking, current security proofs mostly cover only scenarios in which the final key length is determined completely by the public announcements before error correction. 

\subsection{Error correction and error verification}
\label{subsec:step_ECandEV}

Recall that at this point, one of the parties already holds their \textbf{pre-amplification string} that will be held unchanged until the final privacy amplification step. However, the other party still usually does not know the exact value of that pre-amplification string, since their initial raw data would have been subject to some level of noise with respect to each other. In light of this fact, error correction is performed, in which further public communication is used to allow the latter party to generate their own pre-amplification string, which is intended to match the other's with high probability. This may then be followed by error verification, which refers to a process in which Alice and Bob attempt to verify whether their pre-amplification strings indeed match.

\subsubsection{Error correction}

Broadly, error correction procedures can be classified into those that only use one-way communication, and those that use two-way communication. In this work, we will mostly focus on the former, which is often referred to as \term{one-way error correction}, though we briefly discuss the latter at a few points. We now elaborate further on these options.\\

\paragraph*{Direct reconciliation.} 
This is the form of one-way error correction used in most currently implemented protocols for decoy-state BB84. Here, Alice (i.e.~the party who prepared the signal states, not the one who measured them) is the party who holds the pre-amplification string after the preceding step, and the error correction procedure consists of her computing a classical string from her pre-amplification string and sending it to Bob, who then uses it together with all the data on his side to produce a guess for Alice's pre-amplification string. Qualitatively, this is usually achieved by having Alice compute the syndrome of a suitable error-correcting code as the string to send to Bob, though there are some slight variations. For instance, some proof techniques may require Alice to furthermore apply a one-time-pad to this syndrome before sending it to Bob (see \cref{subsec:phaseerror} for further discussion of when this is necessary), which requires a significant amount of preshared secret key to perform, though the resulting \emph{net} gain in shared secret key is usually the same as other proof techniques when the protocol accepts\footnote{However, this approach has a notable drawback in that if the protocol aborts at some point after error correction, then the preshared key bits used to one-time-pad the error correction syndrome are already consumed and cannot be reused, resulting in a net loss of shared secret key.}.
\\

\paragraph*{Reverse reconciliation.} 
This is also a form of one-way error correction, but it instead requires that Bob is the party who holds the pre-amplification string after the preceding step, and hence he sends a classical value to Alice for her to produce a guess for Bob's pre-amplification string. For some protocols, this can potentially yield better key rates than direct reconciliation. We highlight however that the security proof does need to be structured such that it correctly accounts for the fact that Bob is the party who first holds a pre-amplification string; we return to this point in \cref{subsubsec:Hminchainrule}.\\

\paragraph*{Two-way error correction.} 
Some error correction procedures involve communication in both directions between Alice and Bob. An example of this is the Cascade protocol \cite{brassard_secret-key_1994}, in which the parties potentially perform several rounds of communication. However, the number of communicated bits in such protocols is usually significantly higher than in one-way error correction, and this issue need to be treated with care in the security proof.

\begin{remark}
When specifying an error correction procedure for a QKD protocol, one critical point that must be addressed is whether the number of communicated bits is a fixed value or a random variable. For fixed-length protocols, many security proofs can only handle the former case, or cases where there is a fixed upper bound on the number of communicated bits. For variable-length protocols, some proof techniques can handle the latter, though there are technical conditions that must be satisfied. We return to these points in \cref{subsec:errorcorrection} later.
\end{remark}

\subsubsection{Error verification}
\label{subsubsec:step_EV}

After error correction, the error verification procedure usually consists of one party announcing a $\delta$-AU$_2$ hash (see \cref{subsec:2universal}) of their pre-amplification string, with the other party checking to see whether they obtained a pre-amplification string with the same hash, and aborting if that is not the case. Qualitatively, this serves as a check on whether their pre-amplification strings match, since the properties of $\delta$-AU$_2$ hashing ensure that if the pre-amplification strings differ, the probability of their hashes matching is small. We stress however that this is a very different statement from claiming that if their hashes match, the probability of them having distinct pre-amplification strings is small. In the security proof, it would \emph{not} be valid to make the latter claim; fortunately, the former claim is enough to prove that the correctness condition is satisfied, as we discuss in \cref{subsec:correctnessproof}.

One important benefit of error verification is that it allows the error correction procedure (and related procedures in the preceding step) to be designed based on the \emph{honest} behaviour of the protocol. Intuitively, this is because while such an error correction procedure might fail to give Alice and Bob matching values in the presence of dishonest behaviour, the error verification procedure serves to ensure the correctness condition holds regardless of what happened during error correction. 
This feature is particularly useful from the viewpoint of security proofs for fixed-length protocols, because for instance it usually allows the number of communicated bits in one-way error correction to be chosen as a specific fixed value, corresponding to the honest behaviour. This often simplifies the security proofs.\\

\paragraph*{Alternatives to error verification.} In principle, some protocols may not perform an error verification step in the above form. For instance, one possible alternative used in some earlier protocols~\cite{renner_security_2005} was to perform an acceptance test that implicitly checks whether the number of errors between their raw data strings is below some threshold, aborting if it is not. Given suitable error correction procedures in such a scenario, it is still possible to prove that the correctness condition holds; however, this requires more stringent conditions on the error correction procedure (roughly speaking, they have to be analyzed in the context of dishonest behaviour as well). 
\\

\paragraph*{Error verification after privacy amplification.} 
Another variation on error verification is the possibility to perform it \emph{after} the privacy amplification step (see \cref{subsec:step_PA}). Qualitatively, the proposal is that one party could simply announce a fixed subset of their final key bits, after which the other party checks if those bits match. However, there are some subtleties associated with this idea. First, to apply the standard proof of the correctness condition (see \cref{subsec:correctnessproof}), it would require that privacy amplification is performed via a procedure where this \emph{subset} of its output bits still forms a valid universal$_2$ hash (defined in \cref{subsec:2universal}) --- while this property seems likely to hold for many common privacy amplification procedures, a security proof does need to explicitly justify it for the specific chosen procedure.\footnote{It does not suffice to simply require that privacy amplification is performed with universal$_2$ hashing, as this does not in general guarantee that a substring of it still forms a universal$_2$ hash. More broadly, this property might also not straightforwardly hold for privacy amplification based on Trevisan's extractor~\cite{de_trevisans_2012}, which does not simply consist of applying a universal$_2$ hash.} Second, it might appear that the secrecy definition (\cref{eq:secretfixed}) is still satisfied by simply ``regrouping'' the announced bits  into Eve's side-information rather than the final key --- however, this claim is subtly flawed, because in the real state $\rho_{\KA \Efinal | \Oacc}$ the marginal state on these announced bits might not be \emph{perfectly} uniform, and thus a close inspection of the secrecy definition reveals that a more in-depth analysis is needed to show whether it is still satisfied under this ``regrouping''.\footnote{Potentially, a different secrecy definition that minimizes over Eve's ``side-information'' marginal state (see e.g.~\cite[Eq.~(17)]{portmann_security_2022}) might be more easily shown to hold under this modification. However, as discussed in~\cite[Section~III.C.5]{portmann_security_2022}, 
that secrecy definition is not known to be compatible with a composable security framework (though it only differs from the  definition used in this work by at most a factor of $2$), and thus we do not discuss it further here.} A full security proof for this variant of error verification would need to rigorously account for these issues.

\subsection{Privacy amplification}
\label{subsec:step_PA}

At this point, Eve may still have some limited knowledge about Alice and Bob's pre-amplification strings\footnote{This is because the acceptance test usually has to be designed to accommodate some amount of noise even from the honest behaviour, and hence it cannot rule out the possibility that Eve may be ``hiding'' her eavesdropping in that amount of noise.} --- as a very simplistic example, Eve might know a small subset of the bits in those strings. The purpose of privacy amplification is to transform the pre-amplification strings into final keys that are perfectly uniform and entirely decoupled from Eve, up to the amount of ``{$\esecret$}-imperfection'' allowed in the secrecy definition (\cref{def:correctsecret}). To do so, these final keys are usually shorter than the pre-amplification strings; informally, we can say that Alice and Bob ``sacrifice'' some key bits in order to remove the correlation with Eve. We now discuss some procedures that achieve this. \\

\paragraph*{Universal$_2$ hashing.} A possible method for privacy amplification is for one party to choose a random hash function according to a universal$_2$ hashing procedure (see \cref{subsec:2universal}) using some local randomness, which is sometimes termed as the \term{seed} for the privacy amplification procedure. They then publicly announce their choice, after which both parties apply that choice of hash function to their pre-amplification strings. The core theoretical tool that ensures security from this procedure is the \term{Leftover Hashing Lemma} (\cref{subsec:LHL}), which qualitatively states that if the input to the hash function has ``enough entropy'' relative to the length of the output, then the output will be nearly uniform and decoupled from any other side-information.
\\

\paragraph*{Other extractors.} Apart from universal$_2$ hashing, there are other procedures that can be used for privacy amplification, broadly referred to as \term{randomness extractors}. Qualitatively, these procedures achieve similar results to the Leftover Hashing Lemma, in that they produce outputs that are nearly uniform and decoupled from any side-information, as long as the inputs have ``enough entropy'' --- the difference is that the operation performed on the pre-amplification string does not have to be a universal$_2$ hash. One noteworthy example of an extractor that does not simply apply a universal$_2$ hash is \term{Trevisan's extractor}. This has the advantage that it requires less seed randomness as compared to universal$_2$ hashing; however, it can be more elaborate to implement, and some forms of privacy amplification theorems have only been proven for universal$_2$ hashing rather than Trevisan's extractor. \\

\paragraph*{Privacy amplification with variable input lengths.} A technical point arises in privacy amplification regarding the domain of the input to the procedure. In particular, for some protocols the lengths of the pre-amplification strings are not fixed constants, for instance because some bits are literally discarded during the sifting procedure or other processing steps. Other protocols may instead have pre-amplification strings that are not bitstrings, for instance if bits that are sifted out are replaced with some special ``blank'' symbol. In such circumstances, one cannot ``directly'' implement procedures such as Toeplitz hashing, which were only designed to take bitstrings of fixed-length as input. A full protocol specification in such cases must explicitly state the exact procedure used for privacy amplification (for instance, a different family of Toeplitz matrices is used depending on the length of the pre-amplification strings), and the security proof must rigorously analyze this procedure accordingly. We return to this point in \cref{subsubsec:varlength_rawkey}

\section{Concepts in security proofs} \label{sec:concepts}
Having explored the diverse range of protocol choices in QKD, we now turn our attention to the foundational concepts required for the theoretical analysis of QKD protocols. In this section, we will briefly explain some key concepts needed for rigorous security proofs.
\subsection{Source-replacement scheme} \label{sec:SRS}

Broadly, the source-replacement scheme \cite{bennett_quantum_1992,curty_entanglement_2004} is used, as the name suggests, to replace the actual source with a virtual entangled source, which when coupled with a measurement on Alice's subsystem can mimic the actual source. The exact entangled source and measurement used depends on the proof-technique --- some proof techniques (phase error and EUR) use this to reduce Alice's subsystem to a qubit system to simplify the analysis, while others (postselection and EAT-based) instead require an IID (independent and identically distributed) source preparation that need not be a qubit.

An important consideration when using the source-replacement scheme is that information about the pairwise inner products of signal states should, in general, be included in the security analysis\footnote{Intuitively, this is to be expected since we know that QKD cannot be performed with orthogonal states. Thus, somehow we must include the information that the source prepares non-orthogonal (and hence, not perfectly distinguishable) states in our security analysis.}.
Once again, different security proofs handle this differently. The postselection and EAT-based proof techniques encode the overlap structure by adding an additional constraint on the marginal of the entangled state shared by Alice and Bob. The phase error correction and EUR proof techniques instead include this information when connecting the test-round statistics to the phase errors in generation rounds.

\subsubsection{Shield system}

A special case of the source-replacement scheme where Alice prepares a mixed state can be used by introducing a \emph{shield system} \cite{horodecki_general_2009}.
This works as follows. Alice's mixed state preparation can be thought of as first preparing a pure state (purification of the mixed state), and then tracing out part of it to prepare the mixed state. However, this tracing out can be performed after the measurement phase since it commutes with the measurements. Thus, we can simply assume that Alice prepares pure states, part of which are sent to Bob. Note that the shield system is considered to be inaccessible to Eve, although Alice does not perform any operations on it during the protocol. This fact must be taken into account in the security proof, for instance through an additional constraint on Alice's subsystems, including her shield system.

\subsection{Universal and almost-universal hashing}
\label{subsec:2universal}

Informally, hash functions usually serve the purpose of ``compressing'' long input strings into shorter output strings, in such a way that the event of different input values giving the same output value (usually called a \term{collision}) is ``rare'' in some sense. Universal and almost-universal hashing procedures serve as a rigorous way to formalize some aspects of this intuition, as seen from the definition~\cite{carter_universal_1979,stinson_universal_1994}:
\begin{definition}\label{def:2universal}
Given some domain $D$ and finite codomain $D'$, a \term{$\delta$-almost-universal$_2$ ($\delta$-AU$_2$) hashing procedure} consists of a set $\mathcal{H}$ of functions $h:D \to D'$, together with a specific probability distribution $P_\mathrm{HASH}$ over $\mathcal{H}$, such that if we randomly draw a function $h$ according to the distribution $P_\mathrm{HASH}$, we have:
\begin{align}\label{eq:AUdefn}
\begin{aligned}
&\text{For all } x,y \in D \text{ such that } x\neq y,\quad \\
&\Pr[h(x) = h(y)] \leq \delta ,
\end{aligned}
\end{align}
where the probability is evaluated with respect to the distribution $P_\mathrm{HASH}$. 
A special case of $\delta$-almost-universal$_2$ hashing is \term{universal$_2$ hashing}, which simply refers to the case where
\begin{align}
\delta = \frac{1}{|D'|}.
\end{align}
A further special case is   \emph{{\ideal}} universal$_2$ hashing, which refers to the case where we have an equality instead of inequality in the probability of collision; that is:
\begin{align}\label{eq:AUdefnideal}
\begin{aligned}
&\text{For all } x,y \in D \text{ such that } x\neq y,\quad \\
&\Pr[h(x) = h(y)] = \frac{1}{D'}
\end{aligned}
\end{align}
where the probability is evaluated with respect to the distribution $P_\mathrm{HASH}$. 

\end{definition}
We note that Toeplitz hashing (see below), and hashing with uniformly random binary matrices, are both ideal universal$_2$ \cite{mansour_computational_1993}.

\begin{remark}
We strongly emphasize that $\delta$-AU$_2$ hashing is a very distinct concept from using fixed choices of computationally-hard functions (or functions with ``heuristic hardness estimates'', such as SHA-3), and should never be confused with those notions. In particular, we stress that \emph{no computational-hardness assumptions or heuristics} are needed to construct $\delta$-AU$_2$ hashing procedures that satisfy the above definition.\footnote{To achieve this, it was critical that the function $h$ is drawn randomly in $\delta$-AU$_2$ hashing --- notice that if $h$ had been a fixed function and $|D| > |D'|$, then it would have been impossible to find a bound of the form in~\cref{eq:AUdefn} without further assumptions, since we would necessarily have a collision by the pigeonhole principle.}
Hence $\delta$-AU$_2$ hashing allows us to rigorously ensure information-theoretic security, without computational assumptions.
\end{remark}

Slightly informally, we often refer to the overall process of randomly drawing a function $h$ according to $P_\mathrm{HASH}$ and applying $h$ to some classical register as ``applying a $\delta$-AU$_2$ hash'', and we will refer to the output value as the hash.
For most $\delta$-AU$_2$ hashing procedures used in QKD, the distribution $P_\mathrm{HASH}$ is simply a uniform distribution, though the definition we considered does not specifically require this distribution to be uniform. Furthermore, the codomain $D'$ usually just consists of bitstrings of some length $\ell_\mathrm{hash}$, in which case the bound for universal$_2$ hashing reduces to
\begin{align}
\begin{aligned}
&\text{For all } x,y \in D \text{ such that } x\neq y, \quad \\
&\Pr[h(x) = h(y)] \leq \frac{1}{2^{\ell_\mathrm{hash}}} ,
\end{aligned}
\end{align}
i.e.~it is exponentially small with respect to the length $\ell_\mathrm{hash}$ of the output strings. 
We now highlight some other important properties of $\delta$-AU$_2$ hashing.\\

\paragraph*{Existence and efficiency.} There exist explicit constructions of $\delta$-AU$_2$ hashing procedures that are reasonably efficient, with respect to both computational complexity and the randomness required to sample according to $P_\mathrm{HASH}$. (We again stress that no computational-hardness assumptions are needed to rigorously prove that these constructions satisfy the defining property, \cref{eq:AUdefn}.) For instance, when the domain and codomain are both bitstrings of fixed lengths, a commonly used choice for an ideal universal$_2$ hashing procedure is to sample uniformly from the set of \term{Toeplitz matrices}, as described in Ref.~\cite{mansour_computational_1993} (see also Ref.~\cite{kiktenko_lightweight_2020} for a discussion specific to Toeplitz hashing). 
This approach offers the advantage that there exist algorithms for efficiently computing the output of a Toeplitz matrix applied to a bitstring, and the number of random bits required to choose such a matrix is linear with respect to the lengths of the input and output strings. 
\\

\paragraph*{Interpreting the probability.} We emphasize that the probability term in~\cref{eq:AUdefn} is defined \emph{only} with respect to the distribution over $\mathcal{H}$, i.e.~the random choice of hash function. In particular, there is no requirement of any probability distribution over the values $x$ and $y$; the bound holds for any distinct values $x \neq y$. Still, when we wish to apply this result in contexts where $x$ and $y$ are random variables, we can say this implies that conditioned on $x\neq y$, the probability that $h(x)=h(y)$ is small --- note that this statement holds without any assumptions about the distributions of $x$ and $y$. \\

\paragraph*{Applications in QKD.} As noted in \cref{sec:structural_elements}, $\delta$-AU$_2$ hashing is used in many aspects of QKD, for instance in the following steps.
\begin{itemize}
\item Most straightforwardly, it is used in the error verification step, where a $\delta$-AU$_2$ hash is used to attempt to check whether Alice and Bob obtained matching pre-amplification strings. The fact that it ensures a rigorous bound on the ``collision probability'' is highly relevant in analyzing this step, as we describe in \cref{subsec:correctnessproof}.
\item Universal$_2$ hashing is also applied in many commonly used procedures for privacy amplification. Here, the bound on the collision probability is less ``directly'' relevant; the key property that ensures security is the Leftover Hashing Lemma (\cref{subsec:LHL}), though the proof of that lemma relies on that bound in an intermediate step.
\item A slightly strengthened version of $\delta$-AU$_2$ hashing can also serve as a possible method for the construction of the authenticated classical channel, via Wegman-Carter authentication~\cite{wegman_new_1981,portmann_key_2014}, though we do not consider this within the scope of the QKD security proof itself.
\end{itemize}

\subsection{The correctness condition}
\label{subsec:correctnessproof}

By definition, the correctness condition requires us to show that the probability that the final keys differ \emph{and} the protocol accepts (or that it produces a nonzero key length, in the variable-length case) is small. We first note that for protocols with error verification, this event is a ``stricter'' event than the event of the \textbf{pre-amplification strings} differing (which we denote via $\Odiff$) and the error verification step accepting (which we denote via $\OEV$).\footnote{This is because the final keys can only differ if the pre-amplification strings differ, and the protocol can only accept (or produce a nonzero final key length) if the error verification step accepts.} Thus it suffices to upper-bound the probability of the latter event. This turns out to be fairly straightforward for error verification based on $\delta$-AU$_2$ hashing: observe that this gives
\begin{align}
&\Pr[\Odiff \text{ AND } \OEV] \nonumber\\
=& \Pr[\OEV \;|\; \Odiff] \Pr[\Odiff] \nonumber\\
\leq& \Pr[\OEV \;|\; \Odiff] \nonumber\\
\leq& \delta, \label{eq:correctness}
\end{align}
where the third line holds simply because probabilities are upper bounded by $1$, and the fourth line follows from the defining property of $\delta$-AU$_2$ hashing (\cref{def:2universal}). Hence the correctness condition is satisfied with $\ecorr = \delta$.

An important aspect of this argument is that it relies solely on the properties of error verification based on $\delta$-AU$_2$ hashing; in particular, it has no dependence at all on any properties of the preceding error correction step. This implies that for protocols using such an error verification procedure, the error correction procedure can be designed such that it is only guaranteed to succeed with high probability on the honest behaviour (so the completeness condition is fulfilled), rather than having to consider any form of dishonest behaviour at that step, since the correctness condition is ensured by the error verification step alone. Hence this form of error verification usually simplifies the security analysis. Note however that this is not the case for some of the variants discussed in~\cref{subsubsec:step_EV}; a more elaborate analysis may be required for some of those cases.

We highlight that the correctness condition does \emph{not} require one to show that conditioned on the protocol accepting, the probability that the final keys differ is small --- it only requires us to bound the probability of the joint event, rather than this conditional probability. In fact, one can construct explicit examples of attacks by Eve in which conditioned on the protocol accepting, the probability that the pre-amplification strings differ is still large, i.e.
\begin{equation}\label{eq:EVcounterexample}
\Pr[ \Odiff \;|\;  \OEV] \approx 1,
\end{equation}
as we show in \cref{subsubsec:misconception_EV}.
Therefore, it would be a misconception to claim that the correctness condition ensures this probability is small. However, as discussed in Refs.~\cite{part1,portmann_security_2022}, the composable security framework ensures that the correctness condition in the stated form is already enough to certify the desired composable properties of the protocol, despite the lack of a bound on this conditional probability. 

\subsection{Leftover Hashing Lemma}
\label{subsec:LHL}

A critical result for proving secrecy in a QKD security proof is the \term{Leftover Hashing Lemma}~\cite{renner_security_2005,tomamichel_largely_2017,dupuis_privacy_2023} --- while there are many variants, we focus on the one in~\cite[Proposition~9]{tomamichel_largely_2017} for the purposes of this discussion, as it is most compatible with the definitions we introduced in \cref{sec:defs}.
We can summarize the lemma as follows: for any state $\rho_{CQ}$ with classical $C$, if we apply an ideal universal$_2$ hash to $C$ such that the output is a bitstring $K$ of length $\lkey$, and denote the resulting state as $\rho_{KQH}$ where $H$ is a register storing the choice of hash function, then\footnote{In Ref.~\cite{tomamichel_largely_2017} (Proposition~9), a slightly generalized version of this lemma was presented for subnormalized states.
}
\begin{align}
d\left(\rho_{KQH} , \omega^\lkey_{K} \otimes\rho_{QH}\right) \leq 2^{-\frac{1}{2}(\Hmin^{\esmooth}(C|Q)_\rho - \lkey + 2)} + 2\esmooth,
\label{eq:LHL_Hmin}
\end{align}
where 
$
\omega^\lkey_{K} \defvar \sum_{k \in \{0,1\}^{\lkey}} \frac{1}{2^{\lkey}} \ketbra{k}{k}_{K}
$
denotes a uniformly random key of length $\lkey$ on register $K$.
In other words, as long as $\lkey$ is somewhat shorter than the smooth min-entropy $\Hmin^{\esmooth}(C|Q)_\rho$, then the resulting state on ${KQH}$ is close in trace distance to one where $K$ is uniform and independent of $QH$. Critically, the register $H$ is included in the ``side-information'' in this statement, so in the context of QKD protocols it means the choice of hash function can be publicly announced.\footnote{Including $H$ in the ``side-information'' registers also means that the left-hand-side of \cref{eq:LHL_Hmin} is equal to the average of the trace distance values conditioned on each choice of the hash function; some works have stated the Leftover Hashing Lemma in terms of that expression instead.} We also highlight that more recently, a variant of the Leftover Hashing Lemma was developed based instead on the {\Renyi} entropy, which often yields better finite-size performance, though the qualitative implications are similar. 

By inspecting the secrecy definition (\cref{def:correctsecret}), we see that being close to a state where $K$ is uniform and independent of the ``side-information'' is similar in spirit to what is required in that definition. However, we wish to stress that it is \emph{not} rigorous to directly claim that in a QKD security proof, one simply finds a fixed lower bound on the smooth min-entropy for all possible states and then hashes the pre-amplification string to a length slightly shorter than that value. This is due to the fact that there are critical obstacles to finding such a fixed bound, and furthermore there are various discrepancies between \cref{eq:LHL_Hmin} and the secrecy definition --- for instance, in the secrecy definition for fixed-length protocols there is a prefactor of $\Pr[\Oacc]$, and furthermore the trace distance term is evaluated on the states conditioned on accepting. In the next section, we elaborate further on these issues, as well as the actual method to obtain a rigorous security proof.

We highlight a technical subtlety here, namely that some of the leftover hashing lemmas in the literature are only stated for \emph{ideal} universal$_2$ hashing --- though as briefly mentioned above, Toeplitz hashing does indeed have this property. For instance, both \cite[Proposition 9]{tomamichel_largely_2017} and \cite[Theorem 8]{dupuis_privacy_2023} are only stated in terms of ideal universal$_2$ hashing. 
For examples of versions stated in terms of (non-ideal) universal$_2$ hashing, see e.g.~\cite[Corollary~5.6.1]{renner_security_2005} and \cite[Theorem~6]{tomamichel2011leftover} for the smooth min-entropy\footnote{Note that the definition of smooth min-entropy used in \cite{renner_security_2005} is slightly different from the others listed here. More generally, several different definitions have been used across the literature, and it is important to keep track of definition compatibility when combining results from different works.} case; in fact, the latter can be modified to apply to $\delta$-AU$_2$ hashing \cite[Theorem~7]{tomamichel2011leftover}. 
As for the {\Renyi} case, it is analyzed in Ref.~\cite{lars_thesis}, where the only difference compared to the result for ideal universal$_2$ hashing is one bit in the secure output length.

\begin{remark}
As noted in~\cref{subsec:step_PA}, there also exist privacy amplification procedures not based on universal$_2$ hashing alone, such as Trevisan's extractor~\cite{de_trevisans_2012}. However, the qualitative properties of those procedures are mostly similar to those of the Leftover Hashing Lemma. Furthermore, as mentioned there, additional care is necessary when analyzing privacy amplification if the pre-amplification strings are not bitstrings of fixed-length, since this may not technically fit the conditions of the Leftover Hashing Lemma. We briefly discuss this in \cref{subsubsec:varlength_rawkey} later, and a full rigorous treatment of this issue can be found in Ref.~\cite{tupkary_security_2024}.
\end{remark}

\subsection{Estimating the entropy}
\label{subsec:estimating_entropy}

Let us denote the pre-amplification string as $\PAstring$, and Eve's side-information just before privacy amplification as $\EprePA$. 
For brevity, in the following discussion we focus mainly on the Leftover Hashing Lemma based on smooth min-entropy; essentially similar statements hold for the {\Renyi} version from Ref.~\cite{dupuis_privacy_2023}. 

\newcommand{\entbnd}{\kappa_\mathrm{const}}
We first discuss fixed length protocols, i.e.~those which simply accept or abort and produce a key of fixed length whenever they accept.
Given the Leftover Hashing Lemma, it might seem that to obtain a secret key, we would simply need to find some constant $\entbnd$ such that any state accepted in the protocol will be such that Alice's pre-amplification string $\PAstring$ has smooth min-entropy (conditioned on Eve's side-information $\EprePA$) of at least $\entbnd$, and then hash the pre-amplification string to a length somewhat shorter than $\entbnd$. There are, however, two critical obstacles that prevent this from being a rigorous claim, which we now describe --- for ease of reference, we recall here that the secrecy definition for fixed-length protocols (\cref{eq:secretfixed}) requires us to prove an upper bound on
\begin{equation}
\Pr[\Oacc]\, d\left(\rho_{\KA \Efinal | \Oacc} \,,\, 
\omega^{\lfixed}_{\KA} \otimes\rho_{\Efinal | \Oacc}\right).
\end{equation}
We highlight in particular that it involves the states conditioned on accepting, and has a prefactor of the accept probability $\Pr[\Oacc]$.
\\

\paragraph*{Conditioning changes entropies.}
This is a slightly technical point, though it brings up issues that will be relevant in our later discussion of variable-length protocols. Namely, it is the question of \emph{which state} we are bounding the smooth min-entropy of. As noted above, the secrecy definition (\cref{eq:secretfixed}) is formulated in terms of the final state conditioned on the protocol accepting, i.e.~$\rho_{\KA \Efinal | \Oacc}$. This implies that to apply the Leftover Hashing Lemma to bound the trace distance in that definition, we must bound the smooth min-entropy of the pre-amplification string \emph{conditioned on accepting}, i.e.~for the state $\rho_{\PAstring \EprePA| \Oacc}$.\footnote{More precisely, as we discuss later in \cref{remark:entropy_after_conditioning}, one may need to instead analyze the smooth min-entropy of the \emph{subnormalized} state conditioned on accepting, and apply a suitable variant of the Leftover Hashing Lemma.} This is not the same as the state $\rho_{\PAstring \EprePA}$ without conditioning on accepting, and thus it would not be valid to apply the Leftover Hashing Lemma based solely on the smooth min-entropy of $\rho_{\PAstring \EprePA}$.
Still, this is mainly a technical issue that can be handled by using appropriate bounds relating the entropies of $\rho_{\PAstring \EprePA| \Oacc}$ and $\rho_{\PAstring \EprePA}$, which we briefly describe in~\cref{remark:entropy_after_conditioning} later. See also \cref{subsubsec:conditioning} regarding some other issues about conditioning on events.\\

\paragraph*{Impossibility of entropy bounds over all states.}
This obstacle is more fundamental, namely: if we consider the smooth min-entropy of \emph{all} states that could occur in the protocol, it is in fact impossible to find a constant lower bound $\entbnd$ other than the trivial value $\entbnd=0$, even conditioned on accepting.\footnote{This issue can technically be avoided by considering \emph{subnormalized} conditional states, but it requires introducing a rigorous formalism of smooth min-entropies for such states; we briefly discuss this in \cref{remark:entropy_after_conditioning} later.} This arises from the observation that Eve can always choose to perform an attack of the following form: first, she simply intercepts and collects all the qubits Alice sends, without measuring them, while forwarding arbitrary states to Bob. 
For simplicity let us now focus on\footnote{Here we only focus on this example for ease of explanation; the general principle basically holds for other protocols as well.} a fully qubit BB84 protocol in which Alice eventually announces all her basis choices, in which case after this announcement, Eve measures her collected states in those bases. The core property of this attack is that it allows Eve to \emph{always} perfectly learn Alice's {entire} {raw data} string, regardless of anything Alice and Bob do. 

We stress the strength of this property: whenever Eve performs this attack, she knows Alice's raw data with probability~$1$, so she knows it \emph{regardless of any event we might condition on} --- including the event that the protocol accepts. 
(While the probability of the protocol accepting under this attack is extremely small, this does not in any way imply that the probability that Eve knows Alice's raw data is small, even conditioned on accepting. Also, the fact that the accept probability is small has no bearing on whether Eve is ``allowed'' to perform this attack --- she can \emph{always} choose to perform this attack, as long as she is willing to accept the small accept probability.)
In particular, this means that the smooth min-entropy $\Hmin^{\esmooth}(\PAstring |\EprePA)$ is nearly\footnote{It is not exactly zero because of the smoothing, but for a state of this form, the smooth min-entropy is still very close to zero.} zero, regardless of whether we are considering $\rho_{\PAstring \EprePA| \Oacc}$ or $\rho_{\PAstring \EprePA}$. Hence the existence of this attack makes it fundamentally impossible to find a nontrivial constant $\entbnd$ such that any state accepted in the protocol has smooth min-entropy at least $\entbnd$. \\

To overcome these obstacles, the security proof has to be constructed in a more precise fashion. The key observation is that as noted above, in the secrecy definition for fixed-length protocols (\cref{eq:secretfixed}), the trace distance term is multiplied by a prefactor of the accept probability --- while this makes it weaker than an analogous claim without this prefactor, it is still sufficient to yield composable security, in the sense discussed in Refs.~\cite{part1,portmann_security_2022}. (In fact, our discussion above implies that the claim without this prefactor is simply not true, in that the attack where Eve collects all of Alice's states yields a final state where the trace distance term is large, as claimed in \cref{eq:trdistcounterexample} and elaborated on in \cref{subsubsec:securityconditionedonaccept}.) We now sketch out various methods to rigorously prove that this definition can indeed be satisfied. 

\subsubsection{Filtered and unfiltered states}
\label{subsubsec:filtering}

For fixed-length protocols, one security proof approach is to partition the set of all states (that could be produced by Eve's attacks) into two subsets, which we refer to as \term{filtered} and \term{unfiltered} states. 
This terminology is based on that used in Ref.~\cite{renner_security_2005}. We define them as follows:
\begin{definition}
Given some fixed-length QKD protocol and some value $\eAT\in(0,1)$, we say a state is \term{filtered} by the protocol if its accept probability is at most $\eAT$. Any state that is not filtered is referred to as \term{unfiltered}. 
\end{definition}

With this partitioning, we can now construct a security proof as follows. Suppose we find an explicit constant $\entbnd$ such that every unfiltered state yields a pre-amplification string with smooth min-entropy (conditioned on Eve's side-information) of at least $\entbnd$.\footnote{Here we maintain flexibility in whether we are referring to the state $\rho_{\PAstring \EprePA| \Oacc}$ or $\rho_{\PAstring \EprePA}$, i.e.~with or without conditioning on accepting. Different proof techniques may rely on either version --- as mentioned above, it is usually still possible to use the latter case by invoking suitable bounds that relate it to the smooth min-entropy after conditioning on $\Oacc$, which we briefly describe in \cref{remark:entropy_after_conditioning} later.} Given this, we would be able to perform the following case analysis for any state generated in the protocol:
\begin{enumerate}
\item \emph{Either} it is a filtered state, in which case the $\Pr[\Oacc]$ prefactor is upper bounded by $\eAT$, and the trace distance term is trivially upper bounded by $1$,
\item \emph{Or} it is an unfiltered state, in which case the smooth min-entropy is at least $\entbnd$, hence we can apply the Leftover Hashing Lemma to bound the trace distance term (after applying suitable techniques to account for conditioning on $\Oacc$, as mentioned above).
\end{enumerate}
From this we can conclude that the secrecy definition is always satisfied, regardless of which case holds (taking $\esecret$ to be the larger value between $\eAT$ and the bound obtained from the Leftover Hashing Lemma in the second case). 
This proof approach avoids the 
obstacle raised above regarding obtaining a nontrivial constant lower bound $\entbnd$, because it
only needs to hold for the set of unfiltered states, not the set of all states --- specifically, that particular attack would usually fall into the set of filtered states, depending on the exact choice of $\eAT$.

\begin{remark}
We strongly emphasize that at no point in the above argument did we speak of the probability that Eve is performing a particular attack, or the probability that she is generating a filtered or unfiltered state. These probabilities are not well-defined concepts in the security definitions, which are couched in terms of statements that must hold against all possible attacks by Eve, rather than some probability distribution over those attacks. Our argument is merely a case analysis statement that \emph{one} of the above cases must hold, and the secrecy definition is satisfied in either case.
\end{remark}

With this, we see that the goal of the security proof basically reduces to finding a lower bound $\entbnd$ on the smooth min-entropy of any unfiltered state. This is the purpose served by the test rounds and the acceptance test: they define the set of unfiltered states (for a given $\eAT$ value), and our goal is to find a nontrivial lower bound on their smooth min-entropy. Note that it is usually impractical to compute the \emph{exact} value of the minimum smooth min-entropy over all unfiltered states; instead, one usually relaxes the set of unfiltered states to some larger set, and lower bounds the smooth min-entropy over that set instead. This is still a valid proof approach, because the resulting value is still a valid lower bound $\entbnd$. 

There are various ``degrees of freedom'' in this proof approach --- for instance, the choice of $\eAT$ and the protocol's accept condition defines the set of unfiltered states, and furthermore there are various possible choices when relaxing this set to a larger set.
Hence when writing a security proof, one has to make these choices in such a way that a valid and reasonably tight lower bound $\entbnd$ can be practically computed.

\subsubsection{Entropy bounds with accept-probability dependence}
\label{subsubsec:probdependentbounds}

As an alternative to the above case analysis, another proof approach is as follows. Rather than attempting to find a \emph{constant} lower bound $\entbnd$ on the smooth min-entropy or {\Renyi} entropy (conditioned on $\Oacc$) that holds only for states in some restricted set (i.e.~the unfiltered states), somewhat informally speaking we can find a bound that instead has some dependence on the accept probability $\Pr[\Oacc]$, but holds for all states.\footnote{In proofs that involve subnormalized conditional states (see \cref{remark:entropy_after_conditioning} later), we view this dependence as being implicitly mediated by the entropy definitions for subnormalized states.} 
This dependency on $\Pr[\Oacc]$ can potentially avoid the obstacle imposed by the attack described above, which has a very small value of $\Pr[\Oacc]$, i.e.~such a bound is possible in principle as long as it becomes trivial whenever $\Pr[\Oacc]$ goes to zero.

While the probability $\Pr[\Oacc]$ cannot be operationally observed, it is still well-defined, and in some situations it is possible to construct a security proof such that this dependency is correctly ``compensated for'' by the $\Pr[\Oacc]$ dependence in the secrecy definition. Such security proofs have been obtained in e.g.~\cite{dupuis_privacy_2023,kamin_finite-size_2024}, and in the rest of this work we will occasionally comment on which proof techniques allow this approach. 

\subsubsection{Variable-length protocols}

For variable-length protocols, the analysis is more elaborate, so here we only give a very informal summary. Note that the secrecy definition for variable-length protocols qualitatively states that the final state is close to a \emph{mixture} of states where the key is uniform and independent of the side-information. With this in mind, the idea of the security proof is essentially to use the test-round data to construct a ``statistical estimator'' of the smooth min-entropy or {\Renyi} entropy of the pre-amplification string, then hash to a length determined by that statistical estimator. If it can be shown that this statistical estimator bounds the true entropy value with high probability, then under some additional conditions, it can be shown that the variable-length secrecy definition holds. However, formalizing this intuition requires some careful analysis; furthermore, recall that one must account for the fact that conditioning on each value for the output key length can change the entropy value. Hence we defer further discussion to \cref{subsec:variablelength}. 

\subsection{Reduction theorems: optical modes to finite dimensions}

QKD protocols are typically implemented via quantum optical systems, which are described by (tensor products of) infinite-dimensional Hilbert spaces. Thus, when discussing optical implementations of QKD protocols, some subtle problems arise due to the dimensionality of the spaces. For instance, for numerically computing key rates, infinite-dimensional optimisations cannot be given to a computer.
There has been much work to develop tools to reduce the security proofs of these infinite-dimensional protocols to the analysis of ``equivalent'' finite-dimensional protocols.
Although the ideas are fairly similar, we broadly classify these tools into two subgroups:
\begin{enumerate}
    \item Source maps \cite{nahar_imperfect_2023} that help reduce the analysis of infinite-dimensional sources to equivalent finite-dimensional sources.
    \item Squashing maps \cite{beaudry_squashing_2008,tsurumaru_security_2008,tsurumaru_squash_2010,fung_universal_2011,gittsovich_squashing_2014,zhang_security_2021,upadhyaya_dimension_2021, nahar_postselection_2024} that help reduce the analysis of infinite-dimensional detection setups to equivalent finite-dimensional detection setups.
\end{enumerate}

\subsubsection{Source maps}

\begin{figure}[ht]
    \begin{subfigure}{\linewidth}
        \centering
        \includegraphics[scale = 0.5]{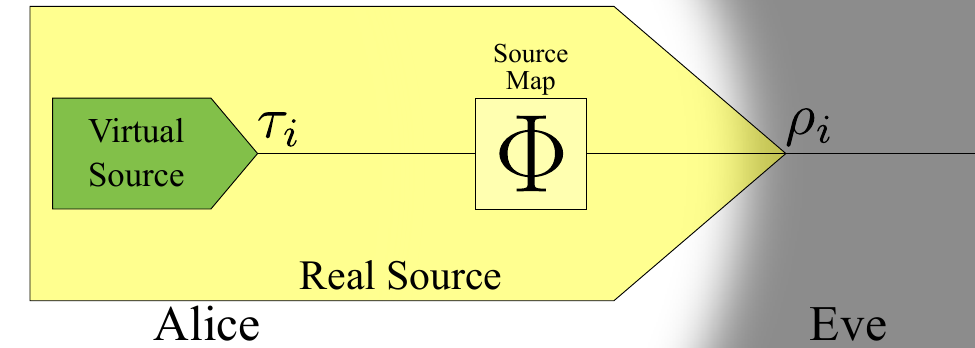}
        \caption{We can model the real source as a virtual source followed by a source map since they both have the exact same output.}  \label{fig:SourceMapBefore}
    \end{subfigure}
    \hfill
    \begin{subfigure}{\linewidth}
        \centering
        \includegraphics[scale= 0.5]{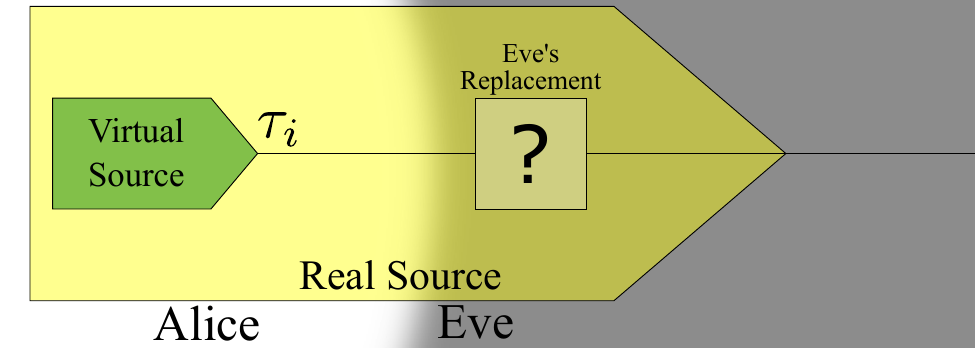}  
        \caption{Once we give Eve control of the source map, she can perform any physical operation on the output of the virtual source, including the source map $\Phi$ if reproducing the real state is optimal for her.}  \label{fig:SourceMapAfter}
    \end{subfigure}
    \caption{The real source can always be replaced by the virtual source in security proofs if they are related via a source map since the virtual source gives Eve more power.}  \label{fig:SourceMaps}
\end{figure}

The main idea behind source maps is fairly intuitive, as described in \cref{fig:SourceMaps}. The real source is modelled as a virtual source followed by a `source map', and the source map is then `given' to Eve. In particular, it is then assumed that Eve is allowed to replace the source-map with any map of her choice. Thus, the security analysis can be restricted to the more convenient virtual source.
Formally, given the set of (infinite-dimensional) states $\{\rho_i\}$ used in the protocol, we find a set of (finite-dimensional) states $\{\tau_i\}$ such that there exists a channel $\Phi$ where $\rho_i = \Phi(\tau_i)$ for all $i$. This channel $\Phi$ constitutes the \emph{source map} which is `given' to Eve, and thus any security analysis that assumes that Alice prepares the states $\{\tau_i\}$ would also hold for the actual protocol where Alice prepares the states $\{\rho_i\}$ as this only gives Eve more power. See \cite[Lemma 8]{nahar_postselection_2024} for a more formal proof of this.

An important source map that is often used for decoy-state protocols is tagging \cite{gllp_security_2004}. When the real source emits block-diagonal states, we can replace it with a virtual source that probabilistically emits either a finite cut-off of the real states, or a `tag' with information, readable by Eve, that reveals the exact state prepared. For instance, a polarization encoded decoy-state protocol utilizes a source that prepares fully phase-randomized coherent states which are block-diagonal in the total number of photons in all polarizations. This source can then be replaced with a virtual source that probabilistically emits a vacuum state, a single-photon state, or a ``tag".

\subsubsection{Squashing maps}
\label{subsubsec:squashing}

A large variety of squashing maps exist to reduce detection setups to finite dimensions. We summarize the majorly used squashing maps (also called ``squashers") in \cref{tab:squashingTypes}. We briefly describe the main idea behind each of them.
\begin{table*}[t]
    \caption{List of squashers}\label{tab:squashingTypes}
    \def\arraystretch{1.5} 
    \setlength\tabcolsep{.28cm}
    \begin{tabular}{p{4cm} p{7cm} p{5cm}}
    \toprule
    \textit{Type of squashing} & \textit{Assumptions/ Conditions} & \textit{Established Applicability}\\
    \toprule
    Simple squashers \cite{beaudry_squashing_2008,gittsovich_squashing_2014} & Protocol-dependent construction --- need to check for each protocol separately (e.g. - equal loss in all detectors for active BB84 setup, no dark counts.) & Proof-technique independent \cite[Lemma 6]{nahar_postselection_2024}\\
    Universal squasher \cite{fung_universal_2011} & No dark counts, equal loss in all detectors, etc. & Phase error correction/EUR-based proof techniques \cite{fung_universal_2011}\\
    Flag-state squasher \cite{zhang_security_2021} & Block-diagonal structure of POVM elements & EAT-based proof techniques \cite[Appendix B]{kamin_finite-size_2024}, IID analysis \cite[Theorem 1]{zhang_security_2021}\\
    Dimension-reduction method \cite{upadhyaya_dimension_2021} & None & EAT-based proof techniques\footnote{Note that although this has been used for CV-QKD protocols, the extension to DV-QKD protocols is straightforward.} \cite[Theorem 2]{primaatmaja_discrete-modulated_2024}, IID analysis \cite{upadhyaya_dimension_2021}\\
    Weight-preserving flag-state squasher \cite[Section IV.B.1.]{nahar_postselection_2024} & Block-diagonal structure of POVM elements & Proof-technique independent \cite[Lemma 6]{nahar_postselection_2024}\\
    \toprule
    \end{tabular}
    \def\arraystretch{1}
\end{table*}

\textbf{Proof-technique independent squashers:} The simple squashers and weight-preserving flag-state squasher use the same general principle \cite{tsurumaru_security_2008,beaudry_squashing_2008,tsurumaru_squash_2010,gittsovich_squashing_2014} depicted in \cref{fig:squashing}, similar to source maps. These squashers allow us to directly replace the infinite-dimensional POVM with a finite-dimensional POVM for a QKD security proof. That is, they work by proving that the security of the QKD protocol with the finite-dimensional POVM implies the security of the QKD protocol with the infinite-dimensional POVM \cite[Lemma 6]{nahar_postselection_2024}. The security proof for the former protocol can then be obtained via any proof technique of one's choice. 
\begin{figure}
    \centering
    \scalebox{1}{\begin{tikzpicture}

\begin{scope}[shift = {(1,0)}]
    \pic {detector={detinf, $\infPOVM$, black}};
    
    \draw (-5,0) -- ([xshift=-0.25cm]detinf.west); 
\end{scope}
        
    \begin{scope}[shift={(-1,-4)}]
        \node[process](SquashingMap){$\Lambda$};
        \pic [right of = SquashingMap,xshift = 2cm]{detector={detfinideal, $\finPOVM$,black}};
        
        \draw (-3,0) -- (SquashingMap);
        \draw (SquashingMap) -- ([xshift=-0.25cm]detfinideal.west);

        \draw[fill=red, opacity=0.2] 
        ([xshift=1.1cm,yshift=0.4cm]SquashingMap.north) 
        .. controls +(right:0.5cm) and +(right:0.5cm) .. 
        ([xshift=1.1cm,yshift=-0.4cm]SquashingMap.south)
        -- ([xshift=-3.0cm,yshift=-0.4cm]SquashingMap.south)
        -- ([xshift=-3.0cm,yshift=0.4cm]SquashingMap.north)
        -- cycle;
        
        \draw[thick,-] ([xshift=1.1cm,yshift =0.4 cm]SquashingMap.north) .. controls +(right:0.5cm) and +(right:0.5cm) .. ([xshift=1.1cm,yshift =-0.4 cm]SquashingMap.south);

        \node[opacity=0.8, text width=3cm, align=center, font=\large] at ([xshift=-0.7cm,yshift=-0.5cm]SquashingMap.west) {\textbf{Eve}};

    \end{scope}
    \node at (0,-2) {\scalebox{2}{$\Big\Updownarrow$}};
    
\end{tikzpicture}}
    \caption{The infinite-dimensional POVM $\infPOVM$ can be modelled as a squashing map $\Lambda$ followed by a finite-dimensional POVM $\finPOVM$. Giving the squashing map $\Lambda$ to Eve lets us restrict our analysis to the finite-dimensional POVM $\finPOVM$.} \label{fig:squashing}
\end{figure}

\textbf{Universal squasher:} This can be thought of as a generalisation of Ref.~\cite{tsurumaru_security_2008,tsurumaru_squash_2010} (and \cref{fig:squashing}), where the exact equivalence is replaced with an approximate equivalence. In contrast to the proof-technique independent squashers, the universal squasher only attempts to preserve \emph{bounds} on the probabilities of relevant events during the squashing. In particular, it bounds the increase in the number of phase errors on replacing the infinite-dimensional measurement with a qubit measurement, which can then be used with the phase error correction/EUR proof-techniques.

\textbf{Flag-state squasher:} This is a generalisation of the simple squasher, that attempts to use the same idea depicted in \cref{fig:squashing}. It attempts to make formal the intuition that although the detection setups are infinite-dimensional, the states typically live `mostly' in a finite-dimensional subspace. Thus, the squashing map preserves the finite-dimensional subspace and fully announces the detection event in the rest of the space, formalised through classical states called `flags'. Such a map would always exist if the POVM is block-diagonal with blocks corresponding to the preserved finite-dimensional subspace, and the rest of the space. However, the reason this squasher is not proof-technique independent is that the squashing map cannot be completely `given' to Eve. An additional bound on the weight of the state in the preserved subspace is required to ensure that this squasher is useful.

\textbf{Dimension-reduction method:} Although this originated as a way of generalising the flag-state squasher to CV-QKD protocols, where the POVM did not possess the required block-diagonal structure, it is more widely applicable to speed up numerical computations. This method can be thought of as a way to generically replace a semidefinite program (SDP) with an SDP where all the optimisation variables live in some smaller subspace. In the context of QKD, this translates to rigorously replacing the (infinite-dimensional) single-round key rate SDP with a finite-dimensional key rate SDP to facilitate numerical computation. Thus, this technique is well-suited to proof-techniques that are easily able to reduce the analysis to a single-round quantity that must be numerically computed such as the EAT-based proof techniques.

Note that only the simple squashers and the weight-preserving flag-state squasher can be used independently of proof-technique. Also note that the flag-state squasher, dimension-reduction method, and the weight-preserving flag-state squasher need a way to bound on the weight outside some finite-dimensional subspace before they can be used\footnote{These methods make formal the following intuition: although the detection setups are technically infinite-dimensional, the states entering the setup live `mostly' in some finite-dimensional subspace. Thus, they require an additional way to bound the weight outside this finite-dimensional subspace.}.

\subsection{Decoy-state methods}\label{subsec:decoy_general}

While the single-photon subspace constitutes the ideal operational regime for standard QKD protocol designs, practical implementations typically rely on weak coherent pulse (WCP) sources for technological feasibility. However, multi-photon pulses are vulnerable to the photon-number splitting (PNS) attack \cite{Lutkenhaus_estimates_1999,Bennett_experimentalquantumcryptography_1992,Brassard_limitationsonpractical_2000,Lutkenhaus_security_2000}, where an adversary can split off a photon from a multi-photon signal and measure it after the basis announcement, thereby gaining full information without introducing errors. Crucially, Eve can exploit the channel loss to selectively suppress the secure single-photon signals (which resists the PNS attack) and preferentially transmit the compromised multi-photon signals.
Consequently, the Poissonian nature of WCP sources makes them a poor approximation of the ideal source, as the unavoidable multi-photon emissions severely impact the secure key rate if not specifically addressed.

Thus, decoy-state QKD protocols \cite{Lo_decoystate_2005,Hwang_qkdhighloss_2003,Wang_beating_2005} are often implemented to mitigate the impact of the PNS attack. To understand the intuition behind this approach, it is instructive to first consider a hypothetical source that randomly emits a state with a definite photon number (as is effectively the case for phase-randomized WCP sources), but additionally announces the photon number of each emission. With access to this explicit photon-number information, Alice and Bob could categorize their detection data by photon number. This would allow them to independently calculate the observed statistics for the single-photon rounds. This prevents Eve from masking a PNS attack by replacing secure single-photon signals with compromised multi-photon ones. Consequently, Alice and Bob can extract the full secret key contribution from the single-photon rounds.

In practice, WCP sources do not announce the photon number of each emission, and therefore do not realize the hypothetical scenario discussed above where photon numbers produced are known exactly. However, the decoy-state method serves as a statistical substitute for this information. The protocol involves modulating the intensity of the WCP source to produce varied photon-number distributions. Importantly, Eve cannot distinguish between an $m$-photon pulse originating from different intensity settings; her interaction is strictly constrained to depend only on the photon number and is independent of the intensity setting chosen by Alice (since the setting is not known to Eve when she is implementing her attack).
This structure allows the decoy-state analysis to be cast as a set of linear constraints\footnote{More recent advancements in decoy-state techniques allow one to state this analysis as an SDP, which are more robust to imperfections \cite{nahar_imperfect_2023} and yield tighter bounds \cite{kamin_improved_2024}. However, for pedagogical clarity we discuss the more commonly used linear program formalization at this point.}.

Broadly speaking, there are two ways to integrate this decoy-state analysis into the security proof. The first, and most common, treats the decoy-state analysis as a precursor to the key rate computation. Here, the linear constraints allow one to find the bounds on single-photon data that are mathematically consistent with the observed data. These estimated statistics are then substituted into a key rate formula derived specifically for the single-photon subspace. The mathematical justification for this modularity depends on the proof technique employed; for instance, by leveraging entropic properties, one can decompose the full key rate formula into a sum of independent terms from each photon-number subspace. Under the pessimistic assumption that Eve possesses full information about multi-photon emissions, the contributions from those subspaces vanish. Consequently, the calculation reduces to a key rate formula or optimization that depends only on the single-photon subspace.

Alternatively, the decoy-state analysis can be integrated directly into the security analysis \cite{kamin_finite-size_2024,kamin2025renyisecurityframeworkcoherent}. Rather than obtaining bounds on the single-photon data separately, the decoy-state data is directly used as constraints in the key rate optimization. While this approach gives better key rates, its development is recent and it can only be used with proof techniques that phrase the key rate as a numerical optimization problem (see also \cref{sec:families}). 

We stress that, at a fundamental level, decoy-state protocols do not change the core task we wish to accomplish: bounding the appropriate entropic quantity, as in \cref{subsec:estimating_entropy}, using the observed data. Decoy-state methods, however, introduce an additional layer of complexity to this process. As the exact finite-size details of the decoy-state analysis depend on the specific proof technique being used, we defer further discussion to \cref{sec:families}.

\section{General common gaps in existing 
proofs} \label{sec:commonGaps}
Having established an overview of the critical concepts necessary for QKD security proofs in the previous section, we now turn our attention to some common gaps in the existing literature. Despite significant progress in the field, certain aspects of security analysis remain incomplete or underexplored. In this section, we will highlight these gaps and discuss their implications.

\subsection{Acceptance conditions}
\label{subsec:gap_AT}

\
As outlined in \cref{subsec:estimating_entropy}, estimating the smooth min-entropy of states conditioned on the protocol accepting is a crucial step in security proofs\footnote{For fixed-length protocols, this acceptance condition is realized by checking whether the observed statistics fall within a predefined acceptance set.}. To achieve this, as described in \cref{subsubsec:filtering}, the set of all possible states is divided into “filtered” and “unfiltered” states. The filtered states are required to satisfy the condition $\Pr[\Oacc] \leq \eAT$, and the entropic bound is only derived for the unfiltered states. Crucially, constructing the set of filtered states must be done carefully, for a well-defined acceptance set, and using appropriate concentration inequalities to ensure $\Pr[\Oacc] \leq \eAT$.

In particular, it does not suffice to simply say that the protocol accepts if the observed frequency distribution matches the expected honest behaviour. This would essentially be what has been termed a \term{unique-acceptance} or \term{single-point-acceptance} protocol, which only accepts if one specific frequency distribution is observed. Such a protocol is \textit{highly} impractical, since it would abort nearly all the time. A practical acceptance set must include some ``tolerance'' to ensure that at least the honest behaviour is accepted with reasonable probability.
While this issue might appear superficial, we highlight that the key rate computations in a number of works are implicitly only valid for a unique-acceptance protocol, even if not specified within the work itself. 

Specifically, a common approach to compute finite-size key rates is to characterize the set of unfiltered states via some ``neighbourhood'' of some frequency distribution $\mbf{F}$. However, if that frequency distribution $\mbf{F}$ consists solely of the honest behavior of the protocol, without a clear specification and analysis of an acceptance set, then the resulting key rate may effectively only be valid for a unique-acceptance protocol.\footnote{More precisely: for the proof structure in \cref{subsubsec:filtering} for fixed-length protocols, any ``useful'' characterization of the set of unfiltered states would require this neighbourhood to be \emph{strictly larger} than the acceptance set. However, as discussed above, any realistic acceptance set must \emph{itself} be a nontrivial neighbourhood of the honest behaviour, and this neighbourhood is defined completely separately from the set of unfiltered states. Failing to take both of these effects into account may result in a security proof that is effectively only valid for a unique-acceptance protocol.}
On the other hand, if that frequency distribution $\mbf{F}$ is a function of the observed values in the protocol, then this inherently describes a variable-length protocol, not a fixed-length one. In the latter case, a rigorous variable-length security proof is required for that protocol, as its security does not straightforwardly follow from fixed-length proofs (see \cref{subsec:variablelength}).

Another common issue arises from the unjustified use of Gaussian concentration inequalities, such as 10-$\sigma$ confidence intervals (see, e.g. Ref.~\cite{bourgoin_experimental_2015}), in characterizing the unfiltered set. This approach is problematic because the underlying random variables may not follow a Gaussian distribution. Thus, while this interval might be chosen very generously, it might still not satisfy the property of $\Pr[\Oacc] \leq \eAT$. Hence, the structure for a proof as outlined in \cref{subsubsec:filtering} is not satisfied anymore.

Furthermore, another potential problem is adding finite-size effects only to parts of the analysis, for example in Ref.~\cite{zhang_improved_2017}, finite-size effects were only considered during the decoy-state analysis. This does not lead to correct key rates, as procedures like the privacy amplification or error verification step introduce further finite-size corrections which needs to be accounted for.

\subsubsection{Choice of testing rounds}
\label{subsubsec:gap_testing}
As already mentioned in \cref{subsubsec:step_testvsgen}, there are generally two options for determining test rounds; One can choose a sample set of fixed size, or  independently and probabilistically assign each round for testing. In the latter case, the total number of test rounds is variable. The most straightforward mistake which could arise here, is that key rate formulas for the former case are used in the latter case, and vice versa.

In any version of the sampling procedure used in a security proof, it is crucial to apply the appropriate concentration inequalities (see  \cref{tab:Concentration Inequalities}) that are valid for the specific scenario and accurately reflect the sampling process. For instance, Serfling's inequality can be directly applied when randomly selecting a fixed-length subset of a classical string for testing. However, it can also be adapted to scenarios where each round is independently and probabilistically selected for testing, as long as the proof justifies that the statistics can be correctly analyzed with such an approach. Furthermore, when applying concentration inequalities, it is also important to clearly specify the random variables involved and verify that they satisfy the required conditions for the chosen inequality. For example, Hoeffding's inequality requires independent random variables, while Azuma's inequality applies to martingales. Ensuring these conditions are met is essential for the validity of the security proof.

Additionally, depending on the proof technique, one may want to reduce certain finite-size correction terms to depend 
only on, say the number of key generation rounds (rather than total number of rounds). One scenario where this is useful is in using the asymptotic equipartition theorem (AEP) \cite{tomamichel_fully_2009} to reduce multi-round state to single round quantities. However, such theorems are typically valid for scenarios with a fixed number of rounds. Hence, the AEP cannot be directly applied if, for example, we restrict the entropy estimation to (the variable number of) key generation rounds only. While this procedure can be made rigorous with special care, we do not go into its details here.

\subsubsection{Iterative sifting}
\label{subsubsec:iterative}
The theoretical analysis of most QKD protocols is undertaken under the setting where a fixed total number of rounds are used. Additionally, all public announcements are only performed after all signals have been sent and received. Therefore, Eve does not have access to public announcements while attacking the quantum channel.

Protocol implementations on the other hand may involve Alice and Bob performing announcements ``on-the-fly". That is to say, they may make announcements for a given round soon after the round is measured but before the end of the prepare-and-measure step\footnote{For example, Bob may announce the rounds he did not detect, so that Alice can immediately delete classical data she stored from those rounds rather than storing them until the end of the prepare-and-measure step.}. Furthermore, they may use these announcements to determine when to stop sending and receiving signals. (For instance, Alice and Bob may send and receive signals until there are at least some number of detection events in a specific basis.) This aspect of protocol implementations and its variations are known as ``iterative sifting". 
The theoretical analysis of such protocols must be handled with care. Many works incorrectly analyze protocols with iterative sifting via methods that are only valid if all announcements take place \textit{after} all signals sent are received. These issues were first pointed out by Ref.~\cite{pfister_sifting_2016}.

Protocols with on-the-fly announcements (but with a fixed number of total rounds) are easily handled in the GEAT framework \cite{metger_generalised_2022}, but then suffer from the sequential attack assumption (see \cref{subsubsec:geat}). They can be straightforwardly handled using MEAT \cite{arqand_marginal_2025}. For analyses within phase error based approaches (see \cref{subsec:eur,subsec:phaseerror}), see Refs.~\cite{wang_phaseerror_2025,mizutani2025protocolleveldescriptionselfcontainedsecurity}. For techniques to additionally handle the scenario where the total number of rounds also depends on these announcements, see Ref.~\cite{tamaki_security_2018}.

\subsection{Abort decisions}
\label{subsec:gap_abort}
\subsubsection{Aborting during authentication} \label{subsubsec:gap_abort_auth}
Many QKD security proofs do not discuss the authenticated classical channel in much detail, as there exist composably secure constructions of such channels.
However, there is a technical issue that should be addressed here. Specifically, most such constructions do not give a channel that \emph{always} successfully transmits the classical message, but rather one that may sometimes abort (informally: whenever Eve has attempted to modify the message). Therefore, a full protocol specification and security proof must discuss this possibility --- clearly, the natural option is for the protocol to abort whenever the authentication procedure aborts, but the security proof should account for this accordingly. 

Moreover, in the setting where the authentication channel is allowed to abort, we note that typical constructions of such channels result in only the \textit{receiving} party learning of the abort, while the sending party remains unaware. This asymmetry implies that Eve can always force an abort on one side while the other party accepts (by exactly mimicking the honest behaviour up until the last message sent between the parties, which she then interferes with to force the receiving party to abort). Note that this scenario is not covered by the security definition in \cref{def:security}, i.e, \cref{def:security} cannot be satisfied for any QKD protocol where authentication is allowed to abort and only the receiving party learns of the abort. 

However, we do not expect this gap to pose a fundamental problem. Specifically, we believe that QKD security proofs established under the assumption that authentication never aborts can be suitably adapted to the setting where the authentication may abort -- potentially resulting in a final state where only one party has accepted (see also \cite[Section VII A]{portmann_security_2022}). However, we note that to our knowledge, the correct security definition
in this setting 
is only explicitly discussed
in Ref.~\cite{part1}. Moreover, the exact modifications needed to adapt the usual protocols and security proofs (established under the assumption that authentication never aborts) to this setting are minimal, and are discussed in Ref.~\cite{inprep_authentication}.  

\subsubsection{Aborting during error verification}
Similarly, the protocol may also abort due to the error verification step, and this also should be taken into account when analyzing the state conditioned on the protocol accepting. This is more frequently addressed as compared to the above point, but is still sometimes overlooked.

\subsection{IID assumptions}

\subsubsection{Collective attacks are not a good approximation to practical restrictions}

A common assumption made in some existing work is that Eve's attack is independent and identically distributed (IID) in each round, i.e., she interacts with Alice's signal states in each round exactly the same way, independent of prior protocol rounds. We emphasise here that this assumption cannot be motivated by limiting Eve to currently practical attacks. That is, with currently available technology, it is easy to implement a non-IID attack.

For example, Eve could measure every odd pulse in the X-basis, and measure every even pulse in the Z-basis. This easily-implementable attack violates the IID assumption. Thus, it is important to prove security against generic attacks.

While there exist proof techniques~\cite{christandl_postselection_2009,nahar_postselection_2024} that extend security proofs against IID collective attacks to coherent attacks, there are some conditions necessary to apply those techniques, and they usually change the security parameter and/or the key rate (see \cref{subsec:postselection} for further discussion). Hence they cannot be generically used to claim that it suffices to analyze the IID case.

\subsubsection{Misuse of symmetry arguments}

It was shown in Refs.~\cite{kraus_lower_2005,renner_information-theoretic_2005}
that the symmetry properties of some protocols, such as the qubit BB84 and 6-state protocols, can be used to reduce the security analysis to a scenario where Alice and Bob hold IID Bell-diagonal states. Therefore, for such protocols it suffices to prove security against collective attacks. 

However, care must be taken regarding the scope of this claim. Specifically, the symmetry analysis used in those works relied heavily on properties of genuine qubit systems, and does not appear to straightforwardly generalize to decoy-state protocols, which involve for instance high-dimensional states consisting of mixtures of different photon numbers. In particular, their results do not imply for instance that photon-number manipulations by Eve have to be IID --- an example of an operation that Eve can certainly perform is to perform a photon-number-splitting attack in some rounds but not others, and the proof techniques in those works are insufficient to show that such possibilities can be excluded without loss of generality. Therefore, it is not correct to use the results of Refs.~\cite{kraus_lower_2005,renner_information-theoretic_2005} to claim that security proofs for decoy-state BB84 protocols can be reduced to collective attacks. 

\subsubsection{Use of concentration inequalities} \label{subsec:concentrationinequalities}
During the process of determining the secure key length, one is often required to bound certain quantities such as the phase error rate. These quantities are random variables and one only has access to a small sample from the test rounds. 
In the fixed-length scenario, one estimates e.g. the error rate based on the acceptance test, and in the variable-length case one derives bounds based on the observed statistics.
For both cases, a common approach to construct intervals in which these quantities lie with high probability is to apply concentration inequalities from classical statistics --- in \cref{tab:Concentration Inequalities}, we present the most commonly used ones along with their assumptions.

We highlight however that there are several critical points to note when applying these concentration inequalities:\\

\paragraph*{Ensuring assumptions are satisfied.}
Each of these concentration inequalities is \emph{only} valid under specific assumptions, thus when applying any of them in a security proof, we stress that it must be rigorously justified that the relevant assumptions are satisfied. For example, as previously stated in \cref{subsubsec:gap_testing}, when applying Hoeffding's inequality, one needs to  make sure that the random variables being considered are indeed independent. 
Serfling's inequality technically requires a subset of fixed-length to be chosen for testing, but protocols may independently decide to test each round. The Serfling statement must then be modified accordingly to fit the protocol.
Finally, Azuma's inequality requires the random variables to be martingales, and hence one must construct appropriate random variables that can be proven to be martingales.\\

\paragraph*{Distinguishing fixed parameters from random variables.} When applying any of these concentration inequalities, it is important to keep track of which quantities are fixed {parameters} (i.e.~constant values that describe the behaviour under a given attack, without any randomness) and which quantities are random variables, so that the concentration inequalities can be validly applied. This is particularly critical if trying to convert an analysis of single protocol rounds to a finite-size analysis of the full protocol. To take a specific example, if we denote the expected detection probability in round $i$ (under some given attack) as $q_i$, then these values $q_i$ are well-defined parameters, but not all of the concentration inequalities in that table are related to those parameters. In particular, Serfling's inequality does not involve those parameters $q_i$ in any way. As for Azuma's inequality, when applying it in security proofs, the martingales constructed are usually not the observed values themselves, but rather they are specific linear combinations of those values and some \term{conditional expectations}. Conditional expectations are random variables, \emph{not} parameters; therefore, they are also not directly related to those parameters $q_i$, and Azuma's inequality does \emph{not} imply (for instance) that the observed detection frequency will lie close to the average of the $q_i$ values.\\

\begin{remark}
One can construct simple counterexamples that demonstrate how failing to track the above points can lead to incorrect conclusions.  For instance, consider $n$ binary random variables $X_i$ with the following simple distribution: with probably $1/2$ we have $X_i=0$ for all $i$, and with probability $1/2$ we have $X_i=1$ for all $i$. In that case, the expected value of each $X_i$ is simply $E[X_i]=1/2$; however, it is clear that the observed sum $\sum_i X_i$ is \emph{always} either $0$ or $n$, so it is \emph{never} close to its expected value $E[\sum_i X_i] = \sum_i E[X_i] = n/2$. Therefore, \emph{no concentration inequality} can claim that the sum $\sum_i X_i$ converges to its expected value with high probability in this scenario, and any such claim that does not impose enough conditions to rule out this scenario is hence necessarily incorrect.\footnote{Regarding the specific inequalities we list here: Hoeffding's inequality and binomial bounds require independent random variables and hence do not apply to this scenario; whereas Serfling's inequality and Azuma's inequality do not discuss convergence to single-round parameters or expected values, and hence only provide results of a rather different nature from this claim.}
\end{remark}

\begin{table*}[t]
    \caption{List of Commonly used Concentration Inequalities.    }
    \label{tab:Concentration Inequalities}
    \def\arraystretch{1.5} 
    \setlength\tabcolsep{.22cm}
    \begin{tabular}{p{4cm} p{5cm} p{7.5cm}}
    \toprule
    \textit{Inequality} & \textit{Conditions} & \textit{Statement} \\
    \toprule
    Hoeffding's Inequality \cite[Theorem 2]{hoeffding_probability_1963} & \(X_1, \dots, X_n\) independent random variables with \(a_i \leq X_i \leq b_i\). & $ \Pr[ \abs{\sum_i (X_i - E[X_i]) } \geq t] \leq 2\exp(-\frac{2t^2}{\sum_i^n\left( b_i - a_i\right)^2}) $ \\
    Binomial bounds & \(X_1, \dots, X_n\) IID Bernoulli random variables, then \(\sum_i X_i \) is a binomial random variable & Concentration inequalities given by cumulative distribution function of binomial random variable \\
    Serfling inequality \cite[Corollary 1.1]{serfling_probability_1974} & \(X_1, \dots, X_n\) drawn randomly without replacement from values \(x_1,\dots,x_N\) for \(n\leq N\) & $\Pr[ \sum_i X_i - n\mu  \geq nt] \leq 2\exp(- \frac{2nt^2}{(1-f_n^{*})(b-a) } ) $, for \(\mu = \sum_i x_i/n, f_n^{*} = \frac{n-1}{N}, a = \min_i x_i\) and \(b = \max_i x_i\). \\
    Azuma's inequality\footnote{Often, the tighter Kato's inequality \cite{kato2020concentrationinequalityusingunconfirmed} is used in place of Azuma's inequality.} \cite{azuma_weighted_1967} & $\{X_i\}_{i=0,1,\dots}$ a martingale and $\abs{X_i - X_{i-1}} \leq c_i $ for all \(i\geq 1\) & $\Pr[ \abs{ X_n - X_0 } \geq t] \leq 2\exp(-\frac{t^2}{2\sum_i^n c_i}) $ \\
    \toprule    
    \end{tabular}
    \def\arraystretch{1}
\end{table*}

\subsection{Error correction} \label{subsec:errorcorrection}

Alice and Bob must implement some amount of error correction to rectify the error in their data.  This invariably leaks information to Eve, and the security proof must correctly account for such information leakage. This leakage must be analyzed carefully. See \cref{subsec:step_ECandEV} for a discussion on the variety of possible implementations, which we recall in brief below. As before, we assume that one party has already obtained the pre-amplification string, and the other party is attempting to obtain the same string. 

\subsubsection{Implementation choices}

\paragraph*{One-way error correction.} Alice holds the pre-amplification string, and sends some number of bits to Bob, who uses this together with his private data to produce a guess for Alice's pre-amplification string. 

\paragraph*{Reverse reconciliation.} A slightly different form of one-way error correction, where Bob holds the pre-amplification string and sends classical data to Alice. The security proof must account for the fact that Bob holds the pre-amplification string in this case; see \cref{subsubsec:Hminchainrule} below. 

\paragraph*{Two-way information reconciliation.} 
More complicated protocols that involve multiple rounds of communication between Alice and Bob may be employed. Depending on the situation, this may involve a scenario where either one of Alice or Bob holds the pre-amplification string, which is then followed by a two-way information reconciliation protocol. Some examples of such protocols are Cascade \cite{brassard_secret-key_1994} and Winnow \cite{buttler_fast_2003}. Note that this may also cover more elaborate scenarios where the pre-amplification string itself is obtained through the interaction between Alice and Bob.

\subsubsection{Technical statement  for entropy based techniques}
\label{subsubsec:Hminchainrule}

For entropy based proof techniques that go through the Leftover Hashing Lemma, the leakage due to error correction is typically analyzed using a chain rule of the form:
\begin{equation} \label{eq:ECchainrule}
    \Hmin^{\esmooth}(\PAstring| \CEC \En)_\rho \geq \Hmin^{\esmooth}(\PAstring|\En)_\rho - \log(|\CEC|),
\end{equation}
where $\PAstring$ denotes the register storing the pre-amplification string, and $\En$ denotes all of Eve's registers excluding the hash-choice for PA, and excluding the communication during error correction and error verification ($\CEC$). Thus, Eve's uncertainty about the pre-amplification string is reduced by 
at most $\log(|\CEC|)$. 
To use this chain rule for the analysis of a general error correction protocol, $\log(|\CEC|)$ is given by \textit{the number of bits needed to ``encode" all possible transcripts} of the error correction protocol. 

We emphasize the fact that in order to combine the above inequality with the Leftover Hashing Lemma, it is important that we are analyzing the smooth min-entropy of the pre-amplification string $\PAstring$ specifically (conditioned on various other registers), not some other strings, such as the raw data or any other ``intermediate'' string values produced during the processing steps. In particular, this means that for protocols with reverse reconciliation (where Bob is the first party to hold the pre-amplification string), this means that typically it should be applied to \emph{Bob's} pre-amplification string, not Alice's.

\begin{remark}
    In order to use this inequality in QKD security proofs, we typically apply it on states conditioned on specific events. For example, for fixed-length protocols one considers the state $\rho_{ | \Oacc}$. Then $|\CEC|$ is the total number of values this register can take conditioned on $\Oacc$. For variable-length protocols, one may consider events corresponding to various observations during the protocol run, and determine the error correction protocol parameters (and therefore $|\CEC|$) based on the observed event. 
\end{remark}

\subsubsection{Computing the cost} 
The computation of $\log(|\CEC|)$ is straightforward, but depends on some important details of the error correction protocol that is implemented.

\begin{enumerate}
    \item \textbf{One-way EC with fixed number of bits:} Since the number of bits leaked $\leak$ is fixed, $\leak$ many bits are enough to encode all possible transcripts.
    \item \textbf{One-way EC with maximum number of bits:} Here the number of bits leaked is less than or equal to $\leak$. It turns out that a $\leak+1$ bit register is enough to encode all possible transcripts\footnote{Since $2 + 2^2 \dots 2^{\leak} \leq 2^{\leak+1}$.}.
    \item \textbf{Two-way EC:} The total number of possible transcripts depends on the exact protocol specifications of the two-way interactions (including number of bits in each round, and total number of rounds). A rigorous analysis of error correction must provide this computation along with the protocol specification. For an example of the subtleties that may arise in such scenarios, see Ref.~\cite{tupkary_using_2023}.
\end{enumerate}

\subsubsection{Using the actual number of bits leaked } \label{subsubsec:usingbitsactuallyleaked}

In many works, it is stated without proper arguments, that one can use the number of bits \textit{actually used} by the error correction protocol in the above calculations. In fact such a claim implicitly suggests that one is implementing a variable-length protocol (see \cref{subsec:variablelength}), and therefore one must prove security for variable-length protocols. Intuitively, such scenarios leak additional information to Eve beyond the calculations described above. To see this, note that Eve can simulate the error correction protocol on all possible combinations of Alice and Bob's data, and rule out combinations that do not lead to the number of bits of error correction actually observed. Thus, even for variable-length protocols, the number of bits to be used for error correction in current security proofs is completely determined by the public announcements during the variable-length decision. 

\begin{remark}
    Formally, the difficulty in analyzing this situation shows up in the observation that in order to only subtract the number of bits actually used from the smooth min-entropy term, one has to condition on the event that ``a certain number of bits observed during error correction". It turns out that obtaining a bound on the smooth min-entropy of the state conditioned on such an event is difficult.  
\end{remark}

\subsubsection{Information leakage in phase error correction frameworks}
Information leakage due to error correction communication is handled differently in phase error correction based proofs. In particular, the argument does not involve the use of entropic chain rules. Many early works based on the phase error correction proof framework required the error correction information to be uncorrelated with the pre-amplification string, for which the protocols were required to one-time pad the error correction announcements. This requirement was later removed in some follow-up works~\cite{tamaki_quantum_2010,du_advantage_2024}; however, note that such proofs have currently only been provided for one-way error correction procedures in which the announced bits are computed by applying a \emph{linear} function to the pre-amplification string (see \cref{subsec:phaseerror}). In comparison, proofs treating error correction announcements via the above chain rule (\cref{eq:ECchainrule}) allow for more general protocols to be analyzed. 

\subsubsection{Error verification and the correctness condition}
\label{subsubsec:misconception_EV}

As previously mentioned in~\cref{subsec:correctnessproof}, care is necessary when interpreting the guarantees of error verification. 
In particular, a common misconception is to claim that conditioned on the hashes matching, the probability of the pre-amplification strings differing is small. This is \emph{not} guaranteed by (for instance) error verification based on $\delta$-AU$_2$ hashing, which only ensures that conditioned on the pre-amplification strings being different, then the probability that their hashes match is small---these two conditional probabilities can have extremely different values, but conflating them is a common mistake.

We now present here an explicit counterexample to the former claim, i.e.~we present an attack such that \emph{even conditioned on the hashes matching}, the pre-amplification strings differ with high probability. The attack is very simple, though showing that it indeed yields the claimed property is slightly elaborate. 
For simplicity, let us focus on protocols where Alice's pre-amplification string is a uniformly random string of $m$ bits corresponding exactly to her signal state choices in $m$ generation rounds, and one-way error correction takes place from Alice to Bob with a string of length upper bounded by $\xi m$ for some constant $\xi<1$.  Now consider a trivial attack where Eve simply intercepts and discards \emph{all} states that Alice sends, and forwards some arbitrary unrelated states to Bob. 

Observe that with this, all registers that Bob holds before error correction (including basis announcements) are completely independent of Alice's pre-amplification string, and thus his probability of guessing that string at that point is at most $2^{-m}$. By chain rules for guessing probability\footnote{Basically, Eq.~\eqref{eq:ECchainrule} with the smoothing parameter $\esmooth$ set to zero, noting that min-entropy has an interpretation in terms of guessing probability~\cite{konig_operational_2009}.}, his probability of guessing Alice's pre-amplification string after receiving the error correction string is then at most $2^{-(1-\xi)m}$.\footnote{Here for simplicity we suppose that for each $m$ an error correction string of some \emph{fixed} length (smaller than $\xi m$) is used; if the string length can vary \emph{up to} $\xi m$, then we would simply obtain a bound of $2^{-(1-\xi)m+1}$ instead, since there are only at most $2^{\xi m+1}$ such strings.} 
Let $\Odiff$ and $\OEV$ respectively denote the events that the pre-amplification strings differ and that error verification accepts, so the probability of interest (e.g.~in Eq.~\eqref{eq:EVcounterexample}) is just $\Pr[\Odiff|\OEV]$. With this we observe that for the described attack,
\begin{align}
\Pr[\Odiff \text{ AND } \OEV] &= \Pr[\OEV |\Odiff]\Pr[\Odiff] \nonumber \\
&\geq \Pr[\OEV |\Odiff](1-2^{-(1-\xi)m}),
\end{align}
and (writing $\neg$ to denote negation of an event)
\begin{align}
\Pr[\OEV]&= \Pr[\OEV |\Odiff]\Pr[\Odiff] \nonumber \\
&+ \Pr[\OEV |\neg\Odiff]\Pr[\neg\Odiff] \nonumber\\
&\leq \Pr[\OEV |\Odiff]\Pr[\Odiff] + (1)(1-\Pr[\Odiff]) \nonumber\\
&\leq 1 +\left(\Pr[\OEV |\Odiff]-1\right) (1-2^{-(1-\xi)m})
\nonumber\\
&= \Pr[\OEV |\Odiff](1-2^{-(1-\xi)m}) + 2^{-(1-\xi)m},
\end{align}
where we used $ \Pr[\OEV |\Odiff] - 1 \leq 0  $ in the third inequality. This gives
\begin{align}
\Pr[\Odiff|\OEV] &= \frac{\Pr[\Odiff \text{ AND } \OEV]}{\Pr[\OEV]} \\
&\geq \frac{\Pr[\OEV |\Odiff](1-2^{-(1-\xi)m})}{\Pr[\OEV |\Odiff](1-2^{-(1-\xi)m}) + 2^{-(1-\xi)m}} \nonumber\\
&= \frac{1-2^{-(1-\xi)m}}{1-2^{-(1-\xi)m} + \frac{2^{-(1-\xi)m}}{\Pr[\OEV |\Odiff]} }.
\end{align}
For simplicity, suppose furthermore that error verification is performed with a \term{strongly universal$_2$} hash~\cite{wegman_new_1981}, which has the property that $\Pr[\OEV |\Odiff] = 2^{-\ell_\mathrm{hash}}$ where $\ell_\mathrm{hash}$ is the hash length. 
Then the last line in the above bound limits to $1$ as $m\to\infty$ (taking $\ell_\mathrm{hash}$ to be constant, or sublinear with respect to $m$). Hence $\Pr[\Odiff|\OEV]$ becomes arbitrarily close to $1$ at large $m$, as claimed.

While we have focused on a simple example here, the overall principle should generalize to other cases. For instance, while we considered strongly universal$_2$ hashing as an example, the same conclusion holds for any error verification procedure in which $2^{-(1-\xi)m}$ is ``vanishingly small'' compared to $\Pr[\OEV |\Odiff]$ at large $m$ (more rigorously: their ratio limits to zero as $m\to\infty$).  
We expect this to indeed be the case for most error verification procedures --- this is because in order to obtain convergence to the asymptotic key rate, the number of bits announced for the acceptance test or error verification usually has to be sublinear in $m$, and it seems unlikely that $\Pr[\OEV |\Odiff]$ can be of comparable magnitude to $2^{-(1-\xi)m}$ in that case.\footnote{A possible exception is if error correction is performed with strings of variable length based on some announced data, such that the maximum possible length is not upper bounded by $\xi m$ for any constant $\xi<1$. However, this forces the protocol to be a variable-length protocol, and thus it must be accompanied by a rigorous variable-length security proof.}

\begin{remark}
As discussed in \cref{subsubsec:step_EV}, some alternatives to the above form of error verification procedure are possible, for instance by implicitly including a step in the acceptance test that estimates the number of errors that occurred, and aborting if the estimate is too high. However, a similar counterexample structure applies --- suppose for instance that the acceptance test aborts if the estimated fraction of errors is higher than $5\%$ of the rounds, and otherwise the protocol proceeds to perform an error-correction code designed to correct some slightly higher fraction of errors, say $6\%$ (these numbers are rather arbitrary and not critical to the core structure of the counterexample). Then Eve could simply choose an attack that \emph{always} flips the bit values on exactly $50\%$ of the rounds, for instance by choosing exactly half the rounds and applying a Pauli-$Y$ rotation on them. Again, under such an attack, it remains the case that \emph{even conditioned on the acceptance test accepting}, the total fraction of errors is still {exactly} $50\%$, as that event had probability $1$ and thus its probability remains unchanged by conditioning on \emph{any} other event. Since the fraction of errors is still $50\%$ even conditioned on the acceptance test passing, which is much higher than the $6\%$ error fraction the error-correction code is designed for, it follows from strong-converse theorems on error correction that it must fail to correct the errors with exponentially high probability.\footnote{To be slightly more detailed, we should account for the fact that the fraction of errors in generation rounds alone can be slightly different from $50\%$, as the errors may not be evenly distributed between test and generation rounds. It turns out this does not affect the conclusion, because the tolerated fraction of errors in test rounds under the acceptance test is less than $50\%$, and thus the fraction of errors in generation rounds conditioned on accepting must actually be \emph{strictly higher} than $50\%$, so our analysis still applies.} 

As briefly noted above, a possible workaround would be to vary the length of the error-correction string based on the estimated number of errors --- in theory, it is possible that such a procedure could genuinely ensure a bound on the probability of the keys differing conditioned on accepting. However, this inherently results in a variable-length protocol, and \emph{must} be accompanied by a rigorous and complete security proof for such variable-length procedures.
Moreover, another point that must be noted in such protocols is that the error-correction code used must be one that has a \emph{proven} bound on the probability it fails to correct the number of errors bounded during the acceptance test. This is generally not the case for many error-correction codes implemented in practice, where the probabilities of failing to correct errors are only empirical estimates that are quite far from proven bounds --- the central benefit of performing error verification using $\delta$-AU$_2$ hashing is that the security proof does not rely on any of those empirical estimates, and only solely requires the rigorously proven properties of $\delta$-AU$_2$ hashing.
Also, such protocols have to genuinely be estimating the number of errors that occurred in the basis used for key generation (sometimes referred to as ``bit errors''). It particular, for such protocols it does \emph{not} suffice to bound the number of ``phase errors'' (in the sense of e.g.~Ref.~\cite{koashi_simple_2005}) alone; they must explicitly obtain a bound on the number of ``bit errors'' somewhere in the analysis, and physically implement an error correction procedure that rigorously accommodates that number of errors. 
\end{remark}

In summary, it would be incorrect for a security proof to claim at any point that $\Pr[\Odiff|\OEV]$ is small (unless the protocol has sufficient additional features to rule out the above counterexamples).
With this in mind, we re-emphasize however that as observed in~\cref{subsec:correctnessproof}, $\Pr[\Odiff|\OEV]$ is not in fact relevant when proving that the correctness condition holds. This is because the correctness condition is not phrased in terms of this conditional probability---it only requires that the joint probability of the keys differing \emph{and} the protocol accepting is small, which can be rigorously bounded using only the guarantees of $\delta$-AU$_2$ hashing, as noted in \cref{eq:correctness}. The framework of composable security then ensures that we have the desired composable properties of the protocol, despite the fact that $\Pr[\Odiff|\OEV]$ can be large.

\subsection{Variable-length protocols} \label{subsec:variablelength}
Many of the existing rigorous security proofs for QKD are restricted to the case of ``fixed-length'' protocols. Such protocols involve a predetermined condition for accepting or aborting the protocol (most critically during the acceptance test, though the protocol may also abort during later steps such as error verification). If the protocol accepts, a final key of fixed length is produced; otherwise, if the protocol aborts, no key is produced. For such protocols, it is necessary for the acceptance conditions to be chosen \textit{before} the protocol is run. 

    In variable-length (also known as ``adaptive-length") protocols, the user is allowed to adjust the amount of error correction to be performed, and the length of the final key that is to be produced, based on the observations made during the protocol. Such protocols are relevant for practical implementations, since they remove the requirement of choosing a well-crafted acceptance condition (which typically requires accurate characterization of the honest behaviour of the channel).

    Proving the security of variable-length protocols requires new arguments, i.e, it does not directly follow from the security proof from fixed-length protocols. A common mistake in QKD literature involves using a security proof for fixed-length protocols in a variable-length implementation. One way this mistake is made is by allowing users to choose their acceptance conditions after looking at the observations.

    For security proofs of variable-length protocols see Refs.~\cite{tupkary_security_2024,nahar_postselection_2024} for an approach involving the use of the postselection technique  \cite{christandl_postselection_2009}, see \cite[Supplementary Note A]{curras-lorenzo_tight_2021}, \cite[Appendix B, F]{tupkary_phase_2024} for a security proof in the EUR framework, and \cite[Chapter 3]{kawakami_security_nodate} \cite{hayashi_concise_2012} for the phase error correction based methods.

\subsection{Miscellaneous}

\subsubsection{Resources used in QKD}

Most QKD security proofs assume that Alice and Bob each locally have access to ideal RNGs. However, an ideal RNG is usually a mathematical abstraction that is not realized in practice. 
One conceptually simple way to address this issue is to use an RNG device that is accompanied by a composable security proof, so that as long as the device satisfies the model assumptions in that security proof, the composable security framework implies that we can simply analyze the QKD protocol as though the RNG is ideal, then add the ``error parameter'' from the composable security analysis of the RNG device. A more elaborate possibility would be to perform the QKD security proof while incorporating an in-depth analysis of the details and imperfections of the real RNG, but this may be fairly challenging. {Passive QKD can serve as an approach to reduce the number of random bits required from the RNGs, but requires the devices to be sufficiently characterized to implement the desired distributions.}

Similarly, when considering the authenticated classical channels, in real QKD implementations one does not truly have access to ideal authenticated classical channels. Again, however, since there exist reasonably simple protocols to construct authenticated classical channels with composable security (given short pre-shared keys), one can first derive the QKD security proof assuming ideal authenticated classical channels, then add the ``error parameter'' from the composable security definition accordingly. (As mentioned in \cref{subsec:gap_abort} above, we re-emphasize that the QKD security proof must still account for the possibility of authentication aborting.)

While the RNGs and the authenticated classical channels can in principle both be constructed with information-theoretic security, for some QKD implementations one might instead wish to use constructions that only have computational security. There exist works considering this possibility and its connection to a concept known as \term{everlasting security} (see e.g.~Ref.~\cite[Section~VIII.D.4]{portmann_security_2022}), but that will not be a focus of this work.

   \subsubsection{Variable-length of pre-amplification string} \label{subsubsec:varlength_rawkey}

    A subtle issue arises due to a gap between theory and implementation of sifting and privacy amplification in QKD protocols. This gap is applicable to proof techniques reducing the analysis to a single-round entropic quantity, such as the postselection technique followed by an IID security proof, and EAT / GEAT based methods. In particular, it is avoided by EUR and phase error correction based methods. 

  \paragraph*{Theory.} In theoretical analysis, sifting is typically modeled as setting the pre-amplification string register to a $0$ or $\bot$ for each round that is discarded, which leads to the pre-amplification string register having a fixed length. Privacy amplification is then implemented by universal$_2$ hashing on this register (of fixed-length) to some $l$ bits. Note that if the discarded rounds are set to $\bot$, then the hashing procedure is no longer applied to binary strings.    

\paragraph*{Implementation.}   In practice, sifting is often implemented  by physically discarding the round, which leads to the pre-amplification string register having a variable number of bits. Then, privacy amplification is done by first looking at the number of bits in the pre-amplification string, and \textit{then} implementing universal$_2$ hashing from those many bits to $l$ bits. This approach is preferred since it allows users to apply hashing on smaller, binary strings.

This difference constitutes a clear gap between the theoretical model and the practical implementation. The gap is further exacerbated by the fact that the implemented protocol does \textit{not} represent a valid universal$_2$ hashing procedure on the pre-amplification string register\footnote{For instance, consider two all-zero strings of different lengths. If we were to look at the number of bits in each string and then apply a universal$_2$ hashing procedure (such as Toeplitz hashing) for that string length to that string, the output would always be all-zero strings in both cases, i.e.~the collision probability would be $1$ even though the input strings are different.}.
    This issue was first pointed out and addressed in Ref.~\cite[Section V]{tupkary_security_2024}, where a variant of the Leftover Hashing Lemma that is applicable to the variable-length pre-amplification strings  case was proven. In particular, it was shown that as long as the positions of the discarded rounds are publicly announced, the theoretical analysis of the following three scenarios are equivalent:
    \begin{enumerate}
        \item Discarded rounds are mapped to $\bot$.
        \item Discarded rounds are mapped to $0$.
        \item Discarded rounds are actually discarded.
    \end{enumerate}
  Thus a security proof using scenario (1) or (2), is applicable to an implementation with scenario (3).

    \paragraph*{} Note that the above problem does not arise in EUR and phase error correction based proof techniques. This is because in these frameworks, the pre-amplification string is only generated from detected signals. If the pre-amplification string is of variable length, then one can separately apply the Leftover Hashing Lemma for each possible length, and obtain a security proof similar to the one for variable-length protocols \cite[Remark 17]{tupkary_phase_2024}. Alternatively, some protocols choose a random subset of fixed length for the pre-amplification string on which privacy amplification is done.

\subsubsection{Conditioning on well-defined events} \label{subsubsec:conditioning}
As discussed earlier in \cref{subsec:estimating_entropy}, when proving rigorous security for QKD protocols, it is crucial to compute entropies and probabilities for states properly conditioned on relevant events. Conditioning can significantly alter the values of these quantities. Furthermore, it is important to ensure that the events we condition on are well-defined. Specifically, a well-defined event in one where there exists (in principle) a classical register that determines whether or not the event occurred.

This issue becomes particularly subtle when using techniques like the EUR, which considers post-measurement states obtained via \textit{different} measurements on the same state (see \cref{subsec:eur} for more details). For instance, one typically needs to bound the entropy of a state conditioned on acceptance (and passing error verification). However, the EUR statement cannot be applied directly to a state conditioned on passing error verification because this event is not well-defined until the state has been measured in one basis. Once measured in this basis, the state obtained by measuring in the  complementary basis (which is needed for the EUR statement), becomes ill-defined. Such issues must be suitably addressed. 

\subsubsection{Infinite Dimensions} \label{subsubsec:infinitedimensions}
As noted earlier, practical implementations involve states and measurements that live in infinite-dimensional Hilbert spaces. Moreover, Eve can always employ infinite-dimensional side registers if she wishes. Thus, the security analysis must be able to handle such scenarios.

Among the proof techniques, the postselection technique (see \cref{subsec:postselection}) is unique in that it has an explicit dependence on the underlying dimensions. Consequently, one must first rigorously apply source maps and squashing maps to reduce the problem to finite dimensions, and then argue that Eve's side-information registers can also be taken to be finite-dimensional without loss of generality. Thus, one first reduces the protocol itself to one with finite dimensions, and then proceeds with the security analysis of the finite-dimensional protocol. This approach is taken in Ref.~\cite{nahar_postselection_2024}.

The other proof techniques — namely the EUR, phase-error correction, and EAT-based methods (see \cref{subsec:eur,subsec:phaseerror,subsec:EAT}) — typically do not have any explicit dimension dependence. Nevertheless, many ingredients of the proof techniques, such as Leftover Hashing Lemmas, the EAT variants themselves, and the various chain rules for entropic quantities are often proved only for \emph{finite}-dimensional systems, even though they hold for arbitrary finite dimensions. This does not automatically guarantee that they remain valid in infinite-dimensional settings, which presents a technical issue that must be resolved rigorously. We emphasize, however, that this is purely a technicality, and the final results do not change.

One possible resolution is to follow the same approach as in Ref.~\cite{nahar_postselection_2024}, reducing the entire protocol to finite dimensions before applying the proof technique of one's choice. Alternatively, one may argue that the technique extends directly to the infinite-dimensional case, since it holds for arbitrary (but finite) dimensions. Ref.~\cite{inprep_BDR3} undertakes both of these approaches.

\subsubsection{Small trace distance conditioned on accept is not achievable} 
\label{subsubsec:securityconditionedonaccept}

Recall from \cref{subsec:securitydefn} that the security 
definition requires us to show that
\begin{equation} \label{eq:securefixedtemp}
\Pr[\Oacc]\, d\left(\rho_{\KA \KB \Efinal | \Oacc} \,,\,  
\tau^{\lfixed}_{\KA\KB} \otimes\rho_{\Efinal | \Oacc}\right) \leq \esecure,
\end{equation}
and in particular that the prefactor $\Pr[\Oacc]$ is essential. We now give 
a concrete example illustrating why the trace distance term alone (i.e.~without this prefactor) cannot be small, even when it is evaluated on the state \emph{conditioned} on acceptance.

Consider a fixed-length qubit BB84 protocol, where Alice and 
Bob use $\nK$ rounds for key generation and $\nT$ rounds for parameter 
estimation. For simplicity\footnote{One can straightforwardly construct similar examples for practical scenarios as well.}, we assume that they have quantum memories, and one party makes the basis choices after all states have been received, and we have $n=\nK+\nT$. They accept if the observed error rate in the test rounds falls 
below a threshold $e_\mathrm{threshold}$, and then proceed with further processing,  error 
correction, error verification, and privacy amplification. This is the 
protocol analysed in, for example, Ref.~\cite{tomamichel_largely_2017}.

Now consider the following attack: Eve intercepts every state emitted by 
Alice, stores it in quantum memory, and forwards any random state 
to Bob in its place. This is a completely valid attack that Eve is free to 
perform (also previously discussed in \cref{subsec:estimating_entropy}). After Alice announces her 
basis choices, Eve measures her stored systems in the corresponding bases 
and thereby recovers Alice's choice of signal states. 
In particular, Eve knows the pre-amplification string $\PAstring$ perfectly, and therefore knows the final key $\KA$ perfectly.

For the output state $\rho_{\KA \KB \Efinal | \Oacc}$ produced under this 
attack, Eve's guessing probability for $\KA$ is exactly~$1$, \emph{even conditioned on acceptance}. In the ideal 
state $\tau^{\lfixed}_{\KA\KB} \otimes \rho_{\Efinal | \Oacc}$, the key 
$\KA$ is uniformly distributed and independent of $\Efinal$, so Eve's 
guessing probability is $\frac{1}{2^{\lfixed}}$. Since the trace distance 
upper-bounds the difference in success probability for any guessing 
strategy, we conclude that\footnote{This can also be seen from the Holevo-Helstrom theorem, which states 
that the trace distance between two states equals the maximum difference 
in probability of any event between those two states, i.e.\ 
$d(\rho,\sigma) = \max_{0 \leq M \leq \mathbb{I}} \mathrm{tr}[M(\rho - \sigma)]$. 
Consider the event where Eve correctly guesses Alice's key $\KA$: under 
$\rho_{\KA \KB \Efinal | \Oacc}$ this can occur with probability $1$, 
while under $\tau^{\lfixed}_{\KA\KB} \otimes \rho_{\Efinal | \Oacc}$ 
this can only occur with probability 
$\frac{1}{2^{\lfixed}}$. Since the trace distance upper-bounds the 
difference in probability for any such event, we immediately 
obtain \cref{eq:lowerboundtrdist}.
}
\begin{equation}
    \label{eq:lowerboundtrdist} d\left(\rho_{\KA \KB \Efinal | \Oacc} \,,\,  \tau^{\lfixed}_{\KA\KB} 
    \otimes\rho_{\Efinal | \Oacc}\right) \geq 1-\frac{1}{2^\lfixed},
\end{equation}
which is close to $1$ for any non-trivial key length. Thus, this trace distance term conditioned 
on acceptance cannot be small. While we have used qubit BB84 as 
a concrete setting, the argument extends immediately to practically any other QKD 
protocol, by again considering an attack where Eve stores all systems sent by Alice and forwards 
random states to Bob.

The security definition in \cref{eq:securefixedtemp} resolves this (in the sense that it can still be satisfied even under this ``intercept-and-forward-garbage'' attack) 
precisely via the $\Pr[\Oacc]$  factor. 
Under this  attack, the states reaching Bob are 
completely uncorrelated with Alice's, so Bob's measurement outcomes are 
uniformly random. Consequently, the observed error rate in the test rounds 
is distributed as $\frac{1}{\nT}\mathrm{Bin}(\nT, \tfrac{1}{2})$, and 
the probability of the protocol accepting is
upper bounded\footnote{Since it can only accept if \emph{both} the acceptance test and error verification pass.} by
\begin{equation}
    \Pr[\Oacc]_{\mathrm{attack}} 
    \leq \Pr\!\left[\mathrm{Bin}\!\left(\nT,\tfrac{1}{2}\right) 
    \leq \lfloor e_{\mathrm{threshold}}\,\nT\rfloor\right].
\end{equation}
For any threshold value $e_{\mathrm{threshold}} < \tfrac{1}{2}$, the 
Hoeffding inequality then gives
\begin{equation}
    \Pr[\Oacc]_{\mathrm{attack}} 
    \leq \exp\!\left(-2\nT\!\left(\tfrac{1}{2} 
    - e_{\mathrm{threshold}}\right)^{\!2}\right),
\end{equation}
which is exponentially small in $\nT$. Since the trace distance 
is at most $1$, with this we see that the overall security requirement can indeed be satisfied at suitable values of $\nT$, even under this attack. (Of course, a full security proof would need to prove the security definition holds against all attacks --- this calculation is just to show that it can indeed hold against this particular attack.)

\subsubsection{Choosing the acceptance threshold} \label{subsubsec:choosingacceptancethreshold}
In practice, the acceptance criterion should also be chosen to ensure that an honest 
implementation --- one where the QBER arises solely from channel noise --- 
accepts with high probability (see \cref{subsec:completeness}). Suppose the expected QBER in the honest 
model is $e_{\mathrm{exp}}$, and we choose an acceptance threshold $e_\mathrm{threshold} > e_{\mathrm{exp}}$\footnote{If $e_\mathrm{threshold} < e_{\mathrm{exp}}$, it is straightforward to show that the protocol will abort with high probability}. Then the probability of acceptance in the 
honest model (assuming for simplicity that error-verification always succeeds) is
\begin{equation}
    \Pr[\Oacc]_{\mathrm{honest}} 
    = \Pr\!\left[\mathrm{Bin}(\nT, e_{\mathrm{exp}}) 
    \leq \lfloor e_{\mathrm{threshold}}\,\nT\rfloor\right],
\end{equation}
and thus by the Hoeffding inequality, we have
\begin{equation}
    1 - \Pr[\Oacc]_{\mathrm{honest}} 
    \leq \exp\!\left(-2\nT(e_{\mathrm{threshold}} 
    - e_{\mathrm{exp}})^{2}\right).
\end{equation}
Setting $e_{\mathrm{threshold}}$ too close to $e_{\mathrm{exp}}$ 
risks frequent abortion of honest runs; setting it too high 
unnecessarily reduces the tolerated QBER and hence the key length. 
In the qubit BB84 case, the secure key length scales roughly as
\begin{equation}
    \ell_\mathrm{fixed} \approx \nK\bigl(1 - 2h(e_{\mathrm{threshold} }+\gamma_{\nK,\nT})\bigr),
\end{equation}
where $h$ is the binary entropy function and $\gamma_{\nK,\nT}$ is a finite-size correction term. 
There is a direct 
trade-off between completeness and key rate that is controlled 
by the gap $\delta = e_{\mathrm{threshold}} - e_{\mathrm{exp}}$. Raising $e_{\mathrm{threshold}}$ 
improves completeness but reduces the key length, since $h$ is 
increasing on $[0, \tfrac{1}{2}]$. These two effects can be 
summarised compactly by the \emph{expected key length} (also briefly mentioned in \cref{subsec:completeness}):
\begin{equation}
\begin{aligned}
    \ell_\mathrm{expected} 
    &\approx \Pr\!\left[\mathrm{Bin}(\nT, e_{\mathrm{exp}}) 
    \leq \lfloor e_{\mathrm{threshold}}\,\nT\rfloor\right] \\
    &\quad\times \nK\bigl(1 - 2h(e_{\mathrm{threshold}} 
    + \gamma_{\nK,\nT})\bigr),
\end{aligned}
\end{equation}
which is the product of the acceptance probability (for some honest behaviour) and the key length 
conditional on acceptance. This quantity has a natural maximum as a 
function of $e_{\mathrm{threshold}}$, and optimizing over it gives a 
principled criterion for threshold selection for fixed-length protocols.  An analogous optimization can be
performed for variable-length protocols, where there is no acceptance
set; instead, one optimizes over choices of the key length function  to
maximize the expected output key length.

Of course, the situation is somewhat more complicated for protocols 
such as decoy-state BB84, where the acceptance 
test (or variable-length decision) involves multiple observed quantities, rather than a single QBER. Nevertheless, the core trade-off remains the same: for acceptance testing, the threshold 
must be set wide enough to absorb statistical fluctuations in an 
honest run, yet tight enough to produce keys upon acceptance, and the expected 
key length is still the natural figure of merit for optimizing this 
choice \cite{tupkary_security_2024,kamin2025renyisecurityframeworkcoherent}.

\section{Families of security proofs} \label{sec:families}
In the previous section, we identified common gaps in existing QKD security proofs. Building on that discussion, we now review the primary finite-size security proofs against coherent attacks for decoy-state BB84 as presented in the literature. While tremendous progress has been made, significant gaps and limitations remain, which we aim to highlight in this section. Rather than examining each proof in detail, we focus on the foundational works that serve as the basis for subsequent analyses and, consequently, inherit the shortcomings of the original.

At a high level, proof techniques can be categorized into those that use the Leftover Hashing Lemma  and those that do not. If the proof technique proceeds via the Leftover Hashing Lemma then it must bound the relevant entropy (see \cref{subsec:estimating_entropy}). The EUR and EAT-based methods obtain this required bound directly against coherent attacks, without relying on any IID-based security analysis (although the key rate expression in the EAT framework involves a single-round optimization).

The phase error correction approach does not rely on the Leftover Hashing Lemma at all, but instead provides an argument based on ideas from quantum error correction. 

In contrast, the postselection technique first reduces the security of the entire protocol to that of IID collective attacks. After this reduction, one may use \emph{any} preferred proof technique to obtain security against IID collective attacks. For example, Ref.~\cite{Cong_sidechannelsecure_2025} first applies postselection to reduce to the IID setting, and then uses a phase error correction argument to complete the proof. We note, however, that if one wishes to use the phase error correction approach at this stage, it is generally preferable to rely directly on the underlying de Finetti reductions to simplify the sampling process, as is done, for example, in Refs.~\cite{tamaki2003unconditionally,matsuura_asymptoticallytight_2025,Shan_improvedfinitekey_2025}, rather than  rely on the ``generic'' lift theorems from Ref.~\cite{nahar_postselection_2024}.

It is important to note that while we point out issues in many of these works, they represent significant advancements in the proof techniques available at the time. Many of the gaps and limitations we discuss were later addressed by subsequent research, which greatly benefited from the simple fact that it was developed at a later stage, with the advantage of building upon prior insights and refinements.

\subsection{Proofs based on entropic uncertainty relations} \label{subsec:eur}
Proofs that rely on the entropic uncertainty relations \cite{tomamichel_uncertainty_2011} utilize the following statement to obtain a bound on the smooth min-entropy of the pre-amplification string: that for any state $\rho_{A^mB^m \En}$, 
\begin{equation} \label{eq:eur}
    \Hmin^{\bar{\epsilon}}(Z_A^m | \En )_{\mathcal{E}_Z(\rho)} + H_\mathrm{max}^{\bar{\epsilon}}(X_A^m | B^m)_{\mathcal{E}_X(\rho)} \geq m c_q
\end{equation}
where $\mathcal{E}_Z(\rho)$ is a channel that measures the $A^m$ system in the $Z$ basis, and $\mathcal{E}_X(\rho)$ is a channel that measures the $A^m$ system in the $X$ basis, and $c_q$ is a parameter that depends on these measurements POVMs\footnote{The $Z$ and $X$ basis here need not be taken literally. One can use the EUR statement for  \textit{any} two POVMs, one corresponding to $Z$ and $X$. (To obtain a useful statement, one must ensure that $c_q>0$ for the two POVMs considered). For the BB84 protocol with ideal source, these correspond to the canonical $Z$ and $X$ basis measurements.}. Here, Alice's measurement POVMs and her reduced state are obtained from the source-replacement scheme (see also \cref{sec:SRS}).

In the actual protocol, the $Z$ measurements are performed, which determines (a part of) the pre-amplification string whose min-entropy we wish to bound.
The max entropy term is then suitably bounded by obtaining an estimate (upper bound) on the number of ``phase errors". This refers to the number of errors corresponding to (fictitious) measurements, where Alice and Bob use a different basis for the measurements than the one in the protocol. Note that one typically only applies the EUR analysis on a part of the  pre-amplification string $\PAstring$. Thus, one must also relate the smooth min-entropy obtained via the EUR analysis to the min-entropy of the pre-amplification string. This is done with the use of suitable chain rules. For a more detailed walk-through of the proof, see \cref{app:eurfinite}.

When working within EUR based security proofs, the following unique factors must be considered

\begin{enumerate}
    \item In order to use the EUR statement, one must properly define the state to which it is applied. Typically, we only apply the EUR statement on single-photon rounds used for key generation (typically, these are the detected rounds where Alice sent a single-photon and both Alice and Bob measured in the same basis). This needs to be explained rigorously, and can be done by reformulating the measurement procedure to consist of several measurement steps. This is somewhat technical, and is undertaken in Refs.~\cite{tomamichel_largely_2017,koashi_efficient_2006}.
    \item As outlined in \cref{subsubsec:conditioning}, we require min-entropy bounds on states conditioned on specific events, which must be well-defined. The use of the EUR statement introduces additional complications to this task, as conditioning on events that presuppose a fixed measurement basis can be problematic. For example, conditioning on the success of error verification poses problems, since this event requires measurements in a particular basis.

    The basic approach to address these issues is to consider states which are conditioned not on the protocol's actual events, but on related, well-defined events. The EUR statement can be rigorously applied on states conditioned on these related events. The resulting entropic bounds can then be translated back into bounds on states conditioned on the actual events in the protocol \cite{wiesemann_consolidated_2024,tupkary_phase_2024}.
    \item In order to obtain an estimate of the phase error rate, most works require the following assumption on the detection setup: that the probability of detection is independent of Bob's basis choice. This is equivalent to demanding all detectors have exactly identical efficiencies and dark count rates. For passive setups, this additionally requires the beam splitter to be perfectly balanced.
    This assumption is known as ``basis-independent loss", and its violation is known as ``basis-efficiency mismatch" or ``detection-efficiency mismatch" in the literature.  Due to this assumption, almost all works using the EUR for practical protocols are not applicable to real-world scenarios for standard QKD. However, this assumption is not a problem for MDI-QKD, where the detectors are completely untrusted and assumed to be in Eve's control. Thus many works utilizing this proof technique appeal to MDI-QKD as a method to remove this assumption,  and close all detector side-channels. 

    Recently, a method to remove this assumption for EUR and phase error correction based security proofs for active basis choice BB84 was obtained in Ref.~\cite{tupkary_phase_2024}, which we discuss in greater detail in \cref{subsubsec:tupkary}.
\end{enumerate}
Note that a highly formal proof using the EUR statement for fixed-length qubit BB84 can be found in Ref.~\cite{tomamichel_largely_2017}.

\paragraph*{Decoy-state methods:}\label{par:decoy_EUR}

Recall from \cref{subsec:decoy_general} that the goal of decoy-state methods is to use
observations obtained from different intensities to bound statistics associated with
the single-photon rounds. The argument for doing so within the entropic uncertainty
relations and phase error correction approaches (see \cref{subsec:phaseerror}) proceeds via concentration inequalities
as follows. We describe the approach taken in
Refs.~\cite{lim_concise_2014,hayashi_security_2014,curty_finitekey_2014}, although the
precise details may differ between various works.

Let $n_{O,\mu_k}$ denote the number of rounds in which the outcome $O$ is observed and
Alice used intensity $\mu_k$. This is a random variable that is observed in the protocol. For instance, $O$ could correspond to a detected round (i.e Bob obtains a detection)
in which both Alice and Bob measured in the $X$ basis. We seek to estimate the random variable $n_{O,m}$ corresponding to the number of rounds where outcome $O$ is observed, and Alice sent an $m$-photon pulse.

Operationally, in each round the protocol proceeds by Alice choosing an intensity
$\mu_k$ with probability $\Pr(\mu_k)$. She then prepares a phase-randomized WCP state which can be described by an emission of photon-number $m$ with probability
\begin{equation}
    \Pr(m \mid \mu_k) = e^{-\mu_k} \frac{\mu_k^m}{m!}.
\end{equation}
Without loss of generality, we may equivalently view Alice as \emph{first} choosing
a photon number $m$ with probability
\begin{equation}
    \Pr(m) = \sum_k \Pr(\mu_k)\Pr(m \mid \mu_k),
\end{equation}
and then choosing an intensity setting $\mu_k$ with conditional probability
\begin{equation}
    \Pr(\mu_k \mid m)
    = \frac{\Pr(\mu_k)\Pr(m \mid \mu_k)}{\Pr(m)},
\end{equation}
where we have used Bayes' rule.

In this picture, if we condition on a fixed sequence of photon numbers across all the rounds, the assignment of an
intensity setting to each signal is independent (though not identically distributed)
across rounds, since it depends only on the photon number of the corresponding pulse.
This allows us to perform an analysis involving Hoeffding's inequality (see
\cref{tab:Concentration Inequalities}) to obtain\footnote{We stress that it is \emph{not} valid to simply apply Hoeffding's inequality to the quantities $n_{O,\mu_k}$ and $n_{O,m}$, as these are not sums of independent random variables. 
Instead, a rigorous analysis can only apply Hoeffding's inequality to random variables that can be properly justified as being independent across rounds. While we do not discuss the details within this sketch, the overall idea is typically to first condition on a fixed value for the entire photon-number sequence, then identify suitable sequences of independent random variables under this conditioning, apply the relevant concentration inequality, and finally remove the conditioning on the photon number. See for example, \cite[Methods]{curty_finitekey_2014} or \cite[Appendix E]{tupkary_phase_2024}} 
\begin{equation}
    n_{O,\mu_k} \approx \sum_{m=0}^{\infty} \Pr(\mu_k \mid m)\, n_{O,m},
\end{equation}
where the approximation reflects finite-size correction terms (see \cref{app:eurfinite}).

By taking suitable linear combinations of these relations, one can derive upper and
lower bounds on the quantities $n_{O,m}$ of interest (typically for $m=0,1$ only).
Ref.~\cite{rusca_finite-key_2018} performs a similar analysis, but instead exploits a special
property of vacuum ($m=0$) signals: that the
number of $X$-basis vacuum rounds is approximately twice the number of such rounds that
result in an error. (This too requires use of suitable concentration inequalities). 

Thus, in both EUR-based and phase error correction-based methods, the decoy-state argument proceeds by applying suitable concentration inequalities to bound the relevant vacuum and single-photon statistics. These bounds are then used to estimate the relevant entropy in EUR-based security proofs, which typically requires the use of entropic chain rules.  The phase error correction approach utilizes different arguments for the same task. 
Notice that this constitutes a two-step procedure: first, one derives bounds on the single-photon statistics, and only subsequently are these bounds used in the entropy estimation.

\subsubsection{Ref.~\cite{lim_concise_2014}} \label{subsubsec:charleslim}
The first security proof for decoy-state BB84 using the entropic uncertainty relations was obtained in Ref.~\cite{lim_concise_2014}. A major follow up work on this result that reduces the number of decoy intensities required for the protocol is  Ref.~\cite{rusca_finite-key_2018}, which we discuss in greater detail in \cref{subsubsec:rusca}.
The following works \cite{guarda_bb84_2023,rusca_finite-key_2018,li_one-decoy_2020,boaron_secure_2018,attema_optimizing_2020,chau_decoy-state_2018,chau_application_2020,wang_tight_2016,zhou_finite-key_2022,foletto_security_2022,zhou_tight_2014,lucamarini_security_2015,wang_finite-key_2018,navarrete_improved_2022,sidhu_finite_2022} base their security analysis on Ref.~\cite{lim_concise_2014}\footnote{We have grouped together works that rely on \cite{lim_concise_2014,rusca_finite-key_2018} due to the similarities in these two works.}.

The proof in Ref.~\cite{lim_concise_2014} goes through the entropic uncertainty relations approach, and has the following shortcomings. Note that despite these shortcomings, the main essence of the result remains true after suitable fixes \cite{tupkary_phase_2024}:
\begin{enumerate}
    \item The proof does not condition on relevant events (see \cref{subsubsec:conditioning}), which presents a technical issue and raises questions about the application of certain chain rules. 
    \item The protocol described is a variable-length protocol, however the security definition for variable-length protocols is not used (see \cref{subsec:variablelength}).
    \item The proof implements a protocol with iterative sifting and announcements, but does not adequately address it in their sampling estimates for the phase error rate (see \cref{subsubsec:iterative}). The estimate on the phase error rate is obtained via Ref.~\cite{fung_practical_2010}, and goes through a Taylor approximation step. Thus the estimate is not a true bound.
    \item The proof claims that the error correction cost can be computed from the number of bits actually leaked, without proper justification (see \cref{subsubsec:usingbitsactuallyleaked}).
    \item It does not properly differentiate between random variables, fixed parameters and observed values of random variables (see \cref{subsec:concentrationinequalities}). Moreover, it does not properly specify the state on which the EUR statement is applied.
\end{enumerate}
Additionally, this work requires the source preparation to be perfect. However, there are several follow up works on the topic of source imperfections, that provide methods to obtain a suitable bound on the phase error rate, even in the presence of source imperfections \cite{curras-lorenzo_security_2024, zapatero2023implementationsecurityquantumkey,gllp_security_2004,tamaki_loss-tolerant_2014,curras2023security,sixto_quantum_2025,sixto2022security,zapatero2021security,pereira2020quantum,navarrete_improved_2022}.
It also considers active basis choice by Bob, and assumes that the probability of detection in Bob's setup is independent of basis. This assumption is a core issue for all EUR and phase error correction (see \cref{subsec:phaseerror}) based proofs for standard QKD, and was removed recently in Ref.~\cite{tupkary_phase_2024}. 
 
\subsubsection{Ref.~\cite{rusca_finite-key_2018}} \label{subsubsec:rusca}
A major improvement to Ref.~\cite{lim_concise_2014} was shown in Ref.~\cite{rusca_finite-key_2018}, by reducing the number of total decoy intensities needed  in the protocol from three to two. 
The analysis in Ref.~\cite{rusca_finite-key_2018} follows that of Ref.~\cite{lim_concise_2014}, and hence inherits all the shortcomings outlined above in \cref{subsubsec:charleslim}. However it adds the following issues:
\begin{enumerate}
    \item The protocol in Ref.~\cite{rusca_finite-key_2018} requires the knowledge of the number of errors in the key generation rounds. These are not announced in the protocol, and the paper does not specify how Alice and Bob obtain this information.

    A straightforward approach would be for Bob to compute the number of errors in the key generation data, assuming that error correction has succeeded. However, this relies on the success of error correction, and the security analysis must rigorously account for this dependency. 

    Alternatively, the error rate could be estimated through sampling by using the observed errors in the testing fraction of $Z$-basis rounds. This introduces additional finite-size effects which must be accounted for. 

\item Moreover, if Bob computes the error rate in the $Z$ basis through error correction, even the acceptance event depends on a specific measurement basis. This introduces additional complications when conditioning on these events, beyond the complications due to conditioning on error verifications outlined in \cref{subsec:eur}. For an approach to resolve this complication, see Ref.~\cite{wiesemann_consolidated_2024}.
\end{enumerate}

\subsubsection{Ref.~\cite{tupkary_phase_2024}} \label{subsubsec:tupkary}
The main focus of Ref.~\cite{tupkary_phase_2024} was to provide a method to estimate the phase error rate \textit{without} requiring the basis-independent loss assumption on detectors (see \cref{subsec:eur}). Along the way, Ref.~\cite{tupkary_phase_2024} points out and addresses several issues in Ref.~\cite{lim_concise_2014} discussed in \cref{subsubsec:charleslim}. In particular, it conditions on relevant events, rigorously proves the security of a variable-length protocol, does not implement iterative-sifting, obtains a phase error rate bound without approximations and determines the error correction cost from public announcements during the variable-length decision. Moreover, it carefully differentiates between random variables and observed values, clarifies the use of Hoeffding's inequality in the decoy analysis, and rigorously specifies the state on which the EUR statement is applied.

However, Ref.~\cite{tupkary_phase_2024} is still not completely general. In particular, it is limited to the decoy-state BB84 protocol with perfect signal states, active basis choice on Bob's side, and detectors that are ``close'' to satisfying the basis-independent loss assumption. While there do not appear to be fundamental reasons why these issues cannot be addressed in the future, these arguments do not currently exist in the literature (see \cite{curras-lorenzo_security_2025} for some recent developments.).

\subsubsection{Ref.~\cite{wiesemann_consolidated_2024}}
A security proof for decoy-state BB84 was recently obtained in Ref.~\cite{wiesemann_consolidated_2024}, which presents a pedagogical treatment of the proof techniques from Lim et al  \cite{lim_concise_2014} and Rusca et al \cite{rusca_finite-key_2018}. The first version of this work inherited all the shortcomings of Refs.~\cite{lim_concise_2014,rusca_finite-key_2018}. This was then modified in later revisions to address most of the issues outlined in \cref{subsubsec:charleslim,subsubsec:rusca}. In particular, the second version considers a fixed-length protocol, avoids iterative announcements, and conditions on relevant events. It uses a finite-size sampling bound that is a true bound (does not go through any approximations), and determines error correction cost based on the announcements and not from the number of bits actually leaked. The main drawback is the assumptions of basis-independent loss on the detectors, active basis choice by Bob, and the assumption of ideal sources. Moreover, Ref.~\cite{wiesemann_consolidated_2024} is missing some key details in some parts of the argument --- it does not properly differentiate between random variables and constants, it does not properly specify the state on which the EUR statement is applied, and it does not explain the application of the Hoeffding's inequality in the decoy analysis.

\subsection{Phase error correction approach} \label{subsec:phaseerror}

The phase error correction based proof technique \cite{koashi_simple_2005,koashi_simple_2009}, sometimes referred to as a proof based on complementarity, is a proof technique that is significantly different from the other QKD proof techniques discussed in this report. This proof technique is based on the original Shor-Preskill \cite{shor_simple_2000}, Lo-Chau \cite{Lo_unconditional_1999}, and the GLLP \cite{gllp_security_2004} proofs.
The phase error correction based proof technique does \emph{not} proceed by bounding the (smooth min-)entropy of the state, and using the Leftover Hashing Lemma (\cref{subsec:LHL}).
Instead, the security statement is related to the probability of a particular (virtual) phase error correction protocol succeeding. This, in turn is related to an upper bound on the number of ``phase errors".

The number of phase errors, in the form presented in~\cite{koashi_simple_2005,koashi_simple_2009} can be qualitatively understood as follows, though we do not aim to present a rigorous definition here.
Suppose we have some state of the form $\rho_{Z_A^mB^m\En}^{(Z)} = \mathcal{E}_Z(\rho_{A^mB^m\En})$, where $A^m$ consists of $m$ qubit systems, and $\mathcal{E}_Z$ is a channel that measures the $A^m$ system in the $Z$-basis.
Now consider instead the state $\rho_{X_A^mB^m\En}^{(X)} = \mathcal{E}_X(\rho_{A^mB^m\En})$ 
where $\mathcal{E}_X$ is a channel that measures the $A^m$ system in the $X$-basis.
The number of phase errors is, informally, quantified by 
how many bits Alice would need to send Bob (holding system $B^m$) in the latter state in order for him to guess the string $X_A^m$ with high probability.

Refs.~\cite{koashi_simple_2005,koashi_simple_2009} showed that if there exists some \emph{constant} $N_\mathrm{const}^\mathrm{err} \in \mathbb{N}$ that bounds the number of phase errors with high probability, 
then hashing\footnote{There are some technical points regarding the hash families considered in this technique; we shall discuss this point shortly.} the string $Z_A^m$ in the state $\rho_{Z_A^m\En}^{(Z)}$ to a length determined by $N_\mathrm{const}^\mathrm{err}$ results in the output being nearly uniform and decoupled from $\En$, roughly analogous to the Leftover Hashing Lemma.\footnote{There are a number of similarities between this approach and EUR-based approaches when constructing a QKD security proof, as discussed further in~\cref{subsubsec:similaritiesBetweenPECandEUR}.}
We highlight however that this result by itself does not constitute a security proof for a QKD protocol. This is firstly because of issues similar to those raised in \cref{subsec:estimating_entropy} --- for instance, one needs to address the question of whether the states being analyzed are conditioned on various events (such as the protocol accepting or the key length being chosen to be a particular value); also, it is usually impossible to find a \emph{constant} upper bound $N_\mathrm{const}^\mathrm{err}$ on the number of phase errors, similar to the problem of finding a constant lower bound $\entbnd$ on the smooth min-entropy. Furthermore, since the proof technique only analyzes hashing the string $Z_A^m$ specifically, to analyze a QKD protocol it must be explicitly justified that the pre-amplification string $\PAstring$ can be written in the form $\PAstring = Z_A^m$, in some state of the form $\mathcal{E}_Z(\rho_{A^mB^m\En})$ such that the number of phase errors can also be suitably analyzed. Approaches to handle these issues have been presented in \cite{hayashi_concise_2012,koashi_efficient_2006,hayashi_security_2014}, though care is needed in interpreting their results; we discuss this further in \cref{subsubsec:phaseerrordecoy}.

There are also some other features unique to phase error correction based proof techniques:
\begin{enumerate}
    \item The original proofs in Refs.~\cite{koashi_simple_2005,koashi_simple_2009} required that the error correction information used in the protocol be one-time padded. This allows the error correction to be dealt with separately from the secrecy condition. More recent works \cite{tamaki_quantum_2010,du_advantage_2024} allows for more sophisticated error correction procedures that remove the need for this one-time padding, though the error correction procedures then need to satisfy certain properties. For instance, Ref.~\cite{tamaki_quantum_2010} only analyzes the case where the error correction procedure consists of applying a linear map to a string Alice holds, then announcing a subset of it for error correction while retaining the rest as the pre-amplification string. Note that proof techniques based on the Leftover Hashing Lemma can account for the error correction information being revealed (without one-time padding) by instead using entropic chain rules as described in \cref{subsubsec:Hminchainrule}.
    
    \item Refs.~\cite{koashi_simple_2005,koashi_simple_2009} only studied a specific hashing procedure; however, subsequent work~\cite{tsurumaru_dual_2012} generalized the result to a broader class of hashing procedures known as almost-dual-universal$_2$ hashing --- we do not discuss the technical details here, but simply note that this includes universal$_2$ hashing as a special case. It does require that the hash family consists of linear functions, though we note this is indeed satisfied by commonly used procedures such as Toeplitz hashing.
\end{enumerate}

The above descriptions were a fairly generic discussion of the phase error correction approach. We now turn more specifically to the case of security proofs for decoy-state protocols. We highlight in particular that it is important for such a proof to explicitly address the point noted above, of describing the pre-amplification string via measurements on qubit systems in a manner compatible with analyzing the phase errors, despite the fact that the decoy-state protocol involves various mixtures of states of different photon numbers.

This proof technique has been applied to the study of device-independent QKD as well~\cite{zhang_complementarityapproachtodevice_2023}. However, some questions have been raised~\cite{tan2023memoryeffectsdevicedependentdeviceindependent} regarding its applicability to scenarios where the devices have memory.

\subsubsection{Refs.~\cite{koashi_efficient_2006,hayashi_security_2014}}
\label{subsubsec:phaseerrordecoy}

The first work to outline an approach for using phase error correction to prove security of decoy-state BB84 against coherent attacks is Ref.~\cite{koashi_efficient_2006}\footnote{Ref.~\cite{hayashi_concise_2012} also provided a security proof for variable-length protocols using phase error correction, using similar ideas, but the analysis was restricted to fully qubit BB84 rather than decoy-state BB84.}.
It thus forms the primary basis for further decoy-state security proofs in the phase error correction approach, though by itself it does not constitute a full finite-size security proof. This is because it gives a key length formula (below Eq. (9)) that depends on the number of single-photon phase errors, and number of zero and single-photon detection events, but it does not perform a finite-size analysis that bounds these quantities. (Qualitatively, it addresses some of the difficulties raised in the preceding section by arguing that the variable-length secrecy definition (\cref{eq:secretvar}) can be satisfied if the variable key length is chosen as a function of the number of phase errors, though it does not address how the latter should be computed or estimated in practice.)

The subsequent work \cite{hayashi_security_2014} performed a more detailed finite-size analysis to bound the relevant quantities, and is the first such attempt at a full decoy-state security proof, serving as a basis for subsequent works such as~\cite{mizutani_finite_2015,mizutani_quantum_2019}. Note that it uses a different argument to obtain the key length formula in Ref.~\cite{koashi_efficient_2006} mentioned above\footnote{Specifically, Eq.~(7) in Ref.~\cite{hayashi_security_2014} corresponds to Eq.~(9) in Ref.~\cite{koashi_efficient_2006}.}; however, the argument appears somewhat incomplete --- specifically, it does not give a detailed justification of how multi-photon pulses can be replaced with rounds with a different ``effective'' phase error rate. 
Furthermore, it also does not provide a detailed explanation of why some statistics gathered in the protocol can be modelled using a hypergeometric distribution. (While it describes how to analyze a hypergeometric distribution in terms of a binomial distribution, it does not appear to justify why the statistics should in fact follow a hypergeometric distribution.)
Similar to many EUR-based security proofs, it also relies on the basis-independent loss assumption.

Note that the gap in Ref.~\cite{koashi_efficient_2006}, i.e., the rigorous finite-size estimation of the single (and zero) photon detection events and single-photon phase error events, is a critical ingredient of EUR based proofs as well, as we shall discuss below. Thus, the estimation from Refs.~\cite{lim_concise_2014,tupkary_phase_2024} could also be used to complete this security proof. Although it is not hard to combine these results from the various existing works, we do not know of any self-contained decoy-state security proof in the phase error correction approach.

\subsubsection{Comparisons with EUR} \label{subsubsec:similaritiesBetweenPECandEUR}

As mentioned earlier, there are some similarities between the EUR-based security proof approaches and the phase error correction-based security proof approaches, though there are also some differences, which we shall also elaborate on here. 
First, we note that 
Ref.~\cite[Theorem 1]{tsurumaru_leftover_2020} showed that a bound on the smooth min-entropy of a classical string conditioned on some side-information gives a bound on the failure probability of a corresponding phase error correction protocol, in an attempt to bridge the gap between Leftover Hashing Lemma based approaches and phase error correction based approaches. 
However, this connection is only made in terms of statements similar to the result in Refs.~\cite{koashi_simple_2005,koashi_simple_2009}, which as mentioned above, does not constitute a full security proof for a QKD protocol (as it is only a statement about some abstract state assumed to satisfy certain properties, rather than the pre-amplification state in a QKD protocol).

In light of this, we highlight that there are some differences between how existing decoy-state security proofs based on the two approaches proceed. In particular, the proofs based on phase error correction involve expressing the \emph{entire} pre-amplification string $\PAstring$ as the output of $Z$-basis measurements on some qubit systems, despite the fact that many of the bits in $\PAstring$ usually correspond to Alice sending vacuum or multi-photon states --- this requires a suitably formulated source-replacement argument. In contrast, existing EUR-based security proofs often use suitable chain rules to decompose the smooth min-entropy of $\PAstring$ into terms corresponding to vacuum, single-photon, and multi-photon events, and only apply the EUR on the single-photon component. In other words, the registers on which (virtual) $Z$-basis or $X$-basis measurements are analyzed are different in the two approaches. While it may be possible to connect the approaches, we do not know of an existing self-contained work that formalizes such a connection.

Still, we note that via the analysis in Ref.~\cite{koashi_efficient_2006}, it is possible to reduce the main task in a phase error correction based security proof to the question of finding a ``statistical'' bound on the number of phase errors in detected rounds corresponding to single-photon emissions by Alice. In existing EUR-based security proofs, one computes bounds on the same quantity in order to bound $H_\mathrm{max}^{\bar{\epsilon}}(X_A^m | B^m)_{\mathcal{E}_X(\rho)}$ on the detected single-photon component. Thus, one major non-trivial step in the usage of these proof techniques is the same --- upper bound the number of phase errors (in rounds where Alice emitted a single photon and it was detected) through statistical arguments. As a result, a lot of the factors that must be considered for EUR based security proofs also apply to phase error correction based proofs, as we now explain.

For instance, one must properly define the state on which the number of phase errors is bounded. Typically, an upper bound on the number of phase errors in the detected single-photon rounds is obtained. This is justified via breaking up the measurement into a two step process where the first step decides whether or not the round will be detected, followed by the rest of the measurement which dictates the final measurement result when the round is detected \cite{koashi_efficient_2006,tomamichel_largely_2017}. While this argument works straightforwardly when all detectors have the same detection efficiency, this must be modified in the case when the different detectors have varying efficiencies, as previously discussed in \cref{subsec:eur}. For passive setups, the effective basis choice is now implemented by the beam splitter, and depends on the photon number of the incoming pulse, which further complicates the analysis \cite{wang_phaseerror_2025,mizutani_passive_2025}.

\subsection{Postselection technique}
\label{subsec:postselection}

The postselection technique is a proof technique that reduces the analysis of coherent attacks to the analysis of IID collective attacks. It is composed primarily of three parts:
\begin{enumerate}
    \item First, given a permutation invariant protocol, the permutation invariance property is used to reduce the analysis from all possible states\footnote{This is the state after the source-replacement scheme and Eve's attack.} $\rho_{A^nB^n}$ that could be shared by Alice and Bob to permutationally invariant states $\bar{\rho}_{A^nB^n}$.
    \item Then, one applies a de Finetti theorem for permutationally invariant states, of the form
    \begin{align} \label{eq:deFinettiTheorem}
        \bar{\rho}_{A^nB^n} \leq g_{n,x} \tau_{A^nB^n} = g_{n,x} \int d\sigma \sigma_{AB}^{\otimes n}.
    \end{align}
    In the above formula, $\tau_{A^nB^n}$ is a particular mixture of IID states, while the $g_{n,x}$ term is defined by the formula $g_{n,x} = \binom{n+x-1}{x-1}$ (where $x$ is a value that depends on the dimensions of Alice and Bob's subsystems), and affects the final security parameters and key length in the security analysis against coherent attacks. More precisely, the above theorem is used to reduce the security of the permutationally invariant state $\bar{\rho}_{A^nB^n}$ to the security of a \textit{purification} of $\tau_{A^nB^n}$, with some costs to the security parameter and key length that depend on $g_{n,x}$. 
    \item Finally, a finite-size security proof is obtained against \emph{all} IID states in the mixture $\tau_{A^nB^n}$. A particularly suitable approach for doing so is the division of all states into ``filtered'' and ``unfiltered" states  previously described in \cref{subsubsec:filtering}. This is then used to prove the security of the purification of $\tau_{A^nB^n}$.
\end{enumerate}

Essentially, the proof can be viewed as ``lifting'' the security proof against IID collective attacks to one against coherent attacks, at the price of some penalties to the key length and security parameter. Note that some works use the underlying de Finetti theorem directly as part of a phase error correction based proof; see, for example, Ref.~\cite{matsuura_asymptoticallytight_2025}. In contrast, Ref.~\cite{nahar_postselection_2024} derives a reduction to the security analysis of IID collective attacks, which can then be combined with any chosen security proof against IID collective attacks. In this work, we use the term postselection technique to refer to the latter approach (coupled with a suitable IID security analysis, which is typically straightforward).

\begin{remark}
Note that in practice, rather than concluding a larger security parameter at the end of the security proof against coherent attacks, one would usually instead fix the desired \emph{final} security parameter $\eps$, and then aim to obtain an IID security proof that achieves a suitably smaller security parameter $\eps'$, such that the final security parameter against coherent attacks is the desired value $\eps$. This would usually come at the cost of lower key rates in the IID part of the proof, since it has to achieve a more stringent security parameter.
\end{remark}

When applying the postselection technique, there are several points that should be noted:
\begin{enumerate}
\item It is critical to ensure that the protocol indeed satisfies permutation symmetry; however, we highlight that various steps of most protocols (such as error correction and privacy amplification) are not ``inherently'' permutation-symmetric, which poses an obstacle in applying the postselection technique. One method to resolve this is to apply a random permutation on Alice and Bob's raw data (see \cref{subsubsec:step_permute}), which always yields the required permutation-symmetry property for the entire protocol, as long as the measurements are described by a channel that commutes with this random permutation.\footnote{A method was proposed in~\cite{beaudry_assumptions_2015} to remove the need for this random permutation. However, it comes at the cost of a further penalty to the key length and security parameter, and some technical issues need to be addressed regarding conditioning on the accept event. It was also claimed in~\cite{scarani_security_2008} that the random permutation is unnecessary; however, the argument contains a gap in that it focuses solely on relating the smooth min-entropy of the states with or without the random permutation. Based on the analysis in~\cite{nahar_postselection_2024}, this is not sufficient by itself to yield a full security proof, and so there is a logical gap that would need to be closed for this claim to be valid.} (As a specific example, if the measurements can be described as an IID tensor product of channels, then they indeed commute with this random permutation.)
\item In the lift to security against coherent attacks, there are \emph{two} effects --- the change in the key length and the change in the security parameter, though as described above, the latter is usually compensated by further decreasing the key length. Some follow-up works based on the postselection technique only considered one of these effects, resulting in incorrect conclusions regarding the key length or security parameter. 
\item The lift to security against coherent attacks cannot directly be applied to optical systems, owing to their infinite dimensions resulting in an infinite cost for the lift.\footnote{Technically, in infinite dimensions defining the integral in \cref{eq:deFinettiTheorem} is also a challenge. We don't stress on this point in our description, as the infinite cost is sufficient to illustrate the problem.} Thus, a squashing map (see \cref{subsubsec:squashing}) must be used \emph{before} the use of the de Finetti theorem. This poses a challenge when adding state-dependent constraints (such as those arising from the flag-state squasher) to the mixture of IID states in $\tau_{A^nB^n}$ which is independent of the state $\rho_{A^nB^n}$ as given by the de Finetti theorem.
\item IID security proofs for PM protocols often require Alice's marginal state to be fixed, reflecting the trusted source preparation (see also \cref{sec:SRS}). Care must be taken when lifting such a proof to security against coherent attacks  with the postselection technique, as it needs a security statement against \emph{all} states in the mixture $\tau_{A^nB^n}$. This is elaborated on in \cref{subsubsec:nahar_postselection}.

\end{enumerate}

We now sketch an outline of how to perform the IID part of the security proof, followed by a discussion of a few significant works on the postselection technique.

\subsubsection{Security proofs against IID collective attacks}
\label{subsubsec:IIDproofs}

The approach outlined here follows the lines of Refs.~\cite{renner_security_2005,george_numerical_2021}. The core observation for fixed-length protocols is that under the IID assumption, it is comparatively easy to study the sets of filtered and unfiltered states as described in \cref{subsubsec:filtering}. More precisely, the reasoning would be roughly as follows (in the following, the state always refers to the entangled Alice-Bob-Eve state arising from the source-replacement analysis): 
\begin{enumerate}
\item Given a specific choice of acceptance test and any value $\eps_{\mathrm{AT}}$, and denoting the single-round state (before measurements) as $\rho_{ABE}$, it is possible to use IID concentration inequalities to construct a set $S_{\eps_{\mathrm{AT}}}$ such that any $\rho_{ABE} \notin S_{\eps_{\mathrm{AT}}}$ is a filtered state. 
\item It therefore suffices to bound the smooth min-entropy for the case where the global state (before measurements) has the form $\rho_{ABE}^{\otimes n}$ for some $\rho_{ABE} \in S_{\eps_{\mathrm{AT}}}$, since this would be a lower bound on the smooth min-entropy for any unfiltered state, as required from the discussion in \cref{subsubsec:filtering}.
\item To find such a bound, one can invoke the \term{asymptotic equipartition property} (AEP)~\cite{tomamichel_fully_2009}, which lower bounds the smooth min-entropy of an IID state in terms of the von Neumann entropy of a single round, along with some ``finite-size correction'' terms. This reduces the task to lower bounding the von Neumann entropy produced in a single round from any state $\rho_{ABE}$ constrained to be in the set $ S_{\eps_{\mathrm{AT}}}$, which can be addressed using convex optimization techniques~\cite{winick_reliable_2018,george_numerical_2021,lin2017security}.
\end{enumerate}
As an alternative to smooth min-entropy, one could also consider the {\Renyi} entropy as noted in \cref{subsec:estimating_entropy}; the overall reasoning remains essentially similar.

\begin{remark}\label{remark:entropy_after_conditioning}
In the above outline, the smooth min-entropy being bounded is that of the state \emph{without} conditioning on the protocol accepting.\footnote{The difficulty with analyzing the state conditioned on accepting is that it may no longer be IID, hence the AEP cannot be applied.} To rigorously address the issue mentioned in \cref{subsec:estimating_entropy} regarding how this conditioning affects entropies, one can use~\cite[Lemma~10]{tomamichel_largely_2017} to relate this to the smooth min-entropy of the \emph{subnormalized} state conditioned on accepting, after which a suitable version of the Leftover Hashing Lemma~\cite[Proposition~9]{tomamichel_largely_2017} can be applied. 

We emphasize however that this requires a rigorous formulation of smooth min-entropy for subnormalized states, as described in that work (see also Ref.~\cite{tomamichel_quantum_2016}).
This involves various technical details that we will not cover within this work, but to sketch out some qualitative properties: the min-entropy (\emph{without} smoothing) of a subnormalized state is defined such that it is larger than the min-entropy of its normalized version, by an amount depending on the normalization factor.\footnote{This is the sense in which it may be possible to find a nontrivial constant lower bound $\entbnd$ on the smooth min-entropy of \emph{subnormalized} states conditioned on accepting, because their entropy is larger by an amount depending on the accept probability.} This definition is then extended to a smoothed version, and it is shown that the resulting quantity satisfies various other properties required for the security proof, such as~\cref{eq:ECchainrule}.

For security proofs that are instead based on {\Renyi} entropies, one can use~\cite[Lemma~B.5]{dupuis_entropy_2020} to relate the {\Renyi} entropies of the states with and without conditioning on accepting. That lemma is slightly easier to use in that it applies to the normalized conditional state, rather than the subnormalized one; however, the lower bound it gives has an explicit dependence on the accept probability, as briefly noted in \cref{subsubsec:probdependentbounds}.
\end{remark}

For variable-length protocols, a proof approach was developed in Ref.~\cite{tupkary_security_2024}, though it specifically required using the {Leftover Hashing Lemma} based on {\Renyi} entropy rather than smooth min-entropy. Informally, the idea was to use the IID assumption to construct a statistical estimator that can be computed from the \emph{observed} frequency distribution in the protocol, and that lower-bounds the final {\Renyi} entropy with high probability. This statistical estimator is then used to choose the length of the final key, in such a way that it can be shown to satisfy the variable-length security definition.

\paragraph*{Decoy-state methods:}\label{par:decoy_ps}

Recall from \cref{subsec:decoy_general} that decoy-state methods aim to bound Eve's influence on single-photon signals. Within the postselection technique, this analysis is performed \textit{after} the reduction to the IID regime. Thus, we can define Eve's single-round (IID) attack channel\footnote{This can be equivalently described in terms of the (IID) state shared by Alice and Bob after Eve's attack, by making use of the source-replacement scheme as described in \cref{sec:SRS}. For pedagogical clarity, we describe the more physically motivated picture that explicitly makes reference to Eve's attack channel $\Phi$.} $\Phi$, acting on Alice's prepared states $\{\rho_{i}^{\mu_k}\}_{i,{\mu}_k}$, Here, the superscript $\mu_k$ denotes Alice's intensity choice, and the subscript $i$ represents the specific encoding (e.g., the bit value and basis choice) prepared by Alice.

The decoy-state protocol thus constrains Eve's attack channel $\Phi$ to satisfy\footnote{Strictly speaking, this relationship is an equality only in the asymptotic limit. Due to finite-size effects, the observed frequencies $\gamma_{j\vert i,\mu_k}$ only approximate the underlying probabilities $p(j\vert i,\mu_k)$. In a rigorous security proof, this approximation is quantified using IID concentration inequalities, which define a valid confidence interval around the observed statistics.}
\begin{equation} \label{eq:decoyConstraints}
    \Pr(j\vert i,\mu_k) \coloneqq \Tr\left[\Phi[\rho_i^{\mu_k}]\Gamma_j \right] \approx \gamma_{j\vert i,\mu_k},
\end{equation}
where $\Gamma_j$ is Bob's POVM element corresponding to outcome $j$, and $\gamma_{j\vert i,\mu_k}$ is the empirical frequency of observing Bob’s outcome $j$ among the subset of rounds in which Alice chose encoding $i$ and intensity choice $\mu_k$.

As discussed in \cref{subsec:decoy_general}, there are two ways to integrate these constraints with the rest of the security analysis. In the first approach, the constraints can be added directly to the convex optimization problem used to calculate the secret key rate. This approach is adopted in \cite{kamin_finite-size_2024,kamin2025renyisecurityframeworkcoherent} within the context of EAT-based proofs (see \cref{subsec:EAT} for more details).
Alternately, the analysis can be performed in a modular two step fashion. 

First, the objective function for the secret key rate is decomposed into components corresponding to specific photon-number emissions $m$. For instance, if the joint state $\rho_{AE}$ is block diagonal in $E$ and we write $
\rho_{AE} = \bigoplus_{m=0}^{\infty} p_m \rho^{(m)}_{AE}$,
then the conditional von Neumann entropy decomposes as
\begin{equation} \label{eq:decoystateblockdiagonal}
H(A|E)_\rho = \sum_{m=0}^{\infty} p_m H(A|E)_{\rho^{(m)}}.
\end{equation}
This allows us to minimize the conditional von Neumann entropy of each photon number separately, since Eve can be assumed to know the photon number of Alice's pulse without loss of generality. Since multi-photon emissions ($m>1$) are assumed to be vulnerable to PNS attacks, their contribution to the key rate is typically lower-bounded by zero. Consequently, the single-round key rate optimization problem collapses to a specific sub-problem that only depends on the single-photon state\footnote{For the tightest key rates, a separate sub-problem corresponding to the vacuum emission is also considered. This process as well as including higher photon-number terms follows similarly to that of the single-photon analysis.}.

Second, the constraints in \cref{eq:decoyConstraints} are used to derive bounds on the statistics of the individual photon-number subspaces. Fundamentally, this takes the form of a set of SDPs \cite{nahar_imperfect_2023,kamin_improved_2024}:
\begin{equation} \label{eq:minGenDecoy}
    \begin{aligned}
        p^{L}(j\vert i,1) = \underset{\Phi}{\textrm{min}\ } &\textrm{Tr}\left[\Phi(\sigma_{i}^{(1)}) \Gamma_j\right] \\
        \textrm{s.t. } &\textrm{Tr}\left[\Phi(\rho_a^{\mu_k}) \Gamma_b\right] \approx \gamma_{b\vert a,\mu_k} \ \forall a,b,\mu_k \\
        & \Phi \textrm{ is CPTP,}
    \end{aligned}
\end{equation}
\begin{equation}\label{eq:maxGenDecoy}
    \begin{aligned}
        p^{U}(j\vert i, 1) = \underset{\Phi}{\textrm{max}\ } &\textrm{Tr}\left[\Phi(\sigma_{i}^{(1)}) \Gamma_j\right] \\
        \textrm{s.t. } &\textrm{Tr}\left[\Phi(\rho_a^{\mu_k}) \Gamma_b\right] \approx \gamma_{b\vert a,\mu_k} \ \forall a,b,\mu_k \\
        & \Phi \textrm{ is CPTP,}
    \end{aligned}
\end{equation}
for each encoding choice $i$ and Bob's measurement outcome $j$. Here, $\sigma_i^{(1)}$ corresponds to Alice's single-photon emission encoded with bit and basis choice $i$. These optimizations yield a rigorous range $\left[p^L(j\vert i, 1),p^U(j\vert i,1)\right]$ for each single-photon statistic $\Pr(j\mid i,1)$. These intervals serve as the specific constraints for the final stage of the security proof, allowing the key rate optimization to be restricted exclusively to the single-photon subspace.

The block-diagonal structure of phase-randomized WCP sources allows these SDPs to be reduced to the more conventionally used linear programs. Since the source is statistically equivalent to a classical mixture of number states, the trace constraints in the SDPs (\cref{eq:minGenDecoy,eq:maxGenDecoy}) decompose into weighted sums of the $m$-photon statistics:
\begin{equation}\label{eq:PSDecoy}
    \gamma_{j\vert i,\mu_k} \approx \sum_{m=0}^{\infty} \Pr(m \mid \mu_k)\, \Pr(j\mid i,m).
\end{equation}
Upper and lower bounds on the single-photon statistics $\Pr(j\mid i,1)$ can then be obtained using standard linear programming techniques.

\begin{remark} \label{remark:twostepdecoy}
    The standard decoy-state analysis typically employs this modular, two-step process using linear programming. As discussed, this relies on two distinct simplifications --- one simplification that treats the decoy analysis as a precursor to the key rate computation, and another simplification from SDPs for the single-photon statistics to linear programs. While each simplification significantly reduces computational overhead, they inevitably introduce looseness in the bounds, resulting in a lower secret key rate \cite{kamin_improved_2024}.
\end{remark}

\subsubsection{Ref.~\cite{christandl_postselection_2009}} \label{subsubsec:christandl_postselection}

This work introduced the idea of the postselection technique as a novel approach to security proofs against coherent attacks.
However, it has a number of limitations we describe below. Note that Ref.~\cite{bunandar_numerical_2020} is essentially based on this approach, and thus suffers from the same limitations (in addition to some of the common gaps explained in \cref{subsec:gap_AT}). Additionally, there are a number of papers (e.g.~Refs.~\cite{lucamarini_efficient_2013,rusca_security_2018}) that provide IID security proofs but do not perform a full analysis against coherent attacks. Care must be taken when lifting to security against coherent attacks to avoid these issues. 

The limitations are listed below:
\begin{enumerate}
\item The analysis is based on considering the model of the QKD protocol after source-replacement and squashing, so the protocol is modelled via a channel that acts on an entangled state between Alice and Bob (with Eve holding a purification) by first performing Alice's source-replaced measurements and Bob's \pending{squashed POVMs}. However, the de Finetti theorem used in this work consisted of $\tau_{A^nB^n}$ being a mixture of \emph{all} IID states. Thus, the argument for lifting the IID security proof to one against coherent attacks implicitly required that the former proves security for \emph{all} possible IID input states to this channel. This poses an issue regarding the following points:
\begin{itemize}
    \item \textbf{Source-replacement scheme:} For PM protocols, the source-replacement scheme imposes a restriction on Alice's marginal state. The corresponding IID security proof only proves security against all IID states satisfying the restriction on Alice's marginal state. Thus, the results from this work are not applicable to lift an IID security proof for a PM protocol to a proof against coherent attacks.
    \item \textbf{Squashing:} Some squashing maps\footnote{See \cref{tab:squashingTypes} for more details on the different squashing maps, and their compatibility with the postselection technique} involve additional constraints to the set of finite-dimensional IID states they prove security for.\footnote{As an example, the proof might use the fact that most of the state is in the single-photon subspace.} However, such a constraint is obtained from the infinite-dimensional IID state. Thus, the postselection technique as presented in this work is not compatible with the use of these squashing maps, since the squashing must be performed prior to using the postselection theorem, otherwise the cost $g_{n,x}$ would be infinite.
\end{itemize}

\item There was a gap in the reasoning used in the last step of their security analysis. Specifically, they claim that if Eve is given an additional $K$ qubits of side-information about the pre-amplification string and the final key length is shortened by $2\log K$ bits, then the trace distance term in the Leftover Hashing Lemma (as presented in \cref{eq:LHL_Hmin}) does not increase. However, this claim does not follow directly from the Leftover Hashing Lemma, and hence the argument is incomplete.\footnote{It is true that the \emph{upper bound} in \cref{eq:LHL_Hmin} of the Leftover Hashing Lemma is nonincreasing under these changes, but this is a distinct statement from the claim that the trace distance term itself is nonincreasing.}
\end{enumerate}
These issues were addressed in a follow-up work~\cite{nahar_postselection_2024}, though at the cost of somewhat larger penalties to the key length or security parameter, and some additional conditions to apply the proof technique. We now discuss that work in more detail.

\subsubsection{Ref.~\cite{nahar_postselection_2024}}
\label{subsubsec:nahar_postselection}

This work addressed the limitations described above as follows.
\begin{enumerate}
\item As mentioned in \cref{subsubsec:christandl_postselection}, the analysis is based on the protocol modeled as a channel that acts on an entangled state between Alice and Bob that is a result of the source-replacement scheme and squashing. The challenges that arise for these cases were dealt with as follows:
\begin{itemize}
    \item \textbf{Source-replacement scheme:} The challenge here was that the source-replacement scheme required Alice to have a fixed marginal state (which was used in the security proof against IID attacks), and needs to be accounted for in the mixture of IID states $\tau_{A^nB^n}$. This was addressed through the usage of a modified de Finetti theorem \cite[Corollary 3.2]{fawzi_quantum_2015} which allowed the mixture to be over all IID states that had the same marginal as $\rho_{A^nB^n}$.\footnote{Note that for this to hold, $\rho_{A^n}$ must also be IID. This is equivalent to assuming that Alice's prepares random states in an IID fashion.} Also, this work improved on the cost $g_{n,x}$ in the de Finetti theorem stated in Ref.~\cite{fawzi_quantum_2015}.
    \item \textbf{Squashing:} Here, the challenge was that more generally applicable squashers (flag-state squasher and dimension reduction method) required additional constraints to be applied to the finite-dimensional IID states. This was addressed through the construction of a new squashing map that was just as generally applicable as the flag-state squasher\footnote{In particular, these squashing maps are applicable to all DV protocols, but not to CV protocols.}, yet did not require any additional constraints. Thus, this squashing map can be used before the de Finetti theorem is applied.
\end{itemize}
\item For fixed-length protocols, this work showed that if the IID security proof proceeds \emph{specifically} by bounding the smooth min-entropy of all unfiltered states (satisfying Alice's reduced-state constraint) as described in \cref{subsubsec:filtering}, then it can be validly lifted to a security proof against coherent attacks, though with larger effects on the security parameter as compared to the initial claim in Ref.~\cite{christandl_postselection_2009}. More generally, for either fixed-length or variable-length protocols, this work showed that such a lift is possible even without that condition \cite[Section V. A.]{nahar_postselection_2024}, but the effect on the security parameter is slightly larger. 
\end{enumerate}
In addition to addressing these limitations, this work also derived some improvements to the finite-size terms appearing in the de Finetti theorem, reducing the impact of those terms on the finite-size key rates. Note that although this work describes a lift for decoy-state protocols, it still requires a finite-size, IID security proof for decoy-state protocols.

\subsubsection{Ref.~\cite{Kamin2025}}\label{subsubsec:kamin_finite-size-dceoy}
Recently, Ref.~\cite{Kamin2025} presented an IID security proof for generic decoy-state protocols, together with a lift to coherent attacks using the postselection technique. The two significant contributions of this work, besides the IID analysis for decoy-state protocols and the explicit key rate computations against coherent attacks from this approach, are a tightened statistical analysis and reduced finite-size corrections. The tightened statistical analysis improves key rates, and is especially relevant in the case of very small security parameters, which are required in the postselection technique. Furthermore, the reduced second-order finite-size term in the key length scales with the square root of number of detected events, rather than the total number of signals, leading to improved finite-size performance.

\subsection{Entropy accumulation based proofs} \label{subsec:EAT}
\begin{figure}[b]
    \centering
    \includegraphics[width=\linewidth]{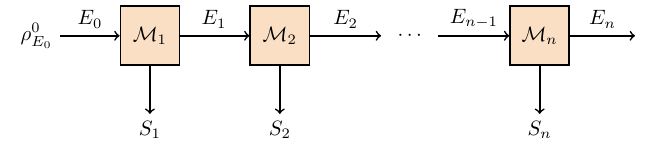}
    \caption{An example of a sequential process representable in the GEAT. Here \(S_i\) can represent the secret registers containing the pre-amplification string, and \(E_i\) is Eve's side-information of round \(i\). The figure was adapted from Ref.~\cite{metger_generalised_2022}.} \label{fig:EAT sequential process}
\end{figure}

There are a variety of accumulation theorems, such as the entropy accumulation theorem (EAT) \cite{dupuis_entropy_2020,dupuis_entropy_2019}, generalized entropy accumulation theorem (GEAT) \cite{metger_generalised_2022,metger_security_2023}  generalized {\Renyi } entropy accumulation theorem \cite{arqand_generalized_2024}, and marginal-constrained entropy accumulation theorems (MEAT) \cite{inprep_vanhimbeeck_tight_2024,fawzi_additivity_2025,arqand_marginal_2025}, which we discuss in greater detail below. Currently, most of these results are focused on fixed-length protocols. Any of these theorems can be utilized for QKD security proofs, as long as the QKD protocol can be abstractly represented as a sequential process satisfying the conditions of the relevant theorem. An example of such a sequential process can be seen in \cref{fig:EAT sequential process}. However, in general one needs to be careful to verify that a sequential process indeed correctly represents a QKD protocol and satisfies the assumptions of the desired accumulation theorem.

Abstractly, entropy accumulation theorems obtain a lower bound on the \(n\)-round entropy (smooth-min and/or \Renyi) of a sequential process in terms of a single-round quantity of the form\footnote{The correction term here depends on the probability of the event $\Oacc$. Technically, the sequential process outputs a string different from $\PAstring$, which goes through a discarding procedure, before resulting in the pre-amplification string  $\PAstring$. However, this discrepancy can be resolved as discussed earlier in \cref{subsubsec:varlength_rawkey}.}
\begin{equation}
    \Hmin^{\esmooth}(\PAstring | \En)_{|\Oacc} \geq n h -\text{corrections},
\end{equation}
where \(\PAstring\) are the secret registers containing the pre-amplification string and \(\En\) is all of Eve's side information except the communication for error correction, error verification and the choice of the hash function for privacy amplification. Here $n$ is the total number of rounds in the QKD protocol, and $h$ is a value that can be computed in terms of a minimization problem only involving single rounds of the protocol.

\paragraph*{Decoy-state methods:}\label{par:decoy_EAT} 
From \cref{par:decoy_ps} recall that, in contrast to the traditional two-step process, where one first computes bounds on the single-photon statistics and then performs the key rate calculation, these two steps can be combined into a single optimization problem (see \cref{remark:twostepdecoy}). Exploiting the block-diagonal structure of WCP signals (see \cref{eq:decoystateblockdiagonal}), this single-shot optimization takes the form:
\begin{equation}\label{eq:single-shot EAT}
    \begin{aligned}
        &\underset{\rho = \bigoplus_m p_m \rho_m, \, \gamma}{\min} \quad f(\rho,\gamma) \\
        &\textrm{s.t. } \Tr\left[\rho_i^{\mu_k} \Gamma_{j} \right] = \gamma_{j|i,\mu_k} \ \forall i,j,\mu_k,
    \end{aligned}
\end{equation}
where the objective function $f$ depends on the specific EAT theorem applied. This single-shot optimization can be relaxed to obtain:
\begin{equation}\label{eq:two-step EAT}
    \begin{aligned}
        &\underset{\rho_1, \, \gamma}{\min} \quad \tilde{f}(\rho_1) \\
        &\textrm{s.t. } \Tr\left[\rho_{i,1}^{\mu_k} \Gamma_{j} \right] \in \left[p^L(j\vert i, 1),p^U(j\vert i,1)\right] \ \forall i,j,\mu_k,
    \end{aligned}
\end{equation}
with $\tilde{f}$ representing a modified version of $f$, and $\left[p^L(j\vert i, 1),p^U(j\vert i,1)\right]$ are obtained in a similar manner as \cref{eq:minGenDecoy,eq:maxGenDecoy}. The primary distinction is that the single-shot approach accounts for correlations between different photon numbers, whereas the two-step process neglects them. 
Note that the combined single-shot optimization (\cref{eq:single-shot EAT} is a tighter result, and the two-step process yields a strictly looser bound (albeit with potentially lower computational cost due to the reduction in optimization variables). The exact magnitude of the improvement due to the combined optimization has not been explicitly quantified. Most decoy-state methods within the EAT framework employ this single-shot analysis, as implementing a separated two-step process is non-trivial in the context of Refs.~\cite{kamin_finite-size_2024,kamin2025renyisecurityframeworkcoherent}. 

Finally, while a single-shot optimization for decoy-state protocols could theoretically be implemented within post-selection or IID analysis, this remains unexplored. We anticipate such an implementation would yield key rate improvements over the methods in Ref.~\cite{kamin_improved_2024}.


\subsubsection{Ref.~\cite{dupuis_entropy_2020}}
This work proves the first version of the EAT for a sequential process (generating the pre-amplification string) that satisfies a technical requirement called the Markov condition. The Markov condition restricts the theorem to protocols that make only ``weakly dependent announcements" \cite{george_finitekey_2022} and limits it to being applied mainly to entanglement-based protocols.
A weakly dependent announcement requires, for example, that a detection event depends only on the total photon number, but not the state. All of the results based on the original EAT perform the reduction to the single-round quantity $h$ by calculating the so-called min-tradeoff function, which is an affine lower bound on the key rate in terms of announcements in the test rounds, e.g. the error rate in the \(X\)-basis. In Ref.~\cite{george_finitekey_2022}, a numerical framework for constructing min-tradeoff functions of entanglement-based protocols in the EAT-framework was presented, and then used to compute key rates for entanglement-based protocols. 

However, the main drawback of the EAT is that it cannot be directly applied to prepare-and-measure protocols. This drawback could in principle be addressed using a ``virtual state tomography'' technique developed in Ref.~\cite{bauml_security_2024}, but thus far this technique has not been applied to decoy-state BB84 specifically, and the resulting performance is uncertain.

\subsubsection{Refs.~\cite{metger_generalised_2022,metger_security_2023}} \label{subsubsec:geat}

Ref.~\cite{metger_generalised_2022} extended the above result to a \emph{generalized} entropy accumulation theorem (GEAT), which considers sequential processes satisfying a different technical requirement called the non-signalling condition (instead of the Markov condition). This allowed it to be applied to prepare-and-measure protocols; however, only under a specific restriction. Namely, it requires that one must assume that while Eve can jointly 
attack all signals, she only interacts with one signal at a time, i.e. applies the attack in sequence. 
We shall refer to this assumption as the ``sequential assumption". Satisfying this assumption in practical implementations imposes strong limits on the repetition rate at which the protocol can be run \cite[Section 5.1]{sandfuchs2025security}. Ref.~\cite{metger_security_2023} uses the GEAT to compute key rates for the B92 protocol, and describes the application of the GEAT to decoy-state calculations, without performing explicit calculations.

Note that in general, the task of numerically finding the min-tradeoff function (required for GEAT calculations) with the best finite-size performance can be very difficult. Thus, most works so far have focused on applying EAT- or GEAT-based security proofs to simple device-independent QKD protocols or qubit protocols, with Ref.~\cite{kamin_finite-size_2024} described below being the only work using it to explicitly compute key rates for decoy-state protocols. Ref.~\cite{arqand_generalized_2024} improves the GEAT by obtaining tighter finite-size key rates, and avoiding explicit min-tradeoff function calculations, which leads to a significantly simpler framework.

\subsubsection{Ref.~\cite{kamin_finite-size_2024}}
This work applies the GEAT \cite{metger_generalised_2022,metger_security_2023} to the qubit BB84 and decoy-state BB84 protocols.
Technically, the usage of the GEAT in this work requires the sequential assumption discussed above, though it was noted in that work that this requirement can be removed 
by instead using the ``marginal-constrained'' versions \cite[Corollary 4.1]{arqand_marginal_2025} we describe next (\cref{subsubsec:MEAT}).
Furthermore, its security proof only applies to fixed-length protocols. 

One technical contribution of this work was that it developed several techniques to optimize the finite-size performance of the GEAT, for instance by optimizing the choice of the min-tradeoff function using techniques somewhat similar to Ref.~\cite{george_finitekey_2022}. Moreover, a tighter formulation for decoy-state protocols is derived, which even in the asymptotic regime should result in higher key rates than the approaches in Ref.~\cite{Kamin2025}.
Finally, in Ref.~\cite[App. B]{kamin_finite-size_2024} it was shown that one can incorporate the flag-state squasher with the GEAT analysis, which is especially relevant for optical detection setups.

\subsubsection{Refs.~\cite{fawzi_additivity_2025,arqand_marginal_2025,kamin2025renyisecurityframeworkcoherent}}
\label{subsubsec:MEAT}
Refs.~\cite{fawzi_additivity_2025,arqand_marginal_2025} derived\footnote{The results in those two works are generalizations of a result in~\cite{inprep_vanhimbeeck_tight_2024} that has thus far only been presented via conference talks, but we note that they both provide standalone proofs that do not depend on that result.} a new \textit{marginal-constrained entropy accumulation theorem} (MEAT) that is particularly well-suited for the analysis of prepare-and-measure protocols. This result enables analysis of both fixed-length and variable-length protocols, while also offering a simpler calculation compared to the GEAT in~\cite{metger_generalised_2022}. Building on this, Ref.~\cite{kamin2025renyisecurityframeworkcoherent} develops a numerical framework for computing key rates for generic decoy-state protocols using the MEAT technique, and applies it to compute key rates for the decoy-state BB84 and the decoy-state 4-6 protocol. It also includes the analysis of certain device imperfections. We note that while the MEAT version in~\cite{arqand_marginal_2025} can in principle be used for handling on-the-fly announcements, Ref.~\cite{kamin2025renyisecurityframeworkcoherent} does not consider this scenario.

\section{Summary of properties of different proof techniques} \label{sec:comparison}
As seen in the previous sections, security proofs for QKD protocols are complex, and rely on a wide variety of mathematical tools and arguments. Given that there exist a variety of proof approaches, in this section we will briefly assess their respective strengths and weaknesses. In doing so, we lay the foundation to make an informed choice of proof techniques  for the decoy-state BB84 protocol (see also Ref.~\cite{devashish_thesis} for a more technical treatment). We stress that this assessment of the various proof techniques reflects the authors' perspective. A choice of the proof technique inevitably involves trade-offs that different individuals may judge differently. Moreover, the characteristics of proof techniques continue to evolve as this remains an active area of research.

In comparing the proof techniques, we consider the following aspects:
\begin{enumerate}
\item \textbf{Performance:} The ability of the proof technique to yield tight key rates in the finite-size regime.
\item \textbf{Robust against device imperfections:} The ability of the proof technique to be modified to address imperfections in the physical hardware, including both IID and non-IID deviations, in contrast to the technique requiring idealized assumptions.
\item \textbf{Scrutiny and Peer Review:} The extent to which the proof approach has been extensively analyzed, scrutinized, and undergone peer review, and whether it is widely accepted in the community.
\item \textbf{Modularity:} The extent to which proof techniques can be adapted to different protocol variations without requiring a complete overhaul of the proof.

\item \textbf{Protocol Choices:} The range of protocol variations that can be accommodated within the proof technique, such as active vs passive detection setups, advantage distillation, noisy preprocessing, on-the-fly announcements, etc.

\item \textbf{Computability of Key Rates:} The ease with which the technique can compute the key rates. In particular, the extent to which the proof technique requires difficult numerical optimizations, in contrast to using analytical formulae.
\end{enumerate}

We include Refs.~\cite{inprep_BDR3,mizutani2025protocolleveldescriptionselfcontainedsecurity} in the discussions below to enable a more complete and current comparison between different proof techniques. As noted in the Introduction, these works aim to provide rigorous, certification-ready proofs and therefore adhere to a higher standard; they are currently under active review by the community.

\begin{table*}[t]
    \caption{Subjective comparison of proof techniques}
    \label{tab:comparison}
    \def\arraystretch{1.3} 
    \setlength\tabcolsep{6pt} 
    \small 
    \begin{tabular}{p{3.1cm} p{3.1cm} p{3.1cm} p{3.1cm} p{3.1cm}}
    \toprule
    \textit{Criterion} & \textit{Entropic uncertainty relation} & \textit{Phase error correction} & \textit{Postselection technique} & \textit{Entropy accumulation theorem} \\
    \toprule
    Performance & Good key rates. & Good key rates. & Pessimistic key rates. & Good key rates. \\
    Robust against device imperfections & Robust against (independent) device imperfections. Some results on correlated imperfections. & Robust against (independent) device imperfections. Some results on correlated imperfections. & Robust against (IID) device imperfections. & Robust against (independent) device imperfections. Limited results on correlated imperfections. \\
     Scrutiny and Peer Review & High degree of scrutiny. & Oldest ``modern'' proof technique. High degree of scrutiny. & High degree of scrutiny. & Very recent work. Not much scrutiny .\\
   Modularity  & The different extensions (e.g.~for robustness against device imperfections, ability to make on-the-fly announcements, etc.) are not modular. Thus, it is not obvious that the extensions can be used simultaneously in the same proof. & The different extensions (e.g.~for robustness against device imperfections, ability to make on-the-fly announcements, etc.) are not modular. Thus, it is not obvious that the extensions can be used simultaneously in the same proof. & Fairly modular and easy to apply. & Fairly modular and easy to apply.  \\
      
    Protocol Choices & Requires certain classical postprocessing. It is unclear how noisy preprocessing and advantage distillation can be incorporated. &  
Requires certain classical postprocessing. 
  Can only handle specific error-correction protocols.
  & Requires a random permutation on the classical data. Cannot handle on-the-fly announcements. & Cannot handle advantage distillation. The theoretical framework itself can directly handle on-the-fly announcements. \\
   
    Computability of Key Rates & Analytical or numerical key rates. Easy to compute. & Analytical or numerical key rates. Easy to compute. & Numerical key rates. Not too hard to compute. & Numerical key rates. Harder to compute than the postselection technique.\\
    \toprule
    \end{tabular}
    \def\arraystretch{1}
\end{table*}

\subsection{Entropic uncertainty relations approach} \label{subsec:comparisoneur}

\begin{enumerate}
    \item \textbf{Performance:} This technique provides good key rates.

    \item \textbf{Robust against device imperfections:} There exist some works that address hardware imperfections using this technique (however, we note that in some cases this requires a quite significant change to the overall structure of the proof). Additionally, due to the close similarity between this technique and phase error correction based methods, this approach can also leverage the numerous results that establish suitable phase error rate bounds in the presence of imperfections \cite{zapatero2023implementationsecurityquantumkey,gllp_security_2004,tamaki_loss-tolerant_2014,curras2023security,sixto_quantum_2025,sixto2022security,zapatero2021security,pereira2020quantum,tupkary_phase_2024,navarrete_improved_2022,wang_phaseerror_2025,mizutani_passive_2025}. Some recent works can also accommodate correlated source \cite{curraslorenzo2026rigorousphaseerrorestimationsecurityframework} and correlated detector imperfections \cite{wang_phaseerror_2025,nahar_phd_2025}.\footnote{These  results on detector imperfections require a protocol modification where detected rounds that are close to a prior detection are discarded} Thus, we consider this proof technique to be robust against a wide range of device imperfections. However, we note that care must be taken when incorporating results from the phase error correction based works into the EUR framework\footnote{Informally, this requires ensuring that the phase error rate bounds derived in the phase error correction framework are consistent with those needed in the EUR-based analysis.}. Furthermore, not all existing works on device imperfections are inherently compatible with one another, thus care must be taken in combining these as well. 
    \item \textbf{Scrutiny and Peer Review:} The proofs are well-established, having undergone extensive peer review over many years \cite{tomamichel_largely_2017,tomamichel2012_tight}. 
     \item \label{item:EURmodularity} \textbf{Modularity:}
    Many proofs using this technique are somewhat modular, with well-stated theorems and lemmas, making them relatively structured and adaptable. However, this modularity is not necessarily retained when considering works that address imperfections and deviations from the ideal BB84 protocol, or protocol variations such as on-the-fly announcements. 


    \item \textbf{Protocol choices:}
    \begin{itemize}
        \item On-the-fly announcements: This technique can be applied to protocols that make certain kinds of announcements during the protocol \cite{tamaki_security_2018}, which has been recently generalized \cite{mizutani2025protocolleveldescriptionselfcontainedsecurity}. However, currently only certain kinds of hardware imperfections can be combined with a restricted class of on-the-fly announcements \cite{wang_phaseerror_2025}.
        \item Classical data processing: This technique must map double-click events to single-click events (in order to not have basis efficiency mismatch), which may prevent us from obtaining optimal key rates. It remains unclear how this proof technique can accommodate noisy preprocessing\footnote{One might attempt to incorporate noisy preprocessing by just changing Alice's POVM  in \cref{eq:eur}, but this does not improve the  $c_q$ parameter, leading to \textit{worse} bounds on the final key rates after accounting for the increased error-correction cost. Thus, this straightforward attempt is not the correct way to obtain good key rate bounds with noisy preprocessing in the EUR approach.} or advantage distillation.
    \end{itemize}

    \item \textbf{Computability of Key Rates:} This proof technique often  gives analytical formulae for key rates, which makes their computation straightforward, although numerical methods also exist \cite{zhou_numerical_2022}.
\end{enumerate}


\subsection{Phase error correction approach} \label{subsec:comparisonphaseerror}

\begin{enumerate}
    \item \textbf{Performance:} This technique provides good key rates.

      \item \textbf{Robust against device imperfections:} The core technique here too involves idealized hardware assumptions, which have been relaxed with subsequent work. Incorporating device imperfections into this proof technique is a subject of active research \cite{curras-lorenzo_security_2024, zapatero2023implementationsecurityquantumkey,gllp_security_2004,tamaki_loss-tolerant_2014,curras2023security,sixto_quantum_2025,sixto2022security,zapatero2021security,pereira2020quantum,tupkary_phase_2024,navarrete_improved_2022}, and there has been a large amount of progress on this front in recent years. Some recent works can also accommodate correlated source \cite{curraslorenzo2026rigorousphaseerrorestimationsecurityframework} and correlated detector imperfections \cite{wang_phaseerror_2025,nahar_phd_2025}. Thus, we view this proof technique as robust to device imperfections. However, as with the EUR technique, not all works on imperfections are inherently compatible with each other, or with diverse protocol choices (see also \cref{item:PECmodularity}).

    \item \textbf{Scrutiny and Peer Review:} The proofs \cite{koashi_simple_2005,koashi_simple_2009,koashi_efficient_2006} are well-established, having undergone extensive peer review over many years.

      \item \label{item:PECmodularity} \textbf{Modularity:} This proof technique is not stated in a modular fashion in the literature. Every modification seems to require a new proof, and it is not obvious if the different proof modifications fit with each other. For example, the current work on device imperfections is inherently incompatible with the work done to incorporate on-the-fly announcements into the proof\footnote{Since both the EUR and phase error correction methods rely on many of the same works for phase error estimation, this observation is also applicable to the EUR proof technique.}. Thus, care must be taken to ensure that security is proved for \emph{all} protocol modifications \emph{simultaneously}. 

    \item \textbf{Protocol choices:}
    \begin{itemize}
        \item On-the-fly announcements: Similar to the EUR, Ref.~ \cite{tamaki_security_2018} presented an approach to accommodate certain kinds of on-the-fly announcements in idealized conditions, which was later extended by Refs.~\cite{wang_phaseerror_2025,mizutani2025protocolleveldescriptionselfcontainedsecurity}. 
        \item Classical data processing: This technique must map double-click events to single-click events (in order to not have basis efficiency mismatch), which may prevent us from obtaining optimal key rates.
        \item Furthermore, the basic proof technique is fairly restrictive in the kind of error correction that can be performed in the protocol (see \cref{subsec:phaseerror} for a more detailed discussion). It is only very recently that more complicated error correction protocols \cite{du_advantage_2024} have been incorporated into this proof technique, and this has not received as much scrutiny as the rest of the proof technique\footnote{Additionally, this proof technique requires that the hash family used is linear and dual universal$_2$, whereas the other three proof techniques use universal$_2$ hashing. However, Toeplitz matrices and random binary matrices are both dual universal$_2$ and universal$_2$.}.
    \end{itemize}

    \item \textbf{Computability of Key Rates:} This proof technique often  gives analytical formulae for key rates, which makes their computation straightforward, although numerical methods also exist \cite{zhou_numerical_2022}.
\end{enumerate}


\subsection{Postselection technique}
\label{subsec:comparisonpostselection}

\begin{enumerate}
    \item \textbf{Performance:} This technique provides pessimistic key rates.

\item \textbf{Robustness Against Device Imperfections:}
Since this proof technique reduces the security analysis to the IID setting, it is naturally compatible with a wide range of IID hardware imperfections \cite{zhang_security_2021,winick_reliable_2018,nahar2025imperfectdetectorsadversarialtasks,nahar_imperfect_2023}. However, its ability to handle non-IID imperfections remains unclear. This is because ensuring permutation invariance of the protocol---a key requirement of this technique---is challenging, even when applying a random permutation to the classical data. As a result, extending this approach to non-IID imperfections is not straightforward.

    \item \textbf{Scrutiny and Peer Review:} The proofs are well-established, having undergone extensive peer review over many years \cite{christandl_postselection_2009}. Critical gaps and extensions to practical scenarios were addressed in recent works \cite{nahar_postselection_2024}. 

      \item \textbf{Modularity: } This proof technique is generally stated in well-defined and self-contained mathematical statements. More importantly, the theorem statements are fairly general and allow for generic protocols. Thus, it is modular --- one needs to separately apply this technique for the lift to coherent attacks, and have protocol-specific proof steps for obtaining the IID security proof. 
    \item \textbf{Protocol choices:}
    \begin{itemize}
        \item Random permutation: This proof technique requires a random permutation to be applied to the classical data after the measurements (and IID behaviour of the hardware). 
        \item On-the-fly announcements: The technique fundamentally starts with a source-replaced state defined over the global protocol execution, and thus it is unclear how it can be modified to incorporate on-the-fly announcements.
        \item Other protocol choices such as active vs passive basis choice, advantage distillation or noisy preprocessing are easily handled within this method. 
    \end{itemize}
   \item \textbf{Computability of Key Rates:} This proof technique has moderate numerical requirements, since informally, one only has to perform key rate computations in the IID case (with suitably modified parameters) \cite{george_numerical_2021,lin2017security,winick_reliable_2018,Kamin2025}.
\end{enumerate}


\subsection{Entropy accumulation approach}
\label{subsec:comparisonentropyacumulation}
There are multiple variations of entropy accumulation, most prominently the original EAT~\cite{dupuis_entropy_2020}, the subsequent GEAT~\cite{metger_generalised_2022}, and very recently some ``marginal-constrained'' versions (which we shall broadly include under the term MEAT, for brevity)~\cite{inprep_vanhimbeeck_tight_2024,fawzi_additivity_2025,arqand_marginal_2025}. Out of these variations, the MEAT appears most suitable for analyzing PM protocols, since the EAT faces difficulties incorporating Alice's marginal state constraints in the source-replacement technique, while applying the GEAT to a PM protocol currently requires it to satisfy a restrictive form of sequential structure that limits its repetition rate. Hence we shall focus on the MEAT for the points discussed below. 

\begin{enumerate}
\item \textbf{Performance:} This technique provides good key rates.
\item \textbf{Robustness against Device Imperfections:} 
The reduction to single rounds in the MEAT only relies on fairly mild core assumptions (mostly regarding memory in the devices), and the analysis of single rounds would mainly rely on finding suitable source maps and squashing maps in the presence of device imperfections, which have been studied in various forms in the IID case. Hence we believe this approach should be robust to many types of imperfections; however, there are currently few works \cite{kamin2025renyisecurityframeworkcoherent} that explicitly perform such an analysis. For correlated source imperfections, see Refs.~\cite{marwah2025provingsecuritybb84source,arqand2024mutualinformationchainrules} for an approach that is currently limited to fairly simple protocols. For correlated detector imperfections, see \cite{nahar2025imperfectdetectorsadversarialtasks,nahar_phd_2025}.

\item \textbf{Scrutiny and Peer Review:} While the general entropy accumulation framework has been established for some time, and the initial ideas behind the MEAT were presented as conference talks in Ref.~\cite{inprep_vanhimbeeck_tight_2024}, 
publicly accessible proofs of the MEAT have only become recently available in Refs.~\cite{fawzi_additivity_2025,arqand_marginal_2025}. Thus, they have not yet received the adequate degree of scrutiny from the community.

\item \textbf{Modularity:} The structure of the proof is reasonably modular, with a self-contained theorem statement reducing the analysis of the overall protocol to the analysis of single rounds, in a manner that mostly does not depend on the specific details protocol.

\item \textbf{Protocol Choices:} The technique applies to a wide range of protocols. It straightforwardly accommodates noisy preprocessing, though it is not directly compatible with advantage distillation. The theoretical framework itself  can directly handle on-the-fly announcements \cite{inprep_BDR3}.

\item \textbf{Computability of Key Rates:} The numerical techniques \cite{kamin2025renyisecurityframeworkcoherent} for computing key rates under this approach appear to be somewhat harder as compared to those for the postselection technique.
\end{enumerate}

\begin{table*}[t!]
    \caption{  Allowed protocol variations for BB84 for each proof technique}
    \label{tab:variationsandtechniques}
    \def\arraystretch{1.3} 
    \setlength\tabcolsep{6pt} 
    \small 
    \begin{threeparttable}
    \begin{tabular}{p{3.1cm} p{3.1cm} p{3.1cm} p{3.1cm} p{3.1cm}}
    \toprule
    \textit{Protocol Variation} & \textit{Entropic uncertainty relation} & \textit{Phase error correction} & \textit{Postselection technique} & \textit{GEAT / MEAT} \\
    \toprule
    Active Detection Setups & \cite{tomamichel2012_tight,wiesemann_consolidated_2024}. & \cite{hayashi_security_2014,koashi_efficient_2006}. &  \cite{Kamin2025}. & \cite{metger_security_2023}. \\
    Active Detection Setup +  Imperfections & \cite{curras-lorenzo_security_2024} (Sources), \cite{tupkary_phase_2024,nahar_phd_2025} (Detectors), \cite{curras-lorenzo_security_2025,nahar_phd_2025} (Combined). &   \cite{curras-lorenzo_security_2024} (Sources), \cite{tupkary_phase_2024,nahar_phd_2025} (Detectors), \cite{curras-lorenzo_security_2025,nahar_phd_2025} (Combined). & Cannot handle.\tnote{d}& \cite{kamin2025renyisecurityframeworkcoherent} (Sources), can be straightforwardly combined with \cite{nahar2025imperfectdetectorsadversarialtasks} to include detectors.  \\
    Passive Detection Setups & 
\cite{wang_phaseerror_2025,mizutani_passive_2025}. & \cite{wang_phaseerror_2025,mizutani_passive_2025}. & \cite{Kamin2025,nahar_postselection_2024}. & \cite{kamin_finite-size_2024,kamin2025renyisecurityframeworkcoherent}. \\
     Passive Detection Setup +  Imperfections \tnote{c} & \cite{curras-lorenzo_security_2024} (Sources), \cite{wang_phaseerror_2025,mizutani_passive_2025} (Detectors). Ref.~\cite{wang_phaseerror_2025} presents an approach to combine, but does not evaluate the resulting expressions due to numerical challenges. &  \cite{curras-lorenzo_security_2024} (Sources), \cite{wang_phaseerror_2025,mizutani_passive_2025} (Detectors). Ref.~\cite{wang_phaseerror_2025} presents an approach to combine, but does not evaluate the resulting expressions due to numerical challenges. & Can handle IID imperfections with straightforward combination of IID security proofs Refs.~\cite{winick_reliable_2018,nahar_imperfect_2023} (Sources) and Ref.~\cite{nahar2025imperfectdetectorsadversarialtasks} (Detectors), but no explicit work has done so.& Ref.~\cite{kamin2025renyisecurityframeworkcoherent} (Sources),  can be straightforwardly combined with \cite{nahar2025imperfectdetectorsadversarialtasks} to include detectors. \\
    On-the-fly announcements & \cite{wang_phaseerror_2025,tamaki_security_2018} (Certain announcements only)  \cite{mizutani2025protocolleveldescriptionselfcontainedsecurity}. & \cite{wang_phaseerror_2025,tamaki_security_2018} (Certain announcements only)  \cite{mizutani2025protocolleveldescriptionselfcontainedsecurity}.  & Cannot handle. &  \cite{metger_generalised_2022,inprep_BDR3}. \\
   Fixed-length vs Variable-length & Both can be handled, and can be obtained from a suitable bound on the phase error rate \cite{tupkary_phase_2024,wiesemann_consolidated_2024,curras-lorenzo_tight_2021}. & Both can be handled, and can be obtained from a suitable bound on the phase error rate \cite{kawakami_security_nodate,hayashi_concise_2012} & Both can be handled, and require slightly different bounds on the single-round entropy \cite{Kamin2025,tupkary_security_2024}. & Both can be handled depending on the EAT theorem used \cite{kamin2025renyisecurityframeworkcoherent,metger_security_2023,inprep_vanhimbeeck_tight_2024}.  \\
    Advantage Distillation & Cannot handle. & Recent work \cite{du_advantage_2024}. & Can handle, although no work has explicitly done so.\tnote{b} & Cannot handle. \\
    \toprule
    \end{tabular}
    \begin{tablenotes}[flushleft]

\footnotesize

\item[a*] For direct vs. reverse reconciliation, one must bound bound the entropy of the pre-amplification string generated by the correct party (or, for the phase error correction approach bound the number of phase errors in the rounds selected for key generation by the correct party). When handling one-way or two-way error correction, entropy-based arguments require bounding the appropriate cost, and similar arguments must also be carried out rigorously for the phase error correction approach (see \cref{subsec:errorcorrection}).

\item[b] Postselection
provides a generic reduction to the IID setting, where advantage
distillation is relatively straightforward to incorporate. For many
protocols the resulting calculation is elementary, though it is less
clear whether numerical approaches can handle it efficiently.

\item[c] A perfectly balanced passive setup with suitable postprocessing can be shown to be equivalent to an active setup.

\item[d] See \cite[Table 1]{nahar2025imperfectdetectorsadversarialtasks} and \cite[Section 5.6]{nahar_phd_2025} for more details. The difficulty lies in using the weight-preserving flag state squasher for active-detection setups, and in bounding the weight in the flag-space, combined with the fact that one has to use the squasher before the postselection technique (in order to have finite-dimensions).  
\end{tablenotes}
    \end{threeparttable}
    
    \def\arraystretch{1}
\end{table*}

\subsection{Protocol Variations}\label{subsec:protocolvariations}

As seen in the previous subsections, different proof techniques
naturally accommodate different sets of protocol variants. We discuss
this in greater detail here, and in particular include
\cref{tab:variationsandtechniques}, which summarizes how the proof
techniques discussed in this work apply across the protocol variants
considered, together with citations to (some of the) relevant literature. The
goal is to give the reader a bird's-eye view of which arguments are
available in each setting, and a glimpse at the complexity of the interaction between proof techniques and protocol design. A few words of caution are in order before consulting it,
however.

The works cited in the table do not necessarily provide a fully rigorous
security analysis of a full QKD protocol (see \cref{sec:commonGaps}). Furthermore, 
 the scope is often limited to, say, bounding the phase-error rate, or
handling a specific imperfection, or treating a restricted class of
protocols. Utmost care must therefore be taken before invoking these
results to justify the security of a given implementation: one must
verify that the entire implemented protocol matches the one analyzed in
a corresponding end-to-end security proof (see
\cref{subsec:focusofreview}), a task which may require combining
results from several works. (Such combinations are sometimes
straightforward but are not always carried out explicitly in the
literature.) Beyond this general caveat, some entries require
non-trivial additional steps even where a technique applies in
principle. For instance, the decoy-state analysis introduces an
additional layer on top of existing qubit analyses that is often not
explicitly included in the cited works. Another example is that most phase error estimation
results and many squashing arguments are only valid under the assumption of no efficiency mismatch. This requires double-click events to
be randomly assigned to single-click outcomes. Whenever possible, we flag these
distinctions within the table itself. We also note that the table covers both qubit BB84 and decoy-state BB84 protocols.

\section{Conclusion}
This review provides a detailed and structured overview of modern proof techniques used in QKD security analysis, with a particular focus on their application to the decoy-state BB84 protocol. By identifying common gaps, misconceptions, and technical challenges in the literature, it clarifies the conceptual foundations and tools required for rigorous and practically relevant QKD security proofs.

Our aim is to serve both as a primer for newcomers entering the field and as a resource for experienced researchers seeking to critically evaluate or build upon existing proof strategies. To our knowledge, this is the most comprehensive review of QKD security proofs to date.

As demonstrated throughout this review, no single publication currently offers a complete and rigorous security proof for a fully specified, practically relevant QKD protocol.  While significant progress has been made and most of the essential tools now exist, they have yet to be combined in a precise and correct manner to yield a certifiable proof. Refs.~\cite{mizutani2025protocolleveldescriptionselfcontainedsecurity,inprep_BDR3} are some recent promising attempts at meeting this high standard, and are currently undergoing scrutiny from the community. Achieving such a proof along with its scrutiny, is critical for advancing QKD from experimental demonstrations to practical deployment. This remains an important open problem for the field.

\section*{Acknowledgments}

This article was written as part of the Qu-Gov project, which was commissioned by the German Federal Ministry of Finance. We thank the Bundesdruckerei --- Innovation Leadership and Team for their support and encouragement. In particular, we thank Holger Eble of the  Bundesdruckerei and Tobias Hemmert of the Federal Office for Information Security (BSI) for providing valuable feedback and comments on drafts of this manuscript.

The writing of this document greatly benefited from interactions with the wider community, which includes numerous insights, discussions, and feedback, over many years. However, any errors or omissions are entirely our own responsibility. We would like to thank Takaya Matsuura for teaching the details of the phase error correction proof approach.
We would like to thank Masato Koashi for further explaining the applications of the phase error correction proof approach, and the source-replacement scheme as applied to this proof technique.
We thank Fred Fung for explaining the universal squasher to us.
We thank Toyohiro Tsurumaru, Victor Zapatero, Margarida Pereira and Guillermo Curr\'{a}s-Lorenzo for helpful discussions about the EUR and phase error correction techniques, and the similarities between them.
We also thank Ian George, Jie Lin and Peter Brown for helpful discussions on the EAT and its improved versions. We thank Christoph Pacher for providing detailed feedback on the first version of this work.

\bibliography{bibliography}

\newcommand{\nX}{n_X}
\newcommand{\nKrv}{\bm{\nK}}
\newcommand{\nXrv}{\bm{\nX}}
\newcommand{\eX}{e^{\mathrm{obs}}_{X}  }
\newcommand{\eXrv}{\bm{\eX}}

\newcommand{\eph}{e^{\mathrm{key}}_X } 
\newcommand{\ephrv}{\bm{\eph}}

\newcommand{\Bound}{\mathcal{B}_\mathrm{ph}}

\newcommand{\epsAT}{\varepsilon_\mathrm{AT}}

\newcommand{\epsATb}{\varepsilon_\text{AT-b}}

\newcommand{\epsATa}{\varepsilon_\text{AT-a}}

\newcommand{\epsATc}{\varepsilon_\text{AT-c}}
\newcommand{\epsdecoy}{\varepsilon_\text{AT-d}}
\newcommand{\epsATsingle}{\varepsilon_\text{AT-s}}
\newcommand{\epsdecoyrusca}{\varepsilon_\text{AT-vac}}

\newcommand{\epsPA}{\varepsilon_{\text{PA}}}
\newcommand{\epsEV}{\varepsilon_{\text{EV}}}
\newcommand{\epsbar}{\bar{\varepsilon}}

\newcommand{\ECost}{ \log(2/\epsEV) }

\newcommand{\smoothmin}[1]{H^{#1}_{\text{min}}}
\newcommand{\smoothmax}[1]{H^{#1}_{\text{max}}}

\newcommand{\floor}[1]{\left\lfloor #1 \right\rfloor} 

\appendix

\section{A finite-size key rate calculation using Entropic Uncertainty Relations} \label{app:eurfinite}

In this appendix we work through a finite-size security proof of
decoy-state BB84 using the EUR framework, illustrating several of the
concepts covered in this work. We stress that this is not meant to be a full rigorous analysis, and is included here purely for pedagogical reasons.  We base our analysis on the
variable-length finite-size proof of Ref.~\cite{tupkary_phase_2024} and largely use the same notation;
the key elements can be found in
Refs.~\cite{tomamichel_largely_2017,tomamichel2012_tight,lim_concise_2014,wiesemann_consolidated_2024}.
We begin with qubit BB84 without loss, then incorporate loss, and
finally extend our analysis to decoy-state BB84.

\subsection{Qubit BB84 without Loss}
Consider first the idealized case where Alice sends perfect qubits, and the channel and Bob's 
detectors are lossless. In this case, every round leads to a detection on Bob's side. Bob implements an active basis choice measurement, and Alice and Bob both choose the $X$ basis with probability $p_X$, and $Z$ basis with probability $p_Z$. The total number of rounds is $n$.

They use $X$ basis rounds for testing, and obtain $\nX$ rounds where they both measured in the $X$ basis. On these rounds, they observe an error rate of $\eX$. They used $Z$ basis rounds for key generation, and obtain $\nK$ rounds where they both measured in the $Z$ basis. The 
central quantity governing security is the \emph{phase error rate}~$\eph$: 
the fraction of key generation rounds on which Alice and Bob would have 
obtained an error had they both measured in the $X$~basis instead of the $Z$~basis. 
This quantity is never directly observed, but its statistical relationship 
to the observed $X$-basis error rate~$\eX$ is the key to the analysis. Based on these observations, Alice and Bob produce a key of output length $\ell(\nX,\nK,\eX)$, and leak $\leak(\nX,\nK,\eX)$ bits during error-correction\footnote{It is natural to split off a part of the $Z$ basis rounds to obtain some estimate on the $Z$-basis QBER. Such a variation can be easily incorporated, however we omit it here for the sake of simplicity.} and $\ECost$ bits during error-verification using two-universal hashing.

We work in the entanglement-based picture obtained via the
source-replacement scheme (see \cref{sec:SRS}), so the full $n$-round protocol is described
by a global state $\rho_{A^n B^n E^n}$ representing Eve's attack.
Classical announcements and measurements on this fixed but unknown state give rise to random variables;
we use the bold font $\bm{x}$ for a random variable and $x$ for a particular value
it takes, and write $\Omega(x)$ for the event $\bm{x}=x$.

The random variables actually observed in the protocol are:
\begin{itemize}
    \item $\nKrv$: the number of rounds in which both Alice and Bob
          measured in the $Z$ basis, retained for key generation.
    \item $\nXrv$: the number of rounds in which both Alice and Bob
          measured in the $X$ basis, used for testing.
    \item $\eXrv$: the observed error rate in the $X$-basis test rounds.
\end{itemize}
We require the following bound on the phase error rate $\Bound(\eX,\nX,\nK)$ such
that
\begin{equation} \label{eq:samplingrequbit}
    \Pr\!\bigl(\ephrv \;\geq\; \Bound(\nXrv,\nKrv,\eXrv)\bigr)
    \;\leq\; \epsAT^2.
\end{equation}
We will derive such a $\Bound$ in the following subsections. For now we treat it as given, and
define the conditional failure probability
\begin{equation} \label{eq:kappaqubit}
\begin{aligned}
   & \kappa(\nX,\nK,\eX) 
    \;\coloneq\; \\
    &
    \Pr\!\bigl(\ephrv \;\geq\; \Bound(\nX,\nK,\eX)\bigr)_{\Omega(\nX,\nK,\eX)},
    \end{aligned}
\end{equation}
so that \cref{eq:samplingrequbit} is equivalent to
\begin{equation} \label{eq:kappaaveragebound}
    \sum_{\nX,\nK,\eX}
    \Pr\!\bigl(\Omega(\nX,\nK,\eX)\bigr)\,\kappa(\nX,\nK,\eX)
    \;\leq\; \epsAT^2.
\end{equation}

The security proof has two parts.
First, we show that a high-probability upper bound on $\ephrv$ implies a
secure key rate via the EUR. Second, we derive such a bound from the
observed data via a Serfling sampling argument. We address these in order.

\subsubsection{Using the EUR for security} \label{subsubsec:EURforsecurity}
We refer the reader to the proof of \cite[Theorem 1]{tupkary_phase_2024} for more details. As stated in \cref{theorem:secrecycorrectness}, it suffices to prove
the secrecy of the protocol; correctness is handled separately (see
\cref{subsec:correctnessproof,subsubsec:misconception_EV}).

Fix a particular observed event $\Omega(\nX,\nK,\eX)$ and let
$\rho_{| \Omega(\nX,\nK,\eX)}$ denote the state conditioned on this event\footnote{In order to properly define this state, one has to break the measurements in the protocol into multiple steps, and modify the timings of various steps for the purposes of the theoretical analysis. For instance, this event can only be defined after Alice and Bob have completed the basis choices for all rounds, have measured all the $X$ basis rounds, and but have \emph{not} measured any of the $Z$ basis rounds (see \cref{subsubsec:conditioning}). }.
Let $Z_A^{\nK}$ be Alice's raw key register, $E^n$ Eve's quantum side
information, and $C^n$ all public classical announcements. Define the
virtual state $\rho^{\mathrm{virt}}_{\mid\Omega}$ as the state that
would result if both Alice and Bob had instead measured their
$\nK$ key-generation rounds in the $X$ basis, producing registers
$X_A^{\nK}$ and $X_B^{\nK}$. This virtual measurement is never
performed in the protocol; it appears only in the security analysis.

The EUR statement \cite{tomamichel_uncertainty_2011} (see \cref{subsec:eur}), followed by data-processing inequalities, relates the smooth 
min-entropy of the measurements in the $\nK$ rounds to the phase error rate on those rounds. In particular, we obtain
\begin{equation} \label{eq:eurtemp}
\begin{aligned}
   & \smoothmin{ \sqrt{\kappa(\nX,\nK,\eX)}}(Z_A^{\nK} | C^n E^n)_{\rho | \Omega(\nX,\nK,\eX)} \\
    &\;+\; 
    \smoothmax{\sqrt{\kappa(\nX,\nK,\eX)}}(X_A^{\nK} | X_B^{\nK})_{\rho^\mathrm{virt} | \Omega(\nX,\nK,\eX)} \;\geq\; \nK,
    \end{aligned}
\end{equation}
where $Z_A^{\nK}$ is Alice's raw key, $E^n$ is Eve's quantum side 
information, $C^n$ collects all public announcements, and 
$X_A^{\nK},X_B^{\nK}$ are the outcomes of the virtual $X$-basis 
measurements on the key-generation rounds. We make a deliberate choice of the smoothing parameter in the above expression. 

Using the fact that \cref{eq:kappaqubit} gives us an upper bound on the phase error rate and \cite[Lemma 7]{tomamichel_largely_2017}, we obtain 
\begin{equation}
\begin{aligned}
   & \smoothmax{\sqrt{\kappa (\nX,\nK,\eX)}}(X_A^{\nK} \mid X_B^{\nK})_{| \Omega(\nX,\nK,\eX)} \\
    &\leq \nK\, 
    h(\Bound(\nX,\nK,\eX)),
    \end{aligned}
\end{equation}
where $h$ is the modified binary entropy function.\footnote{That is, $h(x)$ is the binary entropy function for $x \leq 1/2$, and $1$ otherwise.} Substituting into \cref{eq:eurtemp}, we obtain
\begin{equation}
\begin{aligned}
   & \smoothmin{\sqrt{\kappa(\nX,\nK,\eX)}}(Z_A^{\nK} \mid C^n E^n)_{ | \Omega(\nX,\nK,\eX)} \\
    &\;\geq\; 
    \nK\bigl(1 - h\left(\Bound(\nX,\nK,\eX)\right)\bigr).
    \end{aligned}
\end{equation}
The extra information leaked during error-correction and error-verification can be accounted for using straightforward chain rules (see \cref{subsubsec:Hminchainrule}), to obtain 
\begin{equation}
\begin{aligned}
  &  \smoothmin{\sqrt{\kappa(\nX,\nK,\eX)}}(Z_A^{\nK} \mid \EprePA)_{ | \Omega(\nX,\nK,\eX)} \;\geq\; \\
  &
   \nK\bigl(1 - h\left(\Bound(\nX,\nK,\eX)\right)\bigr) \\
   &- \leak(\nX,\nK,\eX) - \ECost.
    \end{aligned}
\end{equation}
where recall that $\EprePA$ is all of Eve's side information before privacy amplification. This motivates setting the key length to be

\begin{equation} \label{eq:lvaluequbitbb84}
    \begin{aligned}
      &  \ell(\nX,\nK,\eX) = \max\Big\{ 0 , \nK\bigl(1 - h\left(\Bound(\nX,\nK,\eX)\right)\bigr) \\
   &- \leak(\nX,\nK,\eX) - \ECost  - 2 \log(1/\epsPA) \Big \}.
    \end{aligned}
\end{equation}

Then, applying the Leftover Hashing Lemma (see \cref{subsec:LHL}), and ignoring certain technical details related to dealing with conditioning on $\OEV$, and observations leading to zero output key length\footnote{In particular, we first remove the terms corresponding to observations that lead to zero key length, since they do not contribute to the trace distance we wish to bound. The LHL is thus applied only on the terms that correspond to non-zero bits of key length. The conditioning on EV can be appropriate properties of the entropies.}, we obtain
\begin{equation}
\begin{aligned}
  &  \sum_{\lkey}
\Pr[\Ol]\, d\left(\rho_{\KA \Efinal | \Ol} \,,\,  \omega^{\lkey}_{\KA} \otimes\rho_{\Efinal | \Ol}\right) \\
&\lessapprox \sum_{\nX,\nK,\eX} \Pr(\Omega(\nX,\nK,\eX))  \\
& \Big( 2^{-\frac{1}{2} \big(\smoothmin{\sqrt{\kappa(\nX,\nK,\eX)}}(Z_A^{\nK} | \EprePA) - \ell(\nX,\nK,\eX) + 2 \big)} \\
&+ 2\sqrt{\kappa(\nX,\nK,\eX)} \Big) \\
&\lessapprox\sum_{\nX,\nK,\eX} \Pr(\Omega(\nX,\nK,\eX)) \big(  \epsPA + 2\sqrt{\kappa(\nX,\nK,\eX)}  \big) \\
&\leq \epsPA + 2 \sqrt{\sum_{\nX,\nK,\eX} \Pr(\Omega(\nX,\nK,\eX)) \kappa(\nX,\nK,\eX) } \\
&\leq \epsPA + 2\epsAT.
\end{aligned}
\end{equation}
Here, we simply substitute the choice of $\ell(\nX,\nK,\eX)$ from \cref{eq:lvaluequbitbb84}, and use the concavity of the square root function to pull the sum inside the square root in the penultimate step. Thus we obtain that the protocol is $(\epsPA + 2\epsAT)$-secret, and therefore $(\epsPA + 2\epsAT + \epsEV)$-secure.

\subsection{Obtaining bounds on phase error rate} \label{subsec:qubitsampling}
All that remains to complete the qubit BB84 analysis is to establish
the sampling bound (\cref{eq:samplingrequbit}). The key insight is that
the structure of the protocol reduces this to a classical
sampling-without-replacement problem, to which Serfling's inequality \cite{serfling_probability_1974} (see \cref{subsec:concentrationinequalities,tab:Concentration Inequalities})
then applies directly.

Recall that in the source-replacement scheme, Alice and Bob each hold
one half of a bipartite quantum state and subsequently choose a
measurement basis, $Z$ or $X$, independently and uniformly at random
for each round. Crucially, Eve must commit to her entire attack on the
channel \emph{before} these basis choices are made, so the joint
post-attack state $\rho_{A^n B^n E^n}$ is independent of the basis
assignment.\footnote{On-the-fly announcements complicate this picture, and thus we restrict our attention to scenarios where announcements occur after signals are sent and received.}

Now condition on the event $\Omega(\nX,\nK)$, so that exactly
$\nX + \nK$ rounds were measured in the same basis. Consider the virtual
scenario in which \emph{all} of these rounds are measured in the
$X$ basis, but only $\nX$ of these were randomly selected and the error rate was measured. The 
phase error rate corresponds to the error rate in the $\nK$ rounds, while the observed error rate corresponds to the error rate in the $\nX$ rounds. It can be shown that the choice of $\nX$ rounds corresponds to choosing a random subset of $\nX$ rounds out of $\nX+\nK$ positions \cite{tupkary_phase_2024,tomamichel2012_tight}, and thus a direct application of the Serfling bound gives us that
\begin{equation}
    \begin{aligned}
        \Pr(\ephrv > \eXrv +  \gamma^{\epsAT}_\text{serf}(\nX,\nK) )_{| \Omega(\nX,\nK)} \leq \epsAT^2
    \end{aligned}
\end{equation}
where 
\begin{equation} \label{eq:gammaserfdefined}
			\gamma^{\epsAT}_\text{serf}(\nX,\nK) \coloneq \sqrt{\frac{\ln(1/\epsAT^2) (\nX+\nK) (\nX+1) }{ 2 \nK \nX^2} },
		\end{equation}
is the finite-size correction term. The conditioning on $\Omega(\nX,\nK)$ can be removed straightforwardly, to obtain the requirement statement \cref{eq:samplingrequbit}, where
\begin{equation} \label{eq:Bounddef}
    \Bound(\nX,\nK,\eX) = \eX +  \gamma^{\epsAT}_\text{serf}(\nX,\nK) 
\end{equation}

\subsection{Adding in Loss}
We now extend the analysis to the case where Alice still prepares ideal
qubit states, but Bob uses threshold detectors subject to optical loss
and dark counts. We restrict to the symmetric case in which all
detectors share the same efficiency $\eta_\mathrm{det}$ and dark-count rate,
so that there is no basis-efficiency mismatch.

Since the detectors are symmetric across both bases, the qubit squashing
map \cite{gittsovich_squashing_2014} (see \cref{subsubsec:squashing}) applies. This allows us to replace Bob's full threshold-detector POVM to an effective ideal qubit
BB84 measurement, with any double-click outcomes assigned randomly to either detector \footnote{This is necessary to satisfy the basis-efficiency mismatch assumption on realistic detectors}. We may therefore treat Bob as performing an
ideal qubit measurement on a qubit-or-vacuum input, with non-detected 
rounds simply removed from the data.
Moreover, since all detectors share the same efficiency $\eta_{\det}$, the
detector loss acts symmetrically regardless of Bob's basis choice. This allows us to invoke a standard argument \cite{zhang_entanglement_2017}: the lossy
detector is equivalent to a fictitious beamsplitter of transmissivity
$\eta_\mathrm{det}$ followed by a perfect detector, and allowing Eve to control the beamsplitter only gives her more power\footnote{This can also be formalized by thinking of the beamsplitter as a squashing map}. Absorbing this into the channel, Eve now controls a
combined lossy channel of transmissivity $\eta = \eta_\mathrm{chan}\,\eta_\mathrm{det}$,
and Bob's detector is ideal.

After these two steps, the protocol is almost identical to the qubit
BB84 scenario of the previous subsections. The only difference is that
Bob may now obtain a no-detection outcome, corresponding to rounds in
no detectors click. The security analysis can now be reduced entirely to the set of detected rounds.
On this subset, the state shared by Alice and Bob is a genuine qubit state, and
the phase-error sampling bound \cref{eq:Bounddef} and key length formula
\cref{eq:lvaluequbitbb84} apply without modification, with $\nKrv$ and
$\nXrv$ now denoting the number of \emph{detected} rounds used for key
generation and testing  respectively.

\subsection{Extension to Decoy-state BB84 with Loss}

\newcommand{\Bounddecoymin}[1]{\mathcal{B}^\text{decoy}_{\mathrm{min}-#1}}
\newcommand{\Bounddecoymax}[1]{\mathcal{B}^\text{decoy}_{\mathrm{max}-#1}}
\newcommand{\eXmu}[1]{e^{\text{obs}}_{X,\mu_{#1}}}
\newcommand{\eXmurv}[1]{\bm{\eXmu{#1}}}
\newcommand{\eXph}[1]{e^{\text{obs}}_{X,#1}}
\newcommand{\eXphrv}[1]{\bm{\eXph{#1}}}

\newcommand{\nXneqmu}[1]{n_{X_{\neq},\mu_{#1}}}
\newcommand{\nXneqmurv}[1]{\bm{\nXneqmu{#1}}}
\newcommand{\nXneqph}[1]{n_{X_{\neq},{#1}}}
\newcommand{\nXneqphrv}[1]{\bm{\nXneqph{#1}}}
\newcommand{\nXmu}[1]{n_{X,\mu_{#1}}}
\newcommand{\nXmurv}[1]{\bm{\nXmu{#1}}}
\newcommand{\nXph}[1]{n_{X,{#1}}}
\newcommand{\nXphrv}[1]{\bm{\nXph{#1}}}

\newcommand{\nO}{n_O}
\newcommand{\nOrv}{\bm{\nO}}
\newcommand{\nOmu}[1]{n_{O,\mu_{#1}}}
\newcommand{\nOmurv}[1]{\bm{\nOmu{#1}}}
\newcommand{\nOmurvvec}[1]{\bm{\vec{\nOmu{#1}}}}

\newcommand{\nOph}[1]{n_{O,{#1}}}
\newcommand{\nOphrv}[1]{\bm{\nOph{#1}}}

\newcommand{\nKmu}[1]{n_{K,\mu_{#1}}}
\newcommand{\nKmurv}[1]{\bm{\nKmu{#1}}}
\newcommand{\nKph}[1]{n_{K,{#1}}}
\newcommand{\nKphrv}[1]{\bm{\nKph{#1}}}

\newcommand{\allk}{\vec{k}}
\newcommand{\ephph}[1]{e^\text{key}_{X,{#1}}}
\newcommand{\ephphrv}[1]{\bm{\ephph{#1}}}

\newcommand{\decoyparams}{\nXmu{\allk},\nKmu{\allk},\eXmu{\allk}}

We now consider the practical setting where Alice uses a weak
coherent pulse (WCP) source with decoy intensities (see \cref{subsec:decoy_general}) and Bob uses threshold detectors. By comparing the detection
statistics across intensities, one obtains estimates of the statistics on the rounds where Alice sent single photons. Then, the arguments from the earlier section can be applied on these rounds. 

Let $O$ denote a specific outcome of a given round, and let $\nO$ denote the number of rounds that resulted in the outcome $O$. For instance, it could denote that both Alice and Bob measured in the $X$ basis and obtained a detection (in which case $\nO = \nX$).  We use $\nOmu{k}$ denote the number of rounds that resulted in the outcome $O$ where Alice used intensity $\mu_k$. We have access to this information during the protocol, and use $\nOmu{\allk}$ to denote the set of all these statistics. Let $\nOph{m}$ denote the number of rounds that resulted in the outcome $O$ where Alice prepared a state of $m$ photons. We wish to obtain bounds on $\nOph{m}$ using $\nOmu{\allk}$. Such bounds can be obtained using the fact that Alice's choices can be viewed as her first determining photon number, and then assigning an intensity to each round independently, allowing us to use Hoeffding's inequality. Note that this argument must be done carefully, see for instance \cref{subsec:eur,subsec:concentrationinequalities} and \cite[Appendix E]{tupkary_phase_2024} \cite[Methods]{curty_finitekey_2014}. We obtain

\begin{equation} \label{eq:decoy1}
			\Pr( \abs{ \nOmurv{k} - \sum_{m=0}^{\infty} p_{\mu_k | m} \nOphrv{m} } \geq \sqrt{ \frac{\nOrv}{2} \ln( \frac{2}{\epsdecoy^2}) } ) \leq  \epsdecoy^2.
		\end{equation}

For a typical decoy-state BB84 protocol, with $3$ intensities, we apply these arguments $9$ times: For every combination of the intensity and events such as the detected $X$ basis rounds, detected $X$ bounds resulting in error, and detected key generation rounds $K$. 

Then, taking suitable linear combinations of these inequalities, we obtain

\begin{equation} \label{eq:decoyreqstepone}
			\begin{aligned}
				\Pr \Big(  & \eXphrv{1} \geq \frac{	\Bounddecoymax{1} (\nXneqmurv{\allk})  }{  	\Bounddecoymin{1} (\nXmurv{\allk})}
				\quad \lor \\
                & \quad   \nXphrv{1} \leq    \Bounddecoymin{1}( \nXmurv{\allk})   \quad \lor \\
                &\quad 
				\nKphrv{1} \leq   \Bounddecoymin{1}(\nKmurv{\allk}  ) \Big) \leq  9 \epsdecoy^2
			\end{aligned}
		\end{equation}
where the functions $\Bounddecoymax{1}$ can be found in \cite[Section 5.3]{tupkary_phase_2024} \cite[Appendix A]{lim_concise_2014}. Since the phase error rate analysis can now be applied on the single photon rounds, one can combine \cref{eq:decoyreqstepone} with \cref{eq:samplingrequbit} to obtain 
\begin{widetext}
\begin{equation} \label{eq:decoyreqcombined}
			\begin{aligned}
	&\Pr(\ephphrv{1} \geq \mathcal{B}_e(\eXmurv{\allk}, \nXmurv{\allk}, \nKmurv{\allk} ) \quad \lor \quad  \nKphrv{1} \leq \mathcal{B}_1(\nKmurv{\allk} )  ) \leq 9\epsdecoy^2 +\epsAT^2, \\
    &\mathcal{B}_e(\eXmurv{\allk}, \nXmurv{\allk}, \nKmurv{\allk} ) \coloneqq \Bound \Bigg( \frac{	\Bounddecoymax{1} (\nXneqmurv{\allk})  }{  	\Bounddecoymin{1} (\nXmurv{\allk})}   ,  \Bounddecoymin{1}( \nXmurv{\allk})   ,  \Bounddecoymin{1}(\nKmurv{\allk})     \Bigg),\\
    & \mathcal{B}_1(\nKmurv{\allk} ) \coloneq  \Bounddecoymin{1}(\nKmurv{\allk}  )
			\end{aligned}
		\end{equation}
        \end{widetext}
The above equation allows us to lower bound the number of single photon key generation rounds, and upper bound the phase error rate in those rounds.

This motivates, setting the key length for the decoy-state protocol \cite[Theorem 3]{tupkary_phase_2024} to be
\begin{widetext}
\begin{equation} \label{eq:lvaluedecoy}
				\begin{aligned}
					l(\decoyparams) &\coloneq  \max\Bigg(0, \mathcal{B}_1 \left(\nKmu{\allk} \right) \left(1- h \left( \mathcal{B}_e \left(\eXmu{\allk},\nXmu{\allk}, \nKmu{\allk}   \right) \right) \right)  \\
					&- \leak(\decoyparams)					- 2\log(1/2\epsPA) - \ECost \Bigg), 
				\end{aligned}
			\end{equation}
\end{widetext}
for a protocol with $\esecure = 2 \sqrt{9\epsdecoy^2 + \epsAT^2} + \epsEV + \epsPA$.

The application of the Leftover Hashing Lemma, and the use of the Entropic Uncertainty Relations proceeds similar to the case for qubit BB84 from the earlier section, with two
notable differences.

First, the decoy bounds can fail with some small probability. For these cases, we do not apply the EUR or LHL, and instead bound the trace distance by simply noting that these events occur with low probability, and by bounding the probability prefactor term instead.
Second, since Alice's source emits pulses of varying photon number, one
must isolate the entropic contribution from single-photon rounds. This
is done via standard chain rules: given
any partition of the key-generation rounds into single-photon rounds
and the remainder,
\begin{equation}\label{eq:chainrule} \begin{aligned}
   & \smoothmin{\sqrt{\kappa}}({Z_A}^{\nKph{1}}  {Z_A}^{\nKph{\neq 1}}  |  \EprePA)_{\rho | \widetilde{\Omega} }
    \;\geq\; \\
    &  \smoothmin{\sqrt{\kappa}}({Z_A}^{\nKph{1}} |  \EprePA)_{\rho | \widetilde{\Omega}},
  \end{aligned}
\end{equation}
where ${Z_A}^{\nKph{1}}$ denotes the key bits from single-photon rounds and
${Z_A}^{\nKph{\neq 1}}$ the remainder. Note that for technical correctness, one needs to condition on an event $\widetilde{\Omega}$ above that fixes the value of $\nKph{1}$ (see \cref{subsubsec:conditioning}) . We refer the reader to the proof of \cite[Theorem 3]{tupkary_phase_2024} for additional details, and \cite{wiesemann_consolidated_2024} for some special circumstances that arise when handling the one-decoy protocol \cite{rusca_finite-key_2018}. We refer to \cref{fig:EURplots} for the plotted key rate values.

\begin{figure}
    \includegraphics[width=\linewidth]{figRMP.png}
    \caption{Key rates for the decoy-state BB84 protocol for a loss-only channel, with perfect hardware. Alice and Bob choose each basis with equal probability ($=1/2$), and send all three intensities with equal probability ($=1/3$). The intensities are given by $\mu_1=1,\mu_2= 0.1, \mu_3=0.01$, and the protocol is $\esecret=10^{-10}$-secret and $\ecorr=10^{-10}$-correct. Figure is plotted using open-source code from Ref.~\cite{tupkary_phase_2024}.}
    \label{fig:EURplots}
\end{figure}

\end{document}